\documentclass[11pt]{article}
\usepackage[a4paper,bottom=4.2cm,top=1cm,head=3cm,width=18cm,dvipdfm]{geometry}
\evensidemargin=\oddsidemargin

\usepackage[linktocpage=true,bookmarks=true,bookmarksnumbered=true,breaklinks=true,pdfpagemode=Fullscreen,pdfstartview=FitBH]{hyperref}

\usepackage[dotinlabels]{titletoc}
\usepackage{titlesec}

\usepackage{xcolor}
\usepackage{mathtools}
\usepackage{mathrsfs}
\usepackage{amsmath}
\usepackage{amssymb}
\usepackage{bm}
\usepackage{amsfonts}
\usepackage{feynmp}
\usepackage{extarrows}
\usepackage{slashed}
\usepackage{graphicx}
\usepackage{subfig}
\usepackage{dcolumn}
\usepackage{verbatim}
\usepackage{here}
\usepackage{multirow} 

\usepackage{ulem}

\def\bge{\begin{equation}}
\def\ede{\end{equation}}
\def\bga{\begin{aligned}}
\def\eda{\end{aligned}}
\def\bgp{\begin{pmatrix}}
\def\edp{\end{pmatrix}}
\def\bgs{\begin{subequations}}
\def\eds{\end{subequations}}
\newcommand{\beq}{\begin{equation}}
\newcommand{\eeq}{\end{equation}}
\newcommand{\bq}{\begin{equation}}
\newcommand{\eq}{\end{equation}}
\newcommand{\ba}{\begin{array}}
\newcommand{\ea}{\end{array}}
\newcommand{\beqa}{\begin{eqnarray}}
\newcommand{\eeqa}{\end{eqnarray}}
\newcommand{\beqs}{\begin{subequations}}
\newcommand{\eeqs}{\end{subequations}}

\def\dis{\displaystyle}

\def\B{{\mathrm{B}}}
\def\C{{\mathrm{C}}}
\def\D{{\mathrm{D}}}

\def\leqq{\leqslant}
\def\geqq{\geqslant}
\def\[{\left[}
\def\]{\right]}
\def\({\left(}
\def\){\right)}

\def\f{\frac}
\def\Lambdat{\tilde{\Lambda}}
\def\half{\frac{1}{2}}
\def\xx{x_2^{}}
\def\xxx{x_3^{}}
\def\rh{\widehat{r}}
\def\xh{\widehat{x}}

\def\Mp{M_{\mathrm{Pl}}}
\def\ii{\mathrm{i}}

\def\ga{\gamma}
\def\gaga{\gamma\gamma}
\def\cuttb{\tilde{\Lambda}_2^{}}
\def\cuttc{\tilde{\Lambda}_3^{}}
\def\TB{\overline{T}}
\def\UB{\overline{U}}

\def\End{\end{document}}

\begin{document}

 \thispagestyle{empty}
 \renewcommand{\thefootnote}{\fnsymbol{footnote}}
 \setcounter{footnote}{0}
 \titlelabel{\thetitle.\quad \hspace{-0.8em}}
\titlecontents{section}
              [1.5em]
              {\vspace{4mm} \large \bf}
              {\contentslabel{1em}}
              {\hspace*{-1em}}
              {\titlerule*[.5pc]{.}\contentspage}
\titlecontents{subsection}
              [3.5em]
              {\vspace{2mm}}
              {\contentslabel{1.8em}}
              {\hspace*{.3em}}
              {\titlerule*[.5pc]{.}\contentspage}
\titlecontents{subsubsection}
              [5.5em]
              {\vspace{2mm}}
              {\contentslabel{2.5em}}
              {\hspace*{.3em}}
              {\titlerule*[.5pc]{.}\contentspage}
\titlecontents{appendix}
              [1.5em]
              {\vspace{4mm} \large \bf}
              {\contentslabel{1em}}
              {\hspace*{-1em}}
              {\titlerule*[.5pc]{.}\contentspage}

\begin{center}
{\bf {\Large  Probing New Physics of Cubic Higgs Interaction \\[1mm]
via Higgs Pair Production at Hadron Colliders}}

\vspace*{8mm}

{\large Hong-Jian He}\,$^{a}$\footnote{Email: hjhe@tsinghua.edu.cn},\,\,
{\large Jing Ren}\,$^{b}$\footnote{Email: jren@physics.utoronto.ca},\,\,
{\large Weiming Yao}\,$^{c}$\footnote{Email: wmyao@lbl.gov}

\vspace*{2mm}

$^a$\,Institute of Modern Physics and Center for High Energy Physics, \\
Tsinghua University, Beijing 100084, China; \\
Harvard University, 1 Oxford Street, Cambridge, MA 02138, USA; \\
Institute for Advanced Study, Princeton, NJ 08540, USA
\\[1.5mm]
$^b$\,Department of Physics, University of Toronto, Toronto ON Canada M5S1A7
\\[1.5mm]
$^c$\,Lawrence Berkeley National Laboratory, Berkeley CA 94720, USA
\\

\vspace*{25mm}
\end{center}

\vspace*{3mm}

\begin{abstract}
\baselineskip 17pt
\noindent
Despite the discovery of a Higgs boson $h$(125\,GeV) at the LHC Run-1,
its self-interaction has fully evaded direct experimental probe so far.
Such self-interaction is vital for electroweak symmetry breaking, vacuum stability,
electroweak phase transition, and Higgs inflation. It is a most likely place to encode new physics
beyond the standard model. We parametrize such new physics by model-independent
dimension-6 effective operators, and study their tests via Higgs pair production
at hadron colliders. We analyze three major di-Higgs production channels at parton level,
and compare the parameter-dependence of total cross sections and kinematic distributions
at the LHC\,(14TeV) and $pp$(100TeV) hadron collider.
We further perform full simulations for the di-Higgs production channel
$\,gg\to hh \to b\bar{b}\gamma\gamma$\, and its backgrounds
at the $pp$\,(100TeV) hadron collider.
We construct four kinds of benchmark points, and study the sensitivities to probing
different regions of the parameter space of cubic Higgs interactions.
We find that for one-parameter analysis
and with a 3\,ab$^{-1}$ (30\,ab$^{-1}$)\, integrated luminosity,
the $\,gg\to hh \to b\bar{b}\gamma\gamma$\, channel can measure the SM cubic Higgs coupling and
the derivative cubic Higgs coupling to an accuracy of about
\,$13\%$ (4.2\%)\, and  \,$5\%$ ($1.6\%$),\, respectively.
\\[2mm]
PACS numbers: 12.60.Fr, 12.60.-i, 14.80.Bn
\hfill  Phys.\ Rev.\ D (2015), in Press [arXiv:1506.03302]
\end{abstract}


\newpage
\renewcommand{\thefootnote}{\\arabic{footnote}}
\setcounter{footnote}{0}
\setcounter{page}{2}


 \setcounter{footnote}{0}
 \renewcommand{\thefootnote}{\arabic{footnote}}

\baselineskip 18.5pt

\vspace*{5mm}
\section{Introduction}
\label{intro}
\label{sec:1}

The LHC discovery of the light Higgs boson $\,h\,$(125\,GeV) \cite{Higgs1}
has become a historical turning point of particle physics.
The standard model (SM)\,\cite{SM} could provide such a Higgs boson\,\cite{Higgs},
which joins three types of fundamental interactions:
(i) the gauge interactions mediated by spin-1 weak gauge bosons $(W,Z)$;
(ii) the Yukawa interactions with fermions mediated by the spin-0 Higgs boson $\,h\,$;
(iii) and the cubic and quartic Higgs self-interactions $\,h^3\,$ and  $\,h^4\,$.\,
But the type-(ii) and type-(iii) Higgs interactions are largely untested so far,
which provide the most likely place to encode new physics beyond the SM.
The current ATLAS and CMS measurements\,\cite{LHC-1new} find the Higgs boson $\,h\,$(125\,GeV) to appear
SM-like, but only have weak sensitivities to $\,h\tau\bar\tau\,$ and $\,hb\bar{b}\,$
Yukawa couplings, while even the LHC run-2 could not sensitively probe most of other
Yukawa couplings via direct detection\,\cite{peskin}.
Furthermore, the LHC has little sensitivity to probing
the type-(iii) Higgs self-interactions.
It was shown that the high luminosity LHC\,(14\,TeV)
with an integrated luminosity of $3\,\text{ab}^{-1}$ could probe the $\,h^3\,$
coupling to about 50\% accuracy \cite{Yao:2013ika,snowmassH}, and the improved analysis
could reach a sensitivity about $30\%-20\%$ \cite{h3-HL-LHC}.
With the current measurements of Higgs and top quark masses,
the SM Higgs vacuum becomes unstable around $10^{9-11}$GeV \cite{SMvacuum}
and is very sensitive to new physics \cite{VS-NP}.
So new physics is expected to enter the Higgs potential
and modify its self-interactions well below Planck scale\,\cite{SMvacuum,VS-NP}.
The Higgs self-interactions are vital
for the spontaneous electroweak symmetry breaking\,\cite{SM},
the electroweak phase transition\,\cite{EWBG}, and the Higgs inflation\,\cite{HINF}.
Hence, it is important to probe Higgs self-interactions
with precision and pin down the associated new physics deviations from the SM.

The SM Higgs sector is described by the gauge-invariant renormalizable Higgs potential,
\begin{eqnarray}
\label{eq:V}
V=- \mu^2H^\dag H + \lambda(H^\dag H)^2 ,
\end{eqnarray}
where $\,H=(\pi^+,\,\frac{1}{\sqrt{2}}(v+h+\ii\pi^0))^T$\,
is the Higgs doublet and $\,v\simeq 246$\,GeV\, denotes the vacuum expectation value (VEV).
Thus, the Higgs self-interactions take the form,
\beqa
\label{eq:V-int}
V_{\textrm{int}}^{} ~=~
\f{\,\lambda_3^{}\,}{3!}h^3 + \f{\,\lambda_4^{}\,}{4!}h^4 \,,
\eeqa
where at tree-level we have the cubic and quartic couplings of the SM Higgs boson,
$\,\lambda_3^{}=6\lambda v = 3M_h^2/v\,$ and
$\,\lambda_4^{}=6\lambda  = 3M_h^2/v^2$.\,
Hence, given the observed Higgs mass $\,M_h\simeq 125\,$GeV \cite{Higgs1},
the Higgs self-couplings are completely determined in the SM.
One could naively make a shift of Higgs coupling within the SM Higgs potential \eqref{eq:V},
$\,\lambda \to \lambda' = \lambda +\delta\lambda\,$,\, but it causes no observable effect,
because this just redefines the {\it renormalizable} Higgs coupling as $\,\lambda'\,$
no matter what value $\,\delta\lambda\,$ would take.
The nontrivial modification of Higgs couplings could only arise from higher dimensional
effective operators whose effects cannot be absorbed into the dimension-4 SM Lagrangian.

Given that the SM Lagrangian already contains all possible gauge-invariant and
renormalizable operators up to dimension-4 and no new physics is found yet,
the possible leading new physics deviations from the SM can
be generally parametrized by gauge-invariant dimension-6 effective operators
in a model-independent way \cite{BW}.
Since the Higgs potential acts as the core of spontaneous electroweak symmetry breaking
and has escaped from direct measurement so far,
it stands out as a most likely place to encode new physics beyond the SM.
Such new physics will certainly modify the Higgs self-interactions \eqref{eq:V-int},
via dimension-6 operators, which may not only shift the Higgs self-coupling itself
[due to the operator $(H^\dag H)^3$],
but also modify the structure of Higgs self-interactions
(due to the dimension-6 derivative operators).
To sensitively probe such new physics in the cubic Higgs self-coupling,
it is important to study di-Higgs production at high energy hardon colliders
\cite{hh-early,hh-rev}.\footnote{Measuring Higgs quartic coupling would be even more challenging
in the foreseeable future \cite{Plehn-h4}.}

For hadron colliders, the main di-Higgs production channels include gluon fusion production,
vector boson fusion (VBF) production, and top-pair associated production.
Over a wide energy range, the total cross section of di-Higgs production via gluon fusions
is almost 10 times larger than the other channels \cite{hh-rev}\cite{Frederix:2014hta}.
Hence, it provides the dominant di-Higgs production.
The decay mode $\,hh\to b\bar{b}\gamma\gamma$\,
has much cleaner background than others, so it has attracted efforts from both theoretical
and experimental sides \cite{Yao:2013ika}\cite{Barger:2014qva}\cite{ATLASbbgg} for studying the
potential of the high luminosity LHC\,(14TeV) and the future $pp$\,(100TeV) collider.
Other di-Higgs decay modes with larger signal rates are also explored, such as
$\,hh\to b\bar{b}\tau\tau$,\, $hh\to b\bar{b}WW^*\to b\bar{b}2\ell 2\nu$,\,  and
$hh\to b\bar{b}b\bar{b}$,\,  etc \cite{hh-other}.
Due to large backgrounds in these channels, more elaborated strategies like
boosted kinematics are needed.
Another decay mode $\,hh\to WW^*WW^*\to 3\ell3\nu jj$\, was considered
with the use of $m_{T2}^{}$ observable \cite{yan}.
Some rare final states were also explored for $pp$(100TeV) collider \cite{rare}.
In addition,  two more production channels have received recent attentions.
The top-pair associated production $\,pp\to t\bar{t}hh\,$
turns out to be complementary to gluon fusion $\,gg\to hh\,$
with $\,hh\to b\bar{b}b\bar{b}$\, final states \cite{Englert:2014uqa}.
The VBF production channel $\,pp\to hhjj\,$  receives a large contribution
from gluon fusion production in the signal region,
which makes the VBF contribution almost negligible \cite{Dolan:2013rja}.
Most previous studies for the di-Higgs production focused on the SM Higgs potential.
There are recent analyses studying the contributions of dimension-6 operators to
$\,gg\to hh\,$ with $\,hh\to b\bar{b}\gamma\gamma\,$ \cite{Azatov:2015oxa}
and $\,hh\to b\bar{b}\tau\tau$\, \cite{Goertz:2014qta}.
It was noted that certain new operators can modify kinematic distributions of the final states
as well as total cross section.
For an operator that induces $t\bar{t}hh$ coupling, the kinematics could be useful
to increase the sensitivity \cite{Chen:2014xra,Lu:2015jza}.
In general, including these operators with associated new coefficients will enlarge
the parameter space of new physics, and thus make the probe of each individual parameter
in the cubic Higgs interaction harder.
Certain simplifications are needed to reduce the large parameter space.

In this work, we will systematically analyze the new physics contributions of
dimension-6 operators to the di-Higgs productions.
For good physics reasons, we will focus on two rather unique bosonic dimension-6 operators
which contribute to the cubic Higgs coupling and build a 2-dimensional (2d) parameter space.
In particular, we will inspect the new operator that induces derivative cubic Higgs coupling,
and thus has enhanced contributions to high energy processes.
We will derive nontrivial perturbative unitarity constraints on these dimension-6 operators.
Then, we study the di-Higgs production via three major channels for probing cubic Higgs couplings.
For this, we will perform a parton level analysis at the LHC\,(14TeV) and $pp$\,(100TeV) collider.
Finally, we present a full analysis (including Delphes\,3 fast detector simulations)
for the di-Higgs production $\,gg\to hh\,$
with $\,hh\to b\bar{b}\gamma\gamma$\, at the $pp$\,(100TeV) collider.
From this we study the probe of new physics scales associated with the dimension-6 operators.
We also find nontrivial interference between different operators, which can be probed
by using relevant kinematic distributions.

This paper is organized as follows. In section\,\ref{sec:theory},
we discuss the dimension-6 operators relevant to Higgs self-interactions,
and identify the unique operators (\ref{eq:dim6scalar2}) which spans a
2d parameter space. We also motivate these operators by
nonminimal Higgs-gravity interaction.
We further study the perturbative unitarity constraints on the cutoff scales
associated with dimension-6 operators.
In section\,\ref{sec:parton}, we analyze three major di-Higgs production channels
at parton level and compare the parameter-dependence of total cross sections
and kinematic distributions. In section\,\ref{sec:full}, we perform full simulations for
$\,gg\to hh \to b\bar{b}\gamma\gamma$\, at the 100\,TeV hadron collider, and
study the sensitivity to the 2d parameter space for four benchmarks.
We conclude in section\,\ref{sec:conclusion}.
Finally, Appendix\,A discusses the redundancy of dimension-6 operators,
and Appendix\,B summarizes the loop functions of triangle and box diagrams
for the analyses of sections\,\ref{sec:3}--\ref{sec:4}.

\vspace*{2mm}
\section{New Higgs Self-Interactions from Dimension-6 Operators}
\label{sec:theory}
\label{sec:2}

\vspace*{2mm}
\subsection{Identifying Relevant Dimension-6 Operators}
\vspace*{2mm}

The SM Lagrangian is a fairly good effective theory up to gauge-invariant
renormalizable operators of dimension-4. The possible leading new physics
deviations are generally parametrized via dimension-6 effective operators,\footnote{%
If the light neutrinos turn out to be Majorana fermions,
there is a unique dimension-5 effective operator \cite{dim5},
$\,(f_{ij}^\nu/\Lambda_5^{})H^{\alpha}H^{\beta}L_i^{\alpha' T}\hat{C}L_j^{\beta'}
    \epsilon_{\alpha\alpha'}^{}\epsilon_{\beta\beta'}^{}\,$,\,
which provides Majorana neutrino masses and violates lepton number by two units.
Here $\,\hat{C}=i\gamma^2\gamma^0$\, is the charge-conjugation operator,
$(i,j)$ are flavor indices of left-handed
lepton doublet, and $(\alpha,\alpha',\beta,\beta')$ are indices of SU(2) doublets.
This dimension-5 operator is irrelevant to our current study of
the Higgs self-interactions.}
\begin{eqnarray}
\mathcal{L}_{\textrm{eff}}^{}
~=~ \sum_n\frac{f_n^{}}{\,\Lambda^2\,}\mathcal{O}_n^{}\,,
\end{eqnarray}
where $\,\Lambda\,$ characterizes the cutoff scale,
and the dimensionless coupling $\,f_n\,$ is expected to be around
$O(0.1-1)$ for each given operator (unless suppressed by extra symmetry).
The LHC Run-1 data\,\cite{LHC-1new} have constrained the 125\,GeV Higgs boson to be fairly SM-like
and found no new light particle beyond the SM.
Hence, it is well-motivated to use the standard effective theory formulation of
possible new physics effects via dimension-6 operators \cite{BW},
and assume that no other light field exists below its cutoff scale.
The full set of gauge-invariant dimension-6 operators
that modify Higgs self-interactions includes \cite{dim6},
\begin{eqnarray}
\label{eq:dim6scalar}
&&\mathcal{O}_{\Phi,1}^{} =(D^\mu H)^\dag H H^\dag(D_\mu H)\,,
\quad\hspace*{4mm}
\mathcal{O}_{\Phi,2}^{} =\frac{1}{2}\partial^{\mu}(H^\dag H)\partial_\mu(H^\dag H)\,,
\nonumber\\
&&\mathcal{O}_{\Phi,3}^{} =\frac{1}{3}(H^\dag H)^3,\quad
\hspace*{21mm}\,
\mathcal{O}_{\Phi,4}^{} =(D^\mu H)^\dag(D_\mu H)(H^\dag H)\,.
\end{eqnarray}
Among all four operators, $\mathcal{O}_{\Phi,2}^{}$\, and $\,\mathcal{O}_{\Phi,3}^{}$\,
modify scalar sectors only, while $\,\mathcal{O}_{\Phi,1}^{}$\, and \,$\mathcal{O}_{\Phi,4}^{}$\,
also contribute to gauge boson masses and couplings.
The operator $\,\mathcal{O}_{\Phi,1}^{}$ contributes to the mass $m_Z^{}$,
but not to $m_W^{}$.\, Thus, it violates the custodial symmetry and is severely constrained by
the electroweak precision parameter $\,T$\,.\,
For collider searches, it is safe to neglect the effects
of $\mathcal{O}_{\Phi,1}^{}$ \cite{Ellis:2014jta}.
With the equation of motion (EOM), there is redundancy among dimension-6 operators.
As explained in Appendix\,\ref{app:dim6EOM}, the subset operators
$(\mathcal{O}_{\Phi,2}^{},\, \mathcal{O}_{\Phi,3}^{},\, \mathcal{O}_{\Phi,4}^{})$\,
in  (\ref{eq:dim6scalar}) are not independent.
Including the SM Yukawa interactions, another type of dimension-6 operators
$\,\mathcal{O}_{\Phi,f}^{}\,$ become relevant,
\begin{eqnarray}
\label{eq:dim6sf}
\mathcal{O}_{\Phi,f}^{} \,=\, (H^\dag H)\,\overline{L}Hf_R^{} + \textrm{h.c.},
\end{eqnarray}
where $\,L=(f_L^{u},\,f_L^{d})^T\,$ denotes the $SU(2)_L$ doublet, and $f_R^{}$
the $SU(2)_L^{}$ singlet. Among all operators mentioned above,
one operator can be eliminated via EOM.
We choose to drop $\mathcal{O}_{\Phi,4}^{}$ hereafter.
Thus, we have two rather unique bosonic dimension-6 operators
$(\mathcal{O}_{\Phi,2}^{},\, \mathcal{O}_{\Phi,3}^{})$\, relevant to
the present study of Higgs self-couplings.

Next, we inspect the contributions of $(\mathcal{O}_{\Phi,2}^{},\, \mathcal{O}_{\Phi,3}^{},\, \mathcal{O}_{\Phi,f}^{})$\,
to the Higgs self-couplings, as well as the Higgs-gauge and Higgs-fermion couplings.
For later convenience, we define a dimensionless coefficient $\,x_j^{}\,$
and an effective cutoff scale $\,\Lambdat_j^{}\,$
for each operator in (\ref{eq:dim6scalar})-(\ref{eq:dim6sf}),
\begin{eqnarray}
\label{eq:xidef}
x_j^{} \,\equiv\, \frac{f_{\Phi,j}^{}v^2}{\Lambda^2}
\,\equiv\, \textrm{sign}(f_{\Phi,j}) \frac{v^2}{\,\tilde\Lambda_j^2\,}\,,
\hspace*{12mm}
\tilde\Lambda_j^{} \,\equiv\, \f{\Lambda}{\,\sqrt{|f_{\Phi,j}^{}|\,}\,}\,.
\end{eqnarray}

Since $\,\mathcal{O}_{\Phi,3}^{}\,$ is a non-derivative operator,
it only affects Higgs mass and self-couplings. In particular, it modifies
the relation between the observed Higgs mass and cubic Higgs coupling.
The derivative operator $\,\mathcal{O}_{\Phi,2}^{}\,$ induces the following
term for Higgs field,
\begin{eqnarray}
\label{eq:dim6higgs}
\mathcal{O}_{\Phi,2}^{} ~\rightarrow~ \frac{x_2^{}}{\,2v^2\,}
(h+v)^2\partial^{\mu}h\partial_{\mu}h \,,
\end{eqnarray}
with $\,\xx\,$ defined in (\ref{eq:xidef}). This modifies the Higgs kinetic term as,
\begin{eqnarray}
\mathcal{L}_{\textrm{kin}}^{}
\,=\, \frac{1}{2}\left(1+\xx\right)\partial^{\mu}h\partial_{\mu}h \,.
\end{eqnarray}
Thus, we can define the canonical Higgs field via rescaling $\,h\to\zeta h\,$,\,
with the factor,
\begin{eqnarray}
\label{eq:rescaling}
\zeta \,\equiv\, \left(1+\xx\right)^{-\half} .
\end{eqnarray}
This induces a universal modification to all Higgs couplings with SM particles.
After the normalization,
Eq.\,(\ref{eq:dim6higgs}) also generates a derivative cubic Higgs interaction,
$\,\frac{1}{v}\xx\zeta^3h\partial^\mu h\partial_\mu^{}h\,$.\,
In contrast to the SM cubic Higgs coupling, this new derivative interaction vertex
will be enhanced by the center of mass energy in high energy processes,
and thus may have distinctive kinematic feature. The modified cubic Higgs coupling is
\begin{eqnarray}
\label{eq:Chhh}
h-h-h\!: &&
- \ii\frac{\,3M_h^2\,}{v}\zeta\!\left(1-x_3^{}\zeta^2
\frac{2v^2}{3M_h^2}\right)+\ii\frac{x_2}{v}\zeta^3\!\(p_1^2+p_2^2+p_3^2\)
\nonumber\\
&&
= -\ii\frac{\,\zeta\,}{v}\left[\,3\,{{(1+\rh)}}\,M_h^2 - \xh \(p_1^2+p_2^2+p_3^2\)\,\right]
\end{eqnarray}
In the above, $M_h^{}$ is the physical mass of the Higgs boson,
which receives contributions from both the kinetic rescaling factor \eqref{eq:rescaling}
and the dimension-6 operator $\,\mathcal{O}_{\Phi,3}^{}\,$.\,
So, we deduce the Higgs mass formula,
$\,M_h^2 = M_{h0}^2[1- x_3^{}/(2\lambda)]\zeta^2\,$,\,
where $\,M_{h0}^{}=\sqrt{2\lambda\,}v\,$ is the SM Higgs mass.\,
We see that $M_h^{}$ depends on $\{\zeta, x_3^{}\}$.\,
For convenience, we replace $(\xx,\, \xxx)$ by another two independent inputs
$\,(\rh ,\, \xh )$\, which parametrize the modifications of cubic Higgs coupling
with different kinematic properties,
\begin{eqnarray}
\label{eq:rlaxc}
\rh \,\equiv\,-\xxx\,\zeta^2\frac{\,2v^2\,}{\,3M_h^2\,}\,,
\quad~~~
\xh \,\equiv\, \xx\, \zeta^2 \,.
\end{eqnarray}
%
With $\,\xh\,$,\, the rescaling factor can be rewritten as $\,\zeta=(1-\xh )^{1/2}$.\,
We also note that the operators
$\,(\mathcal{O}_{\Phi,2}^{},\,\mathcal{O}_{\Phi,3}^{})\,$ do not affect the $W$ mass
at tree-level, so the Higgs VEV is determined by the Fermi constant $G_F^{}$ as in the SM,
$\,v=\(\sqrt{2}G_F^{}\)^{-1/2}\!\simeq 246\,$GeV.\,
The modification to Higgs-gauge boson coupling only arises from rescaling the Higgs field,
\beqs
\label{eq:CVVhiggs}
\begin{eqnarray}
\label{eq:CVVh}
V_\mu-V_\nu-h\!: && \ii\frac{2m_V^2}{v}\,\zeta\,\eta^{\mu\nu}, 
\\
\label{eq:CVVhh}
V_\mu-V_\nu-h-h\!: && \ii\frac{2m_V^2}{v^2}\,\zeta^2\eta^{\mu\nu} ,
\end{eqnarray}
\eeqs
where $\,V=W,Z\,$.\,
In unitary gauge, the Higgs-fermion dimension-6 operator (\ref{eq:dim6sf}) generates following term,
\begin{eqnarray}
\mathcal{O}_{\Phi,f}^{}\,\rightarrow\,
\frac{x_f^{}}{\,2\sqrt{2}v^2\,}(v\!+\!h)^3 \bar{f}f \,.
\end{eqnarray}
This contributes to fermion mass,
$\,m_f^{}=\frac{v}{\sqrt{2}}\left(y_f^{\text{sm}}-\frac{1}{2}x_f^{}\right)$,\,
where $\,y_f^{\text{sm}}\,$ is the SM Yukawa coupling.
At the same time, it modifies the $f-\bar{f}-h$ Yukawa coupling.
Replacing $\,y_f^{\text{sm}}\,$ by $m_f^{}$, we deduce the following effective Yukawa coupling,
\begin{eqnarray}
\label{eq:Cffh}
\bar{f}-f-h\!: ~~~~~
-\ii\frac{\,\zeta\,m_f^{}\,}{v}\!\left(1-\frac{x_f^{}\,v}{\,\sqrt{2}\,m_f^{}\,}\right)
\end{eqnarray}
This operator also induces a dimension-5 vertex $\,h^2\bar{f}f\,$,
\begin{eqnarray}
\label{eq:Cffhh}
\bar{f}-f-h-h\!: ~~~~~
\ii\zeta^2\frac{3x_f^{}}{\sqrt{2}v\,} \,.
\end{eqnarray}
It contributes to the gluon fusion production
$\,gg\to hh\,$ with triangle quark loop. Since top quark is most relevant in practice,
it is natural to set $\,f=t\,$ for the present analysis.

The dimension-6 operators (\ref{eq:dim6scalar}) and (\ref{eq:dim6sf})
are subject to constraints from measurements of single Higgs production at the LHC.
The current data put the best bound on Higgs-gauge couplings  (\ref{eq:CVVh})
\cite{Ellis:2014jta}.
For a future $e^+e^-$ Higgs factory with 250\,GeV collision energy,
the sensitivity to $\,e^+e^-\to Zh$\, cross section is expected to be
$\,\delta\sigma/\sigma= \mathcal{O}(0.5\%)$ \cite{CEPC} with a 5ab$^{-1}$ integrated luminosity.
This is a direct probe of the modification of Higgs-gauge couplings and thus constraints
$|\xh|$ at $1\%$ level \cite{HiggsFactory}.
The operator ${\cal O}_{\Phi,3}^{}$ will contribute to
the $\,e^+e^-\to Zh\,$ cross section via one-loop corrections \cite{eeh3}.
From this, the sensitivity to $\lambda_3^{\textrm{sm}}$ is estimated
to be about 35\% at the $e^+e^-$ Higgs factory with a 5ab$^{-1}$ integrated luminosity.
Besides, many other dimension-6 operators can contribute to the gauge boson kinetic terms
and thus the wavefunction renormalization. This will further shift Higgs-gauge couplings,
and make the constraint on individual operators much weakened.
For top Yukawa coupling, the LHC run-2 has weak sensitivity
to probing the deviations in (\ref{eq:Cffh}).
The precision of a high-luminosity LHC (HL-LHC) is expected
to be around 10\% \cite{Brock:2014tja}.


Dihiggs production at high energy hadron colliders is an important way
to measure the cubic Higgs coupling.
The dimension-6 operators (\ref{eq:dim6scalar})-(\ref{eq:dim6sf}) contribute in
different di-Higgs production channels.
For gluon fusion and top-pair associated production, three operators
$\mathcal{O}_{\Phi,3}^{}$,\, $\mathcal{O}_{\Phi,2}^{}$\, and
$\mathcal{O}_{\Phi,t}^{}$ are relevant.\,
For the two operators that modify Higgs self-interactions, $\mathcal{O}_{\Phi,3}^{}$
contributes to the SM cubic Higgs coupling by a simple shift
(without affecting its Lorentz structure),
and is commonly studied in the di-Higgs production literature.
On the other hand, $\mathcal{O}_{\Phi,2}^{}$ induces derivative cubic Higgs coupling
(\ref{eq:Chhh}), and is rarely studied for di-Higgs production.
This operator contributes to the Higgs-gauge coupling via Higgs wavefunction renormalization
and thus may receive constraint from measuring single Higgs productions
(via vector boson fusion or Higgs-gauge-boson associated production) at colliders.
But, since other dimension-6 operators also contribute to the Higgs-gauge couplings
with interferences and possible cancellations, there is no unique constraint on
$\,\mathcal{O}_{\Phi,2}^{}$\,.\footnote{One could expect other possible precision constraints
on $\,\mathcal{O}_{\Phi,2}^{}$\, from such as the muon anomalous magnetic moment
$\,g_\mu^{}\!-2$\, at two-loop.\, Again, other new physics operators such as the
dimension-5 Pauli term $\,F_{\mu\nu}\bar{\psi}\sigma^{\mu\nu}\psi\,$
(with $\psi$ being muon field)
can also contribute to $\,g_\mu^{}\!-2$\, at tree-level and become dominant. Hence, there is
no unique constraint on $\,\mathcal{O}_{\Phi,2}^{}$\, from $\,g_\mu^{}\!-2$\, at two-loop.}\,
Hence, it is important to directly probe the derivative cubic Higgs coupling
induced by $\mathcal{O}_{\Phi,2}^{}$ via di-Higgs production,
which has distinctive kinematic features from other non-derivative operators.
For the present study, we will focus on the new physics contributions
to the Higgs self-couplings in di-Higgs production,
and drop the fermionic operator $\mathcal{O}_{\Phi,t}^{}$ (which was considered before
\cite{Chen:2014xra}\cite{Lu:2015jza} and is irrelevant to Higgs self-interactions)\footnote{%
In principle, $\mathcal{O}_{\Phi,t}^{}$ could be discriminated from
$\mathcal{O}_{\Phi,2}^{}$ and $\mathcal{O}_{\Phi,3}^{}$
by further performing a combined analysis of three di-Higgs production channels via
gluon fusion, VBF production, and top-pair associated production.
With these and the single Higgs production $gg\to h$,\, we may also discriminate
another operator $\,G^{a\mu\nu}G_{\mu\nu}^aH^\dag H$\, (which does not modify the
Higgs self-coupling).  It is possible that some other dimension-6 operators
may contribute to the backgrounds as well, but without any special cut or selection
they are expected to be much smaller than the SM backgrounds
(from the dimension-4 operators of the SM).  For clarity of the
current analysis, we assume that these additional operators are negligible.}.\,
With these considerations, we define our parameter space by identifying
the two rather unique bosonic dimension-6 operators,
\begin{eqnarray}
\label{eq:dim6scalar2}
\mathcal{O}_{\Phi,2}^{} \,=\, \frac{1}{2}\partial^{\mu}(H^\dag H)\partial_\mu(H^\dag H)\,,
\quad\quad
\mathcal{O}_{\Phi,3}^{} \,=\, \frac{1}{3}(H^\dag H)^3 .
\end{eqnarray}
In the following subsection\,\ref{subsec:NMCmodel}, we will further motivate the operator
$\,\mathcal{O}_{\Phi,2}^{}$\, from the Higgs-gravity interaction.
Then, we derive generic perturbative unitarity bound on $\mathcal{O}_{\Phi,2}^{}$
in Sec.\,\ref{subsec:Unitarity}.

\vspace*{3mm}
\subsection{Motivation from Higgs Gravitational Interaction}
\vspace*{2mm}
\label{sec:2.2}
\label{subsec:NMCmodel}

The world is apparently described by a joint effective theory of the SM and general relativity (GR)
up to accessible energy scales so far. It is important to probe the interface between the SM and GR.
With the LHC discovery of a light Higgs boson $h$\,(125GeV), there is a unique dimension-4
operator at this intersection, namely, the nonminimal interaction between the Higgs doublet $H$
and the Ricci scalar-curvature $\mathcal{R}$ \cite{NMC0},
\begin{eqnarray}
\label{eq:RHH}
S^{}_{\xi} 
\,=\, \int\!\! d^4x\,\sqrt{-g\,}\,\xi H^\dag H \mathcal{R} \,,
\end{eqnarray}
where $\,\xi\,$ is a dimensionless coupling.
With the proper normalization of graviton propagator,
it is clear that under perturbative expansion the coupling $\,\xi\,$
is always associated with the suppression factor $\,1/M_{\text{Pl}}^{2}\,$.\,
Hence, $\,\xi\gg 1\,$ can be well consistent with perturbative calculation.
The current LHC constraint on this coupling
is actually rather weak, and
$\,\xi\,$ can be as large as $O(10^{15})$ \cite{Atkins:2012yn}\cite{NMC}.
Nontrivial constraints from perturbative unitarity were derived before \cite{NMC}.
The operator \eqref{eq:RHH} has many physical applications
such as the Higgs inflation \cite{HINF}, gravitational dark matter \cite{Ren:2014mta},
and collider signatures \cite{NMC}.
Including this operator, we write the joint effective Lagrangian of the SM and GR,
\begin{eqnarray}
  \label{Action_Jordan}
  \hspace*{-3mm}
  S_{\text{J}}^{} = \int\!\!d^4x\,
  \sqrt{-g^{(J)}} \bigg[\!\left(\frac{1}{2}M^2+\xi H^\dag H\right)\mathcal{R}^{(J)}
   -\sum_j\!\frac{1}{4}F_{\mu\nu j}^aF^{\mu\nu a}_j+(D_\mu H)^\dag(D^\mu H) - V(H)\bigg] ,~~~
\end{eqnarray}
where $\,\mathcal{R}^{(J)}$\, is the Ricci scalar corresponding to the Jordan frame metric
$\,g_{\mu\nu}^{(J)}$,\, and $\,F^a_{\mu\nu i}=(W^a_{\mu\nu},\,B_{}^{\mu\nu})\,$ are
gauge field strengths of the electroweak gauge group $SU(2)_L^{}\otimes U(1)_Y^{}$.\,
In \eqref{Action_Jordan}, we can readily include the SM fermionic Lagranian
$\,{\cal L}_F^{}\,$ as well, though it is not relevant to the discussion below.
For practical applications, it is convenient to make a Weyl transformation for metric field, $\,g_{\mu\nu}^{(E)}=\Omega^2g_{\mu\nu}^{(J)}$,\, with the factor
\begin{eqnarray}
\label{WeylT}
\Omega^2 \,=\, \frac{\,M^2\!+2\xi H^\dag H\,}{\Mp^2} \,.
\end{eqnarray}
After changing variable, we write down the action with new metric $g_{\mu\nu}^{(E)}$,
\begin{eqnarray}
  \label{eq:S-Einstein}
  S_{\text{E}}^{} \!&=&\! \int\!\!d^4x\,\sqrt{-g\,}
  \bigg[  \frac{1}{2}\Mp^2 \mathcal{R}-\sum_j^{}\frac{1}{4}F_{\mu\nu j}^aF^{\mu\nu a}_j
          +\frac{3\xi^2}{\Mp^2\Omega^4}\big(\partial_\mu^{}(H^\dag H)\big)^2
 \nonumber \\
   \!&&\! +\frac{1}{\Omega^2}(D_\mu H)^\dag(D^\mu H)-\frac{1}{\Omega^4}V(H)\bigg].
\end{eqnarray}
For simplicity, we drop the superscript (E) for all geometric quantities associated with
$\,g_{\mu\nu}^{(E)}$\, here. Since the nonminimal interaction term is transformed away and
the gravity sector becomes normal, the new metric is called Einstein frame.
In this case, all effects of $\,\xi\,$ appear in matter sector and are represented
by a series of higher dimensional effective operators{\color{red}{.}}
Expanding these $\xi$-induced terms to leading order, we can deduce two relevant
dimension-6 Higgs operators
$\,\mathcal{O}_{\Phi,2}^{}$ and $\mathcal{O}_{\Phi,3}^{}$\,
from \eqref{eq:S-Einstein},
\begin{eqnarray}
\label{eq:cut1-cut2}
\frac{3}{\Lambda_{\xi 1}^2}\big(\partial_\mu^{}(H^\dag H)\big)^2
+\frac{4\lambda}{\Lambda_{\xi 2}^2}\big(H^\dag H\big)^3 ,
\end{eqnarray}
associated with two different cutoff scales,
\begin{eqnarray}
\label{eq:cut12-xi}
\Lambda_{\xi 1}=\frac{\,\Mp\,}{\xi},\quad
\Lambda_{\xi 2}=\frac{\,\Mp\,}{\sqrt{\xi\,}\,} .
\end{eqnarray}
Among dimension-6 operators in (\ref{eq:dim6scalar}), $\Lambda_{\xi 1}^{}$ is related to $\,\mathcal{O}_{\Phi,2}^{}$\,
with $\,f_{\Phi,2}^{}/\Lambda^2 = 6/\Lambda_{\xi 1}^2$,\,
which is generated due to the third term of Eq.\,\eqref{eq:S-Einstein}.
Expanding the $1/\Omega$\, factors in \eqref{eq:S-Einstein} will induce
$\,(\mathcal{O}_{\Phi,3}^{},\, \mathcal{O}_{\Phi,4}^{},\,\mathcal{O}_{\Phi,f}^{})$,\,
with a cutoff characterized by
$\,\Lambda_{\xi 2}^{}={\Mp}/{\sqrt{\xi}\,}\,$.
For the operator $\,\mathcal{O}_{\Phi,3}^{}$, we have
$\,f_{\Phi,3}^{}/\Lambda^2 = 12/\Lambda_{\xi 2}^2$\,.\,
The other two operators $\mathcal{O}_{\Phi,4}^{}$ and
$\mathcal{O}_{\Phi,f}^{}$\, are induced from $1/\Omega$\, expansion
with the following coefficients,
\beqa
- \frac{2}{\,\Lambda^2_{\xi 2}\,}O_{\Phi,4}^{}
+ \frac{4 y_f^{}}{\,\Lambda^2_{\xi2}\,}O_{\Phi,f}^{} \,,
\eeqa
where $\,y_f^{}\,$ is the SM Yukawa coupling of the fermion $f$.\,
The effective theory with such Higgs-gravity interactions can be viable
for a wide range of $\,\xi$\,.\, To be relevant to collider physics,
we need $\,\xi\gg 1\,$ \cite{NMC}\footnote{As we clarified before\,\cite{NMC},
in this effective theory formulation, we do not concern any
detail of the UV completion above the cutoff $\Lambda_{\xi 1,2}^{}$.\,
There are many well-motivated TeV scale quantum gravity theories on the market.
For instance, extra dimensional models with compactification scale at
$\,\Lambda_{\xi 1}^{}=O(\text{10TeV})$\,
will reveal the Kaluza-Klein modes at energies above this scale,
and other related UV dynamics may show up above this cutoff as well.},\,
which implies
$\,\Lambda_{\xi 1}^{2}\ll\Lambda_{\xi 2}^{2}$.\,
Hence, in this effective theory, the operator $\,\mathcal{O}_{\Phi,2}^{}\,$
will give dominant contributions, while other operators $\,\mathcal{O}_{\Phi,3}^{}$\,
and $\,(\mathcal{O}_{\Phi,4}^{},\,\mathcal{O}_{\Phi,f}^{})$
are negligible.

\vspace*{3mm}
\subsection{Constraints from Perturbative Unitarity}
\label{subsec:Unitarity}
\vspace*{2mm}

In this subsection, we derive perturbative unitarity bound on the parameter space
of dimension-6 operators defined in (\ref{eq:dim6scalar2}).
We analyze the longitudinal weak boson scattering and top-Higgs scattering in high energy regime.
We find that their scattering amplitudes are largely enhanced
by $E^2$ and $E^1$ contributions from the derivative cubic Higgs couplings,
and would eventually violate perturbative unitarity with the increase of scattering energy.
This places an upper bound on the validity range of perturbation expansion of the effective
theory, above which certain nonperturbative dynamics or new physics have
to set in.\footnote{Since the joint effective theory of SM+GR
is nonrenormalizable and its UV completion is unknown,
any naive partial resummation within this effective theory itself cannot give reliable
unitarity restoration\,\cite{NMC}. Hence, the perturbative unitarity bound is important
for such nonrenormalizable effective theories.}\,
For the current analysis, we will derive perturbative unitarity bounds
for both types of processes.
Since the energy dependence of $\,gg\to hh\,$ amplitude is rather mild,
it cannot place better bounds than the processes mentioned above,
and thus needs no consideration here.

\begin{figure}[!h]
  \centering%
  \subfloat{%
    \includegraphics[width=14.5cm]{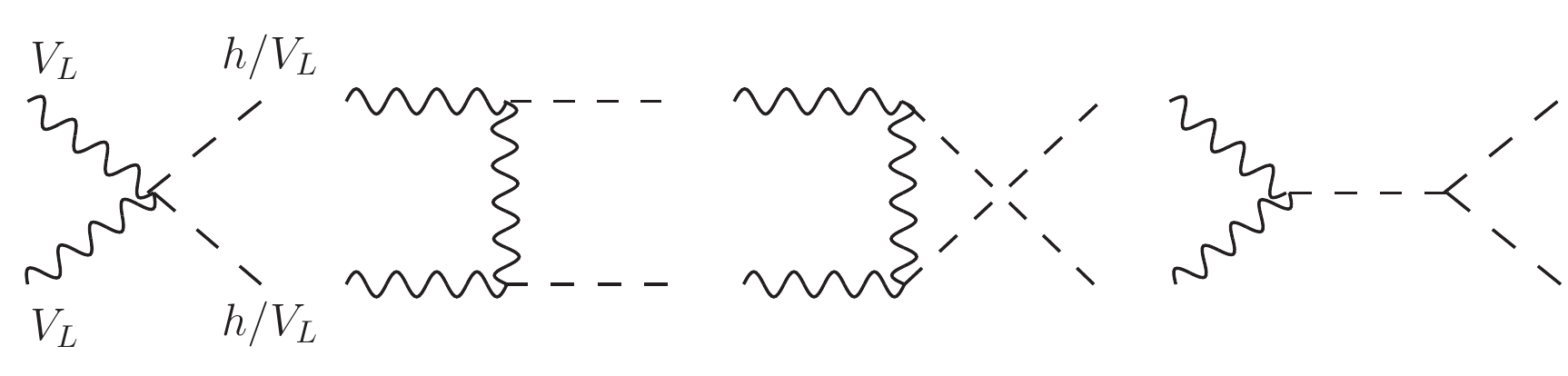}}
\caption{Longitudinal weak boson scattering processes,
$V_LV_L\to h h\,(V_LV_L)$,\, where $V=W^\pm,Z^0$. The crossing channels
also give gauge-Higgs boson scattering.}
\label{fig1:VVhh}
\end{figure}

\begin{figure}[t]
  \centering%
    \includegraphics[height=7cm,width=9cm]{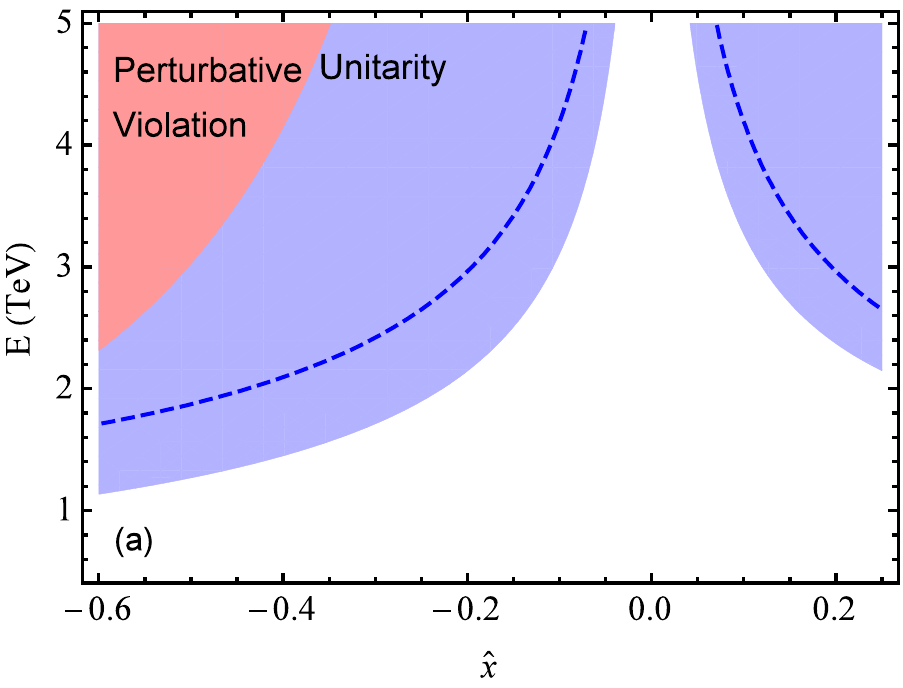}\\[-2mm]
    \includegraphics[height=6.5cm,width=16.5cm]{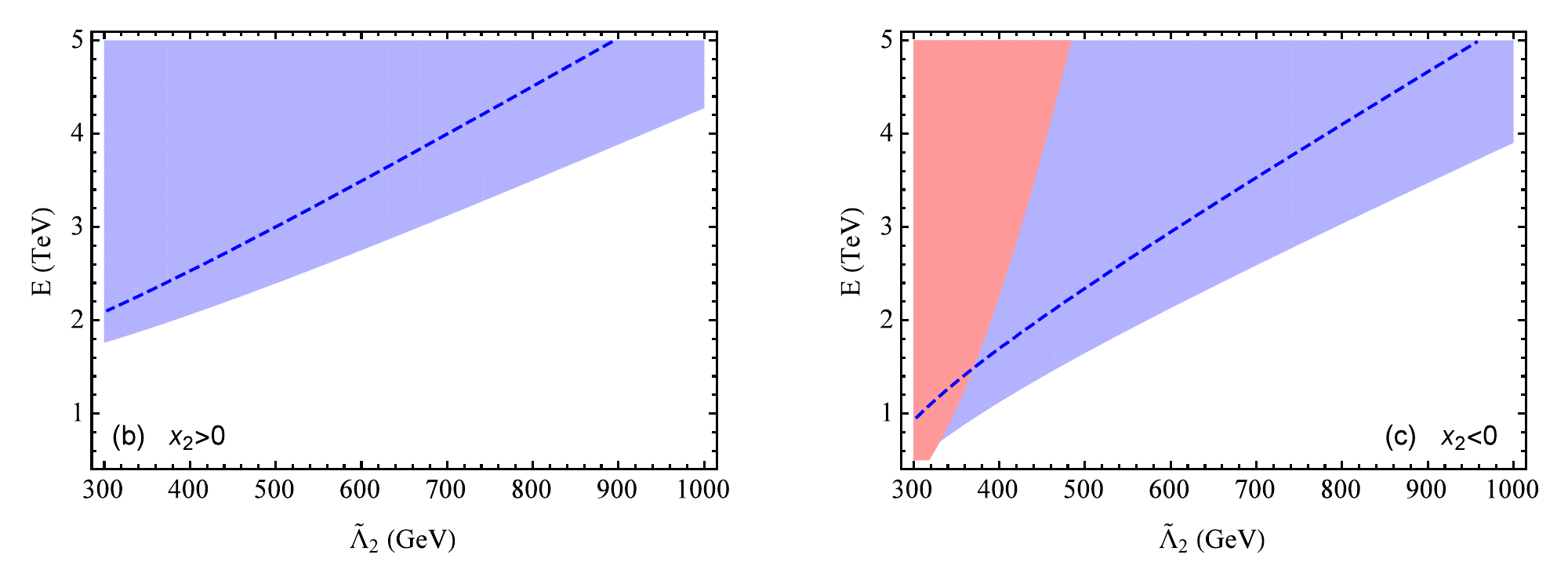}
\vspace*{-4mm}
\caption{Perturbative unitarity violation region from weak boson scattering (blue)
and top-Higgs scattering (red) as a function of $\,\xh\,$ in plot-(a),
and a function of $\,\tilde\Lambda_2^{}\,$ in plots (b) and (c)
for $\,x_2>0\,$ and $\,x_2<0\,$,\, respectively.}
\label{fig:PUbound}
\label{fig:2}
\end{figure}

Fig.\,\ref{fig1:VVhh} depicts the longitudinal weak boson scattering
$V_LV_L\to h h\,(V_LV_L)$\,
and the gauge-Higgs boson scattering in the crossing channels.
The new physics of dimension-6 operators modifies the Higgs-gauge coupling
and the Higgs self-couplings, which can induce nonzero $\mathcal{O}(E^2)$ enhancement
in the scattering amplitudes \cite{NMC}. Since dimension-6 operators are gauge-invariant,
the longitudinal-Goldstone boson equivalence theorem (ET)\,\cite{ET} can be established \cite{NMC}.
Hence, the same $E^2$ enhancement must show up in the corresponding Goldstone boson scattering
amplitudes. To derive the optimal unitarity constraints on dimension-6 operators,
we perform a coupled channel analysis of all electrically neutral channels for
Goldstone boson and Higgs boson scatterings, with initial/final states
$\{\left|\pi^+\pi^-\right\rangle,\, \frac{1}{\sqrt{2}}\left|\pi^0\pi^0\right\rangle,\,
 \frac{1}{\sqrt{2}}\left|hh\right\rangle,\, \left|\pi^0 h\right\rangle \}$.\,
We compute the relevant leading scattering amplitudes at $\mathcal{O}(E^2)$\,,
\begin{eqnarray}
\label{eq:pipiAmp}
\mathcal{T}[\pi^+\pi^-\!\to \pi^+\pi^-] &\!=\!&
\xh\,\frac{\,(1\!+\!\cos\theta)E^2\,}{2v^2},
\quad
\nonumber\\[0mm]
\mathcal{T}[\pi^+\pi^-\!\to \pi^0\pi^0] &\!=\!&
\xh\,\frac{\,E^2\,}{v^2} \,,
\nonumber\\
\mathcal{T}[\pi^+\pi^-\!\to hh] &\!=\!&
\mathcal{T}[\pi^0\pi^0\!\to hh]\,=\, \xh \,(1\!-\!\xh )\frac{\,E^2\,}{v^2} \,,
\\
\mathcal{T}[\pi^0 h\to \pi^0 h] &\!=\!&
-\xh \,(1\!-\!\xh )\frac{\,(1\!-\!\cos\theta)E^2\,}{2v^2} \,,
\nonumber\\[0mm]
\mathcal{T}[\pi^0\pi^0\!\to \pi^0\pi^0] &\!=\!&  \mathcal{O}(E^0) \,,
\quad~~~~
\mathcal{T}[hh\to hh]\,=\,\mathcal{O}(E^0) \,,
\nonumber
\end{eqnarray}
where $\,E\,$ is the center-of-mass energy and $\,\theta\,$ denotes the scattering angle.
With these, we compute the corresponding partial wave amplitudes,
\begin{eqnarray}
\label{eq:aldef}
a_\ell^{}(E) \,=\,
\frac{1}{32\pi}\int^1_{-1}\!\!d \cos\theta\,
P_\ell^{}(\cos\theta)\mathcal{T}(E,\theta) \,.
\end{eqnarray}
We perform a coupled channel analysis for the in/out states
$\{\left|\pi^+\pi^-\right\rangle,\, \frac{1}{\sqrt{2}}\left|\pi^0\pi^0\right\rangle,\,
 \frac{1}{\sqrt{2}}\left|hh\right\rangle,\, \left|\pi^0 h\right\rangle \}$,\,
Then, we can derive the following $\,4\!\times\! 4\,$ matrix for
the $s$-wave amplitudes at $\,\mathcal{O}(E^2)\,$,
\begin{eqnarray}
\label{eq:a0W}
a_0^{}(E)\,=\, \frac{\xh\, E^2}{\,32\pi v^2\,}
\left(\begin{array}{cccc}
1 & \sqrt{2} & \sqrt{2}(1\!-\!\xh ) & 0\\[1mm]
\sqrt{2} & 0 & 1\!-\!\xh & 0 \\[1mm]
\sqrt{2}(1\!-\!\xh) & 1\!-\!\xh & 0 & 0 \\[1mm]
0 & 0 & 0 & -(1\!-\!\xh)  
\end{array}
\right) .
\end{eqnarray}
For a sizable $|1\!-\!\xh |$,\,
the scattering amplitudes with Higgs in initial/final states
have dominant contributions. We deduce the following eigenvalues,
\begin{eqnarray}
a^{\textrm{diag}}_0(E) \,=\,
\frac{\xh\, E^2}{\,32\pi v^2\,}\,
\textrm{diag}\!\left(
1\!+\!\sqrt{1\!+\!3(1\!-\!\xh )^2},\, 1\!-\!\sqrt{1\!+\!3(1\!-\!\xh )^2},\, -(1\!-\!\xh ),\, -1
\right),
\end{eqnarray}
and impose the $s$-wave unitarity condition $\,|\textrm{Re}\,a_0^{}|<\frac{1}{2}$\,
on the maximal eigenvalue.
Thus, we derive the perturbative unitarity bound on the scattering energy,
%
\begin{eqnarray}\label{eq:PUB1}
E ~<~ \Lambda_{\textrm{U}1}^{} \,=\,
\frac{\sqrt{16\pi}\,v}{\,[\,|\xh |(1\!+\!\sqrt{1\!+\!3(1\!-\!\xh )^2}\,)\,]^{1/2}\,} \,.
\end{eqnarray}
We plot this bound $\,\Lambda_{\textrm{U}1}^{}\,$ as a function of $\,\xh\,$
in Fig.\,\ref{fig:PUbound}(a), where the blue region (including the overlap with red region)
denotes perturbative unitarity violation. We also show the dependence of unitarity bound
on the effective cutoff $\,\tilde\Lambda_2^{}\,$ of the dimension-6 operator
$\mathcal{O}^{}_{\Phi,2}$ in plots (b) and (c)
for $\,\xx >0\,$ and $\,\xx <0$\,,\, respectively.
For small $\,|\xh |$,\, we find
$\,\Lambda_{\textrm{U}1}^{}\approx \sqrt{16\pi/3\,}\,\tilde\Lambda_2^{}\,$
at leading order. As mentioned earlier,
$\,\xh\,$ could be constrained by measurements of Higgs-gauge coupling in single Higgs production
due to its contribution to the rescaling of Higgs kinetic term.
But, given the contributions from other dimension-6 operators to the Higgs-gauge coupling
and their possible large cancellations with that of $\,\mathcal{O}^{}_{\Phi,2}$,\,
the Higgs-gauge coupling could be SM-like while $\,\xx\,$ is
more or less free from this constraint.
In this case, $\mathcal{O}^{}_{\Phi,2}$ still receives general perturbative unitarity bound
from high energy scattering processes involving its induced derivative Higgs self-couplings,
even though Higgs rescaling effect may be negligible.
Thus, we derive the corresponding unitarity bound by turning off
the Higgs rescaling effect in (\ref{eq:a0W}),
\begin{eqnarray}
\label{eq:PUB1p}
E ~<~ \Lambda'_{\textrm{U}1} \,=\,
\frac{\sqrt{\,16\pi}\,v\,}{\,3^{1/4}_{}|\xh |^{1/2}_{}\,} \,.
\end{eqnarray}
We depict the upper bound \eqref{eq:PUB1p} by the blue dashed curve in
Fig.\,\ref{fig:PUbound}(a)-(c).
We see that $\,\Lambda'_{\textrm{U}1}$ turns out to be weaker than the
bound $\,\Lambda_{\textrm{U}1}$.\,
In the later analysis of di-Higgs production via vector boson fusion, we will
be conservative and select signal events by imposing the weaker bound
$\,\sqrt{\hat{s}}<\Lambda'_{\textrm{U}1}$.\,

\begin{figure}[!h]
  \centering%
    \includegraphics[width=13cm]{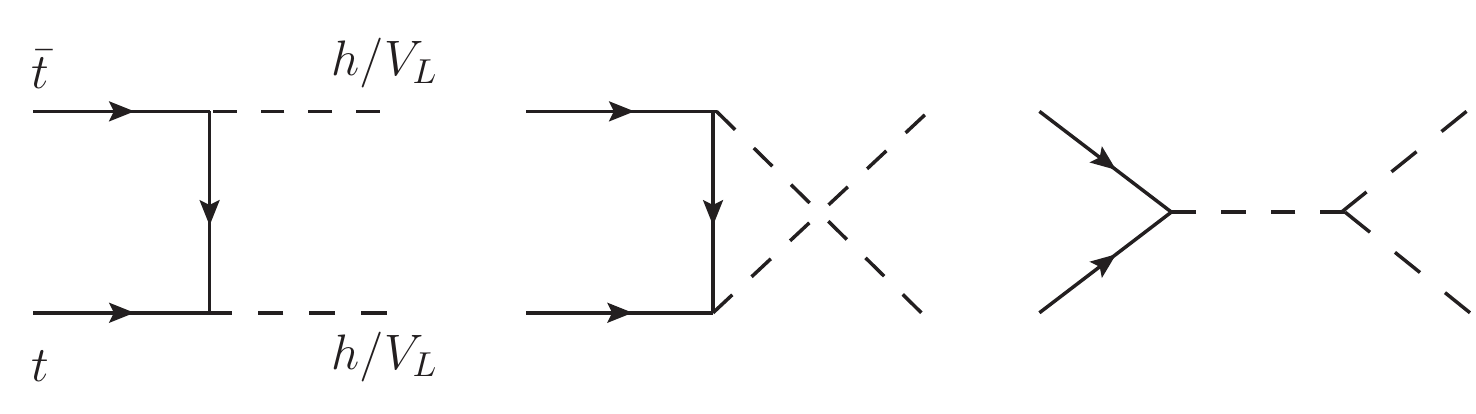}
\caption{Feynman diagrams for $\,\bar{t}t\to h h\,(V_L^{}V_L^{})$\, scattering,
where $V=W^\pm,Z^0$.}
\label{fig:tthh}
\end{figure}

Fig.\,\ref{fig:tthh} presents Feynman diagrams for \,$\bar{t}t\to h h\,(V_L^{}V_L^{})$\, scattering,
where $V=W^\pm,Z^0$.
In high energy limit, the leading amplitudes from dimension-6 operator $\mathcal{O}^{}_{\Phi,2}$
are enhanced by $E^1$ terms. According to equivalence theorem,
we compute the leading amplitudes with final state $V_L^{}V_L^{}$ replaced by
the corresponding Goldstone bosons.
Among all contributions, the amplitudes with $t/u$-channel quark-exchange and the SM Yukawa coupling approach constant in high energy limit. Only the $s$-channel Higgs-exchange
with cubic derivative Higgs coupling in (\ref{eq:Chhh}) gives the $\mathcal{O}(E^1)$
asymptotical behavior and may violate perturbative unitarity.
To derive the optimal bound, we define the spin-0 and color-singlet
helicity state of top-quark pair, i.e.,
$\left|\bar{t}t\right>_s^{}=\dis
 \frac{1}{\sqrt{2N_c}}\sum_{a=1}^{N_c}
 \(\left|\bar{t}_+^{a}t_+^{a}\right>-\left|\bar{t}_-^{a}t_-^{a}\right>\)$\,
\cite{Dicus:2004rg}.\,
Thus, we compute the scattering amplitudes at the leading $\mathcal{O}(E^1)$\,,
\begin{eqnarray}
\mathcal{T}[\left|\bar{t}t\right>_s^{}\!\!\to \left|\pi^+\pi^-\right>]
&\!\!=\!\!& \mathcal{T}[\left|\bar{t}t\right>_s^{}\!\!\to \left|\pi^0\pi^0\right>]
\,=\, -\sqrt{6}\,\xh\, \zeta^2 \frac{\,m_t^{} E\,}{v^2} \,,
\nonumber\\
\mathcal{T}[\left|\bar{t}t\right>_s^{}\!\!\to \left|hh\right>]
&\!\!=\!\!& -\sqrt{6}\,\xh\, \zeta^4 \frac{\,m_t^{} E\,}{v^2},
\quad~~~~
\mathcal{T}[\left|\bar{t}t\right>_s^{}\!\!\to \left|\pi^0h\right>] \,=\, \mathcal{O}(E^0)\,,
\hspace*{7mm}
\end{eqnarray}
where $\,E\,$ is center of mass energy.
To optimize the unitarity bound,
we can further define an $O(4)$ singlet final state
$\,\left|S\right>=\frac{1}{\sqrt{8}}
 \(2\left|\pi^+\pi^-\right>+\left|\pi^0\pi^0\right>+\left|hh\right>\)\,$.\,
Hence, we derive,
\beqa
\mathcal{T}[\left|\bar{t}t\right>_s^{}\!\!\to \left|S\right>]
\,=\, -\xh (1-\xh)(4-\xh)\frac{\,\sqrt{3}m_t^{}E\,}{2v^2} \,.
\eeqa
Using (\ref{eq:aldef}), we compute the partial wave amplitude and impose
the $s$-wave unitarity condition
\,$|\textrm{Re}\,a_0^{}|<\frac{1}{2}$\,.\,
With these we deduce the perturbative unitarity bound on scattering energy $\,E\,$,
\begin{eqnarray}
\label{eq:PUB2}
E ~<~ \Lambda_{\textrm{U}2}^{}
= \frac{16\pi v^2}{\,\sqrt{3}m_t^{}\,}
  \frac{1}{\,|\xh (1\!-\!\xh)(4\!-\!\xh)|\,} \,.
\end{eqnarray}
We plot the upper bound $\,\Lambda_{\textrm{U}2}^{}$  in Fig.\,\ref{fig:PUbound}
(red contours)
as a function of $\,\xh\,$ and $\,\tilde\Lambda_2^{}\,$,\, respectively.
For small $|\xh|$, we derive
$\,\Lambda_{\textrm{U}2}^{}\approx 4\pi\tilde\Lambda_2^2/\sqrt{3}m_t^{}\,$
at leading order. It is clear that the bound from top-Higgs scattering is much
weaker than that of the weak boson scattering.

\vspace*{4mm}
\section{Higgs Pair Production at Hadron Colliders}
\label{sec:parton}
\label{sec:3}

In this section, we study di-Higgs production for the effective theory defined
in Eq.\,(\ref{eq:dim6scalar2})
at both the LHC\,(14\,TeV) and future $pp$\,(100TeV) collider.
There are two new parameters $(\xx,\, \xxx)$,\,
which may be reparametrized as $(\xh,\, \rh)$ in Eq.\,(\ref{eq:rlaxc}) for convenience.
The major di-Higgs production channels at high energy hadron collider include
gluon fusion production ($g g \to h h$), top-pair associated production ($p p \to t\bar{t}hh$),
and VBF production ($p p \to h h j j$).
In the following, we analyze these production channels at parton level,
and compare their differences in total cross sections and in kinematical distributions
over the parameter space of $(\xh,\, \rh)$.\,

\begin{figure}[t]
  \centering%
    \includegraphics[height=6.5cm,width=7.8cm]{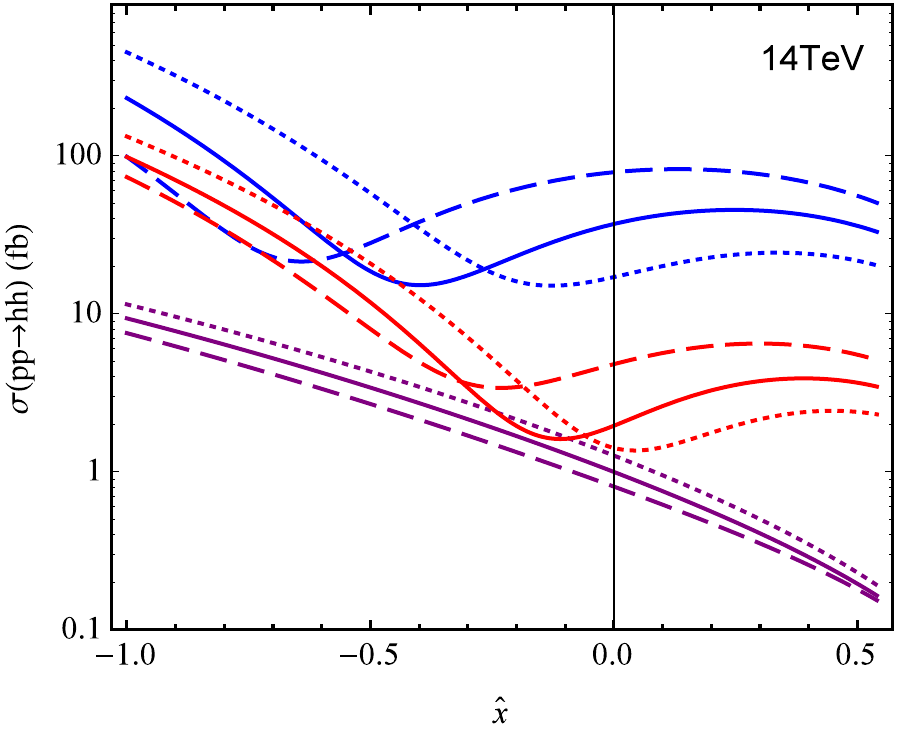}\,\,\,
    \includegraphics[height=6.5cm,width=7.9cm]{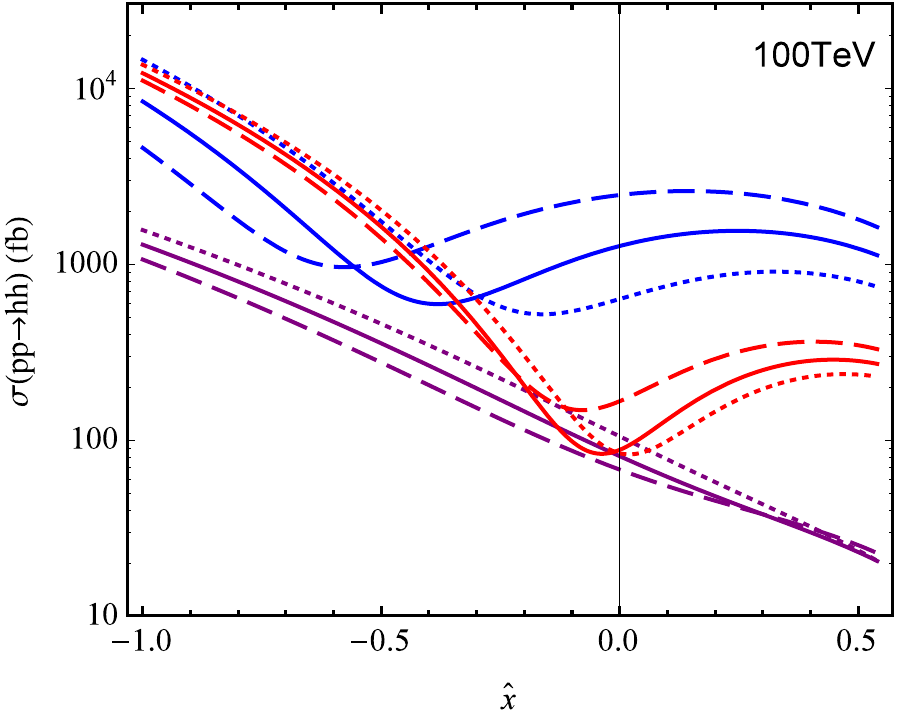}
\caption{Cross sections of di-Higgs production via gluon fusion (blue),
top-pair associated production (purple) and vector boson fusion (red)
at the LHC\,(14TeV) (left plot) and $pp$(100TeV) collider (right plot).
For each production channel, the (dashed, solid, dotted) curves depict cross sections
as functions of $\,\xh\,$ under three inputs of $\,\rh =(-1,\, 0,\, 1)$.}
\label{fig:XSgghhxc}
\label{fig:4}
\end{figure}

With the modified cubic Higgs couplings \eqref{eq:Chhh} from dimension-6 operators,
we derive the differential cross section for gluon fusion production,
\begin{eqnarray}
\label{eq:gghhXC}
\frac{d\hat{\sigma} (gg\to hh)}{d\hat{t}} \,=\, \frac{G_F^2\alpha_s^2}{\,512(2\pi)^3\,}
\,\zeta^4\!\left[\left|
\left((1\!+\!\rh\,)\,\frac{3m_h^2}{\hat{s}-m_h^2}
-\xh\,\frac{\,\hat{s}+2m_h^2\,}{\hat{s}-m_h^2}\right)
F_\triangle^{}+F_\Box^{}\right|^2+|G_\Box^{}|^2\right]\!,
\hspace*{7mm}
\end{eqnarray}
where $(\hat{s},\, \hat{t})$ are partonic Mandelstam variables, and
$(F_\triangle^{},\, F_\Box^{},\, G_\Box^{})$
are loop functions given in Appendix\,\ref{app:loopfactors},
The new contributions from $\,\xx$ (\,$\xh$\,) arise in two ways.
The first is an overall rescaling factor $\,\zeta^4\,$ of the cross section,
and the second is contributed by the derivative cubic Higgs coupling.
The parameter $\,\xxx$\, only appears in $\,\rh\,$,\,
which shifts the SM cubic Higgs coupling.
We generate signal events by MadGraph\,5 \cite{MG5}.\footnote{To include
the effect of finite top mass,
we use the model file $\textrm{SMEFT}\_\,\textrm{FF}\_\,\textrm{bt}$
for events generation. The relevant code is available at https://cp3.irmp.ucl.ac.be/projects/madgraph/wiki/HiggsPairProduction.}\,
The QCD corrections can be significant\,\cite{Dawson:1998py}, but they are insensitive
to the structure of cubic Higgs coupling,\footnote{As shown in Ref.\cite{Grober:2015cwa},
for various dimension-6 operators relevant for gluon fusion production,
the correction to the $K$-factor is around several per cent.}\,
so we normalize the cross section at $\,(\rh,\,\xh)=(0,\,0)\,$
to the SM NLO prediction\,\cite{Frederix:2014hta} and implement the same $K$-factor
for full parameter space of $\,(\rh,\,\xh)$.\,
For gluon fusion, we have $\,K=(2.27,\, 1.44)$\, for $\sqrt{s}=(14,\,100)$\,TeV.
But, for analyzing the {\it ratio} of the cross section over that of the SM,
it is rather insensitive to the $K$-factor.
We perform numerical fits for the total cross sections over the range
$\,-1\leqq \rh \leqq 1$\, and $\,-1\leqq\xh \leqq 0.5$\,
at both LHC\,(14TeV) and $pp$\,(100TeV) collider,
\beqs
\label{eq:GFXsecfit}
\begin{eqnarray}
\label{eq:GFXsecfit-a}
\left.\frac{\sigma(gg\to hh)}{\sigma(gg\to hh)_{\textrm{sm}}^{}}\right|_{14\textrm{TeV}}^{}
&\!\!\!\!\!=\!\!& (1\!-\!\xh )^2
\left(1-0.83\,\rh +3.7\,\xh+0.29\,\rh^2\!+4.2\,\xh^2\!-2.0 \,\rh\,\xh\right),~~~~~
\\[3mm]
\label{eq:GFXsecfit-b}
\left.\frac{\sigma(gg\to hh)}{\sigma(gg\to hh)_{\textrm{sm}}^{}}\right|_{100\textrm{TeV}}
&\!\!\!\!\!=\!\!& (1\!-\!\xh)^2\left(1-0.72\,\rh +3.6\,\xh
+0.22\,\rh^2\!+4.3\,\xh^2\!-1.7 \,\rh\,\xh\right).
\hspace*{11mm}
\end{eqnarray}
\eeqs
This shows that the fitted cross section ratio is not sensitive to the variation of
collision energy from $\sqrt{s}=14$\,TeV to $\sqrt{s}=100$\,TeV.
This is mainly due to the \,$m_f^2/\hat{s}$\, suppression in the loop functions
$\,F_\triangle^{}\,$ and $\,F_\Box^{}\,$ under high energy limit
[cf.\ Eq.\,\eqref{eq:Fd-Fb-bigE}].
Expanding (\ref{eq:gghhXC}) around the SM values $(\rh,\,\xh)= (0,\,0)$,\,
we derive the $\,\rh\,$ dependence,
$\,d(\sigma/\sigma_{\textrm{sm}}^{})/d \rh \,\simeq -(0.7-0.8)$\,.\,
For the parameter $\,\xh\,$,\,
the prefactor \,$(1\!-\!\xh)^2=\zeta^4$\, in Eq.\,(\ref{eq:GFXsecfit})
comes from rescaling factors of Higgs fields $\,hh\,$ in the final state,
while $\,\xh\,$ in the second parentheses is contributed by the derivative cubic Higgs coupling.
We note that these two contributions have some cancellation.
For $\,\xh >0\,$ ($\,\xh <0\,$), the contribution from derivative coupling
interferes constructively (destructively) with the SM part of $(\rh,\,\xh)=(0,\,0)$,\,
while the total cross section is suppressed (enhanced) by the Higgs rescaling factor
\,$(1\!-\!\xh)^2$\,.\, The blue curves in Fig.\,\ref{fig:XSgghhxc} depict
the gluon fusion cross sections at $pp$(14TeV) and $pp$(100TeV).
The (dashed,\,solid,\,dotted) curves present the cross sections varying with $\,\xh\,$,\,
under inputs $\,\rh =(-1,\,0,\,1)\,$,\, respectively.
From Fig.\,\ref{fig:XSgghhxc}, we see that
the di-Higgs production cross sections from gluon fusion
exhibit a minimum in the $\,\xh < 0\,$ region, and the location of this minimum
varies with the input value of $\,\rh\,$.\,

\begin{figure}[t]
  \centering%
    \includegraphics[height=6.5cm,width=8cm]{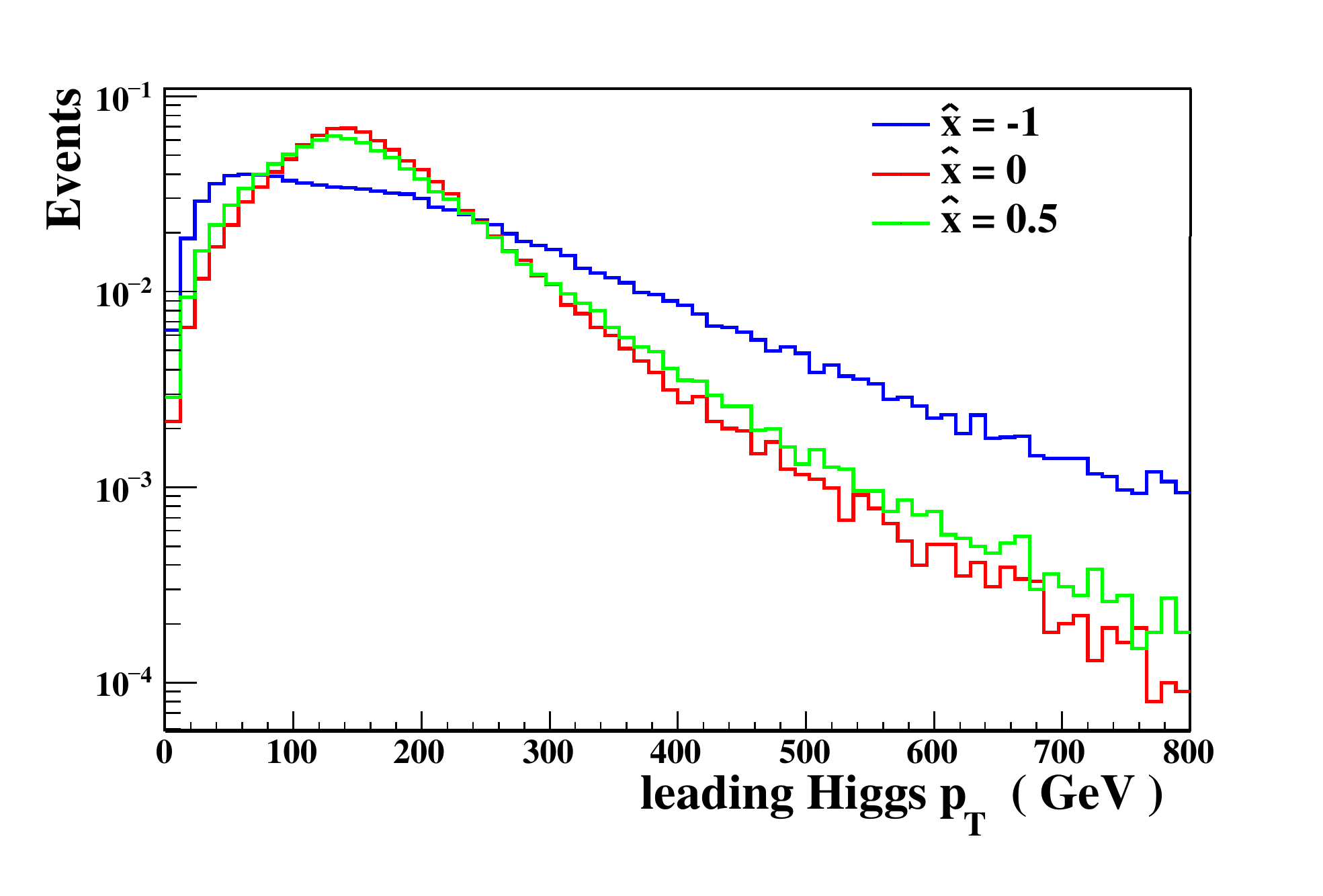}
    \includegraphics[height=6.5cm,width=8cm]{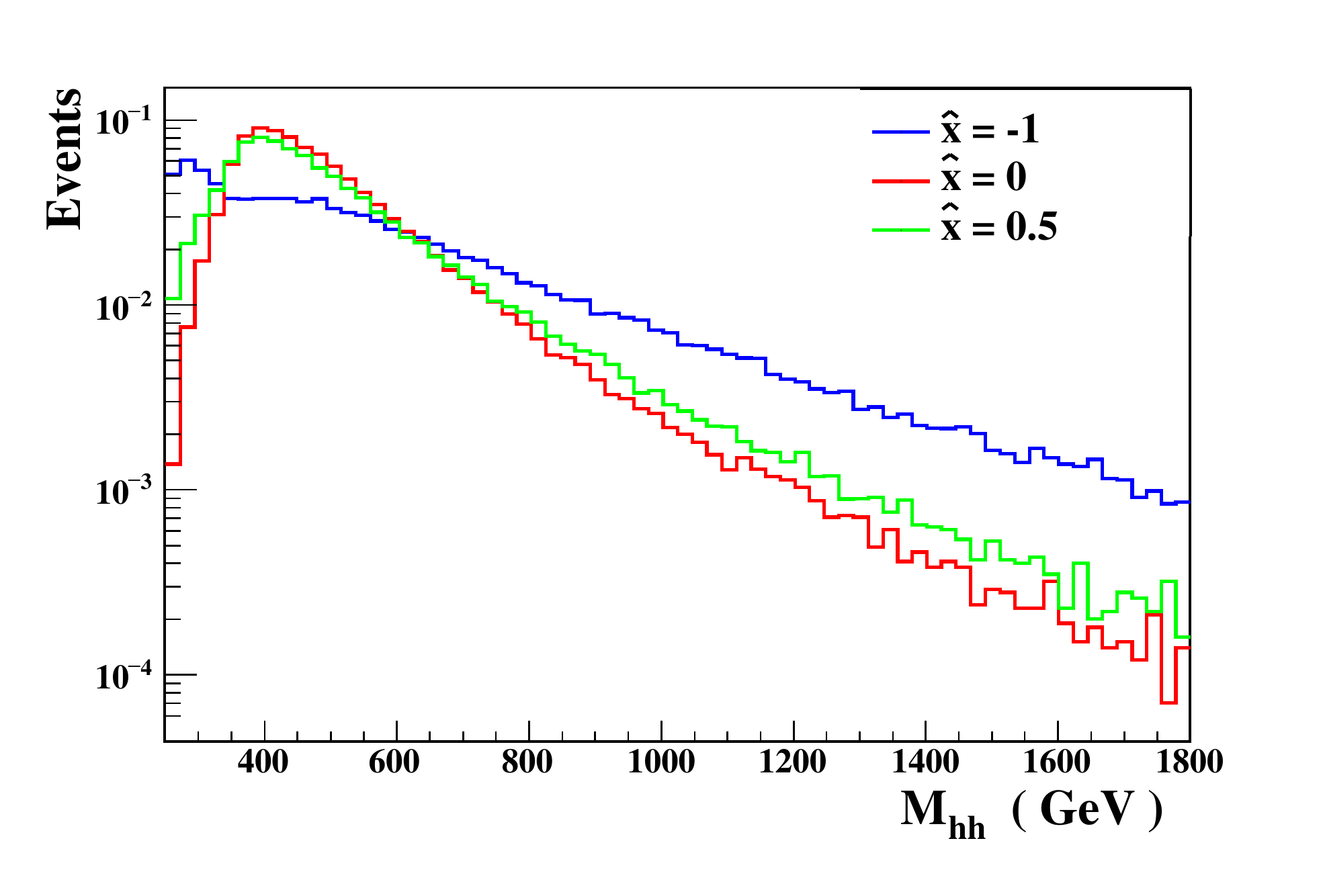}
  \\[-4mm]
    \includegraphics[height=6.5cm,width=8cm]{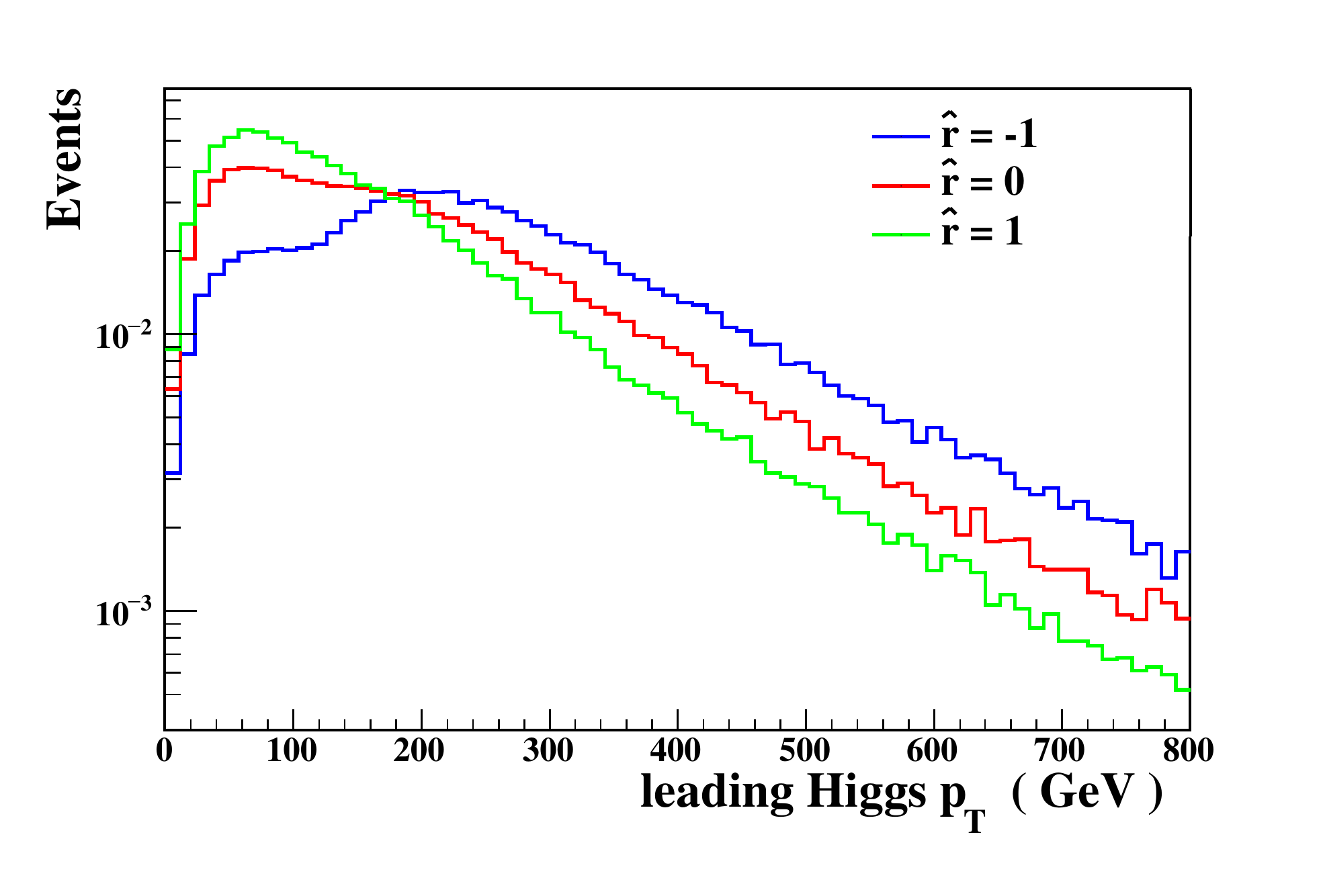}
    \includegraphics[height=6.5cm,width=8cm]{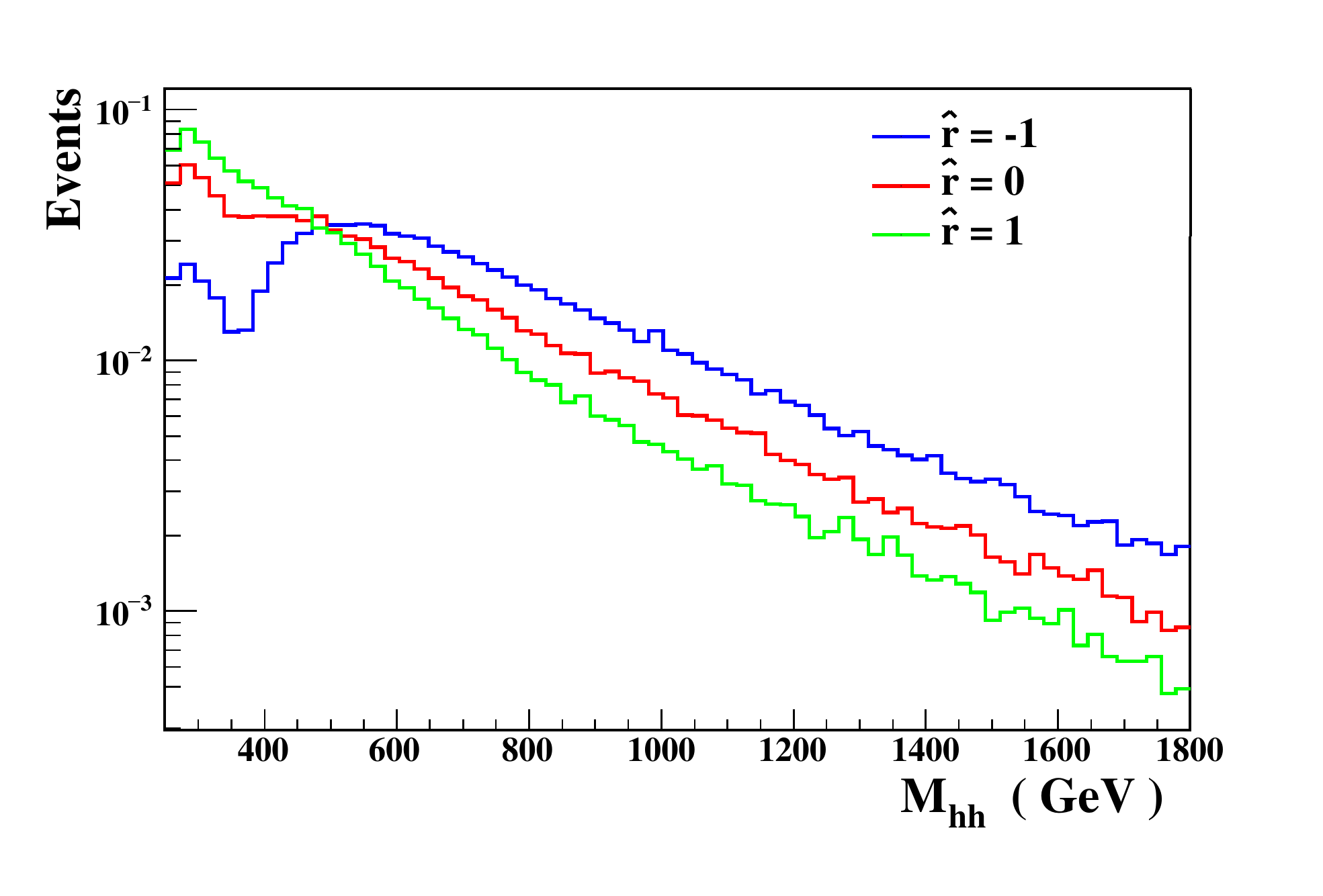}
\vspace*{-4mm}
\caption{Parton level distributions of $\,gg\to hh\,$ for the leading Higgs $p_T^{}$
(1st column) and the invariant-mass $M_{hh}^{}$ (2nd column) at $pp$(100TeV).
In the first row, we input $\,\rh =0$,\, and the (blue,\,red,\,green) curves
correspond to $\,\xh =(-1,\,0,\, 0.5)$.\, In the second row,
we input $\,\xh =-1$,\, and the (blue,\,red,\,green) curves
correspond to $\,\rh =(-1,\, 0,\, 1)$.
}
\label{fig:gghhparton}
\label{fig:5}
\end{figure}

In Fig.\,\ref{fig:gghhparton}, using MadAnalysis-5 package\,\cite{MA5},
we present the normalized kinematic distribution of
final state Higgs bosons at $pp$(100\,TeV) collider.
The first column display the leading Higgs $p_T^{}(h)$
distributions; while the second column depict
the $M_{hh}^{}$ invariant-mass distributions of the Higgs pair.
The shapes of distributions at the LHC(14TeV) and $pp$\,(100TeV) collider
have some similarity since the cross section only has mild energy dependence.
In the first row of Fig.\,\ref{fig:gghhparton}, we have input
$\,\rh =0$\,,\,  and the (blue,\,red,\,green) curves correspond to
$\,\xh =(-1,\, 0,\, 0.5)$;\,
while the second row has $\,\xh=-1\,$,\,
and (blue,\,red,\,green) curves correspond to $\,\rh =(-1,\, 0,\, 1)$.\,
For the parameter range $\,\xh <0$\,,\, there is large cancellation between
the SM box-loop diagram and the triangle-loop diagram with $s$-channel Higgs and
new derivative cubic Higgs coupling over the intermediate momentum range.
This makes the distribution more sensitive to $\,\rh\,$.\,
In particular, if we turn off the SM cubic Higgs coupling by setting $\,\rh =-1$\,,\,
the events are mostly populated in large $\,p_T^{}\,$ and $\,M_{hh}^{}\,$ regions,
as shown by the blue curves in the second row of Fig.\,\ref{fig:gghhparton}.
For $\,\xh >0$\, and \,$\rh >-1$\,,\, all contributions add to each other constructively,
and the normalized distributions do not significantly change.\footnote{In passing,
Ref.\cite{DicusHS} studied interference between the SM cubic Higgs coupling and
other SM contributions in a few di-Higgs production channels,
with focus on the variations of collision energy and parton distribution function.}

%


\begin{figure}[t]
\centering%
    \includegraphics[height=6.5cm,width=8cm]{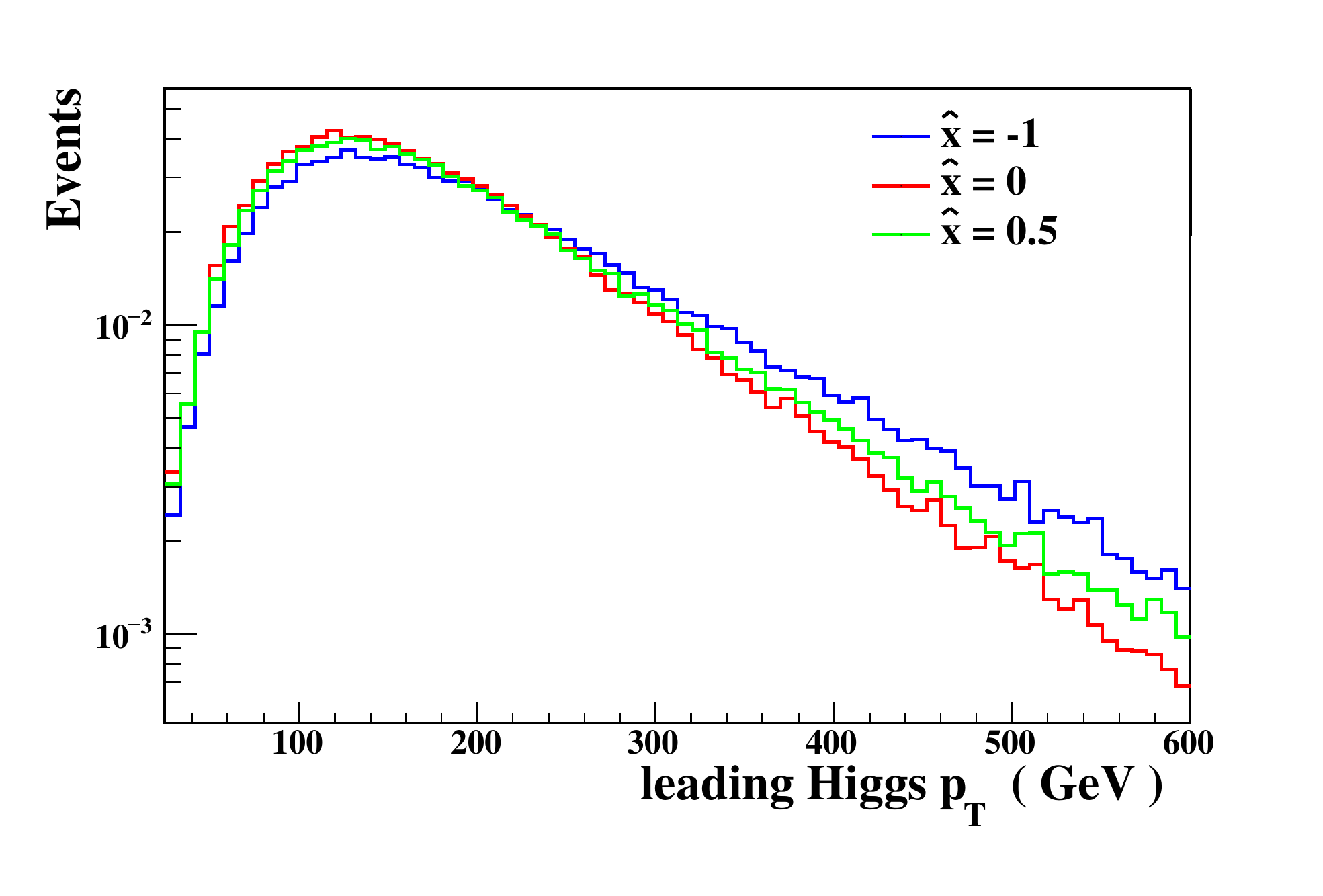}
    \includegraphics[height=6.5cm,width=8cm]{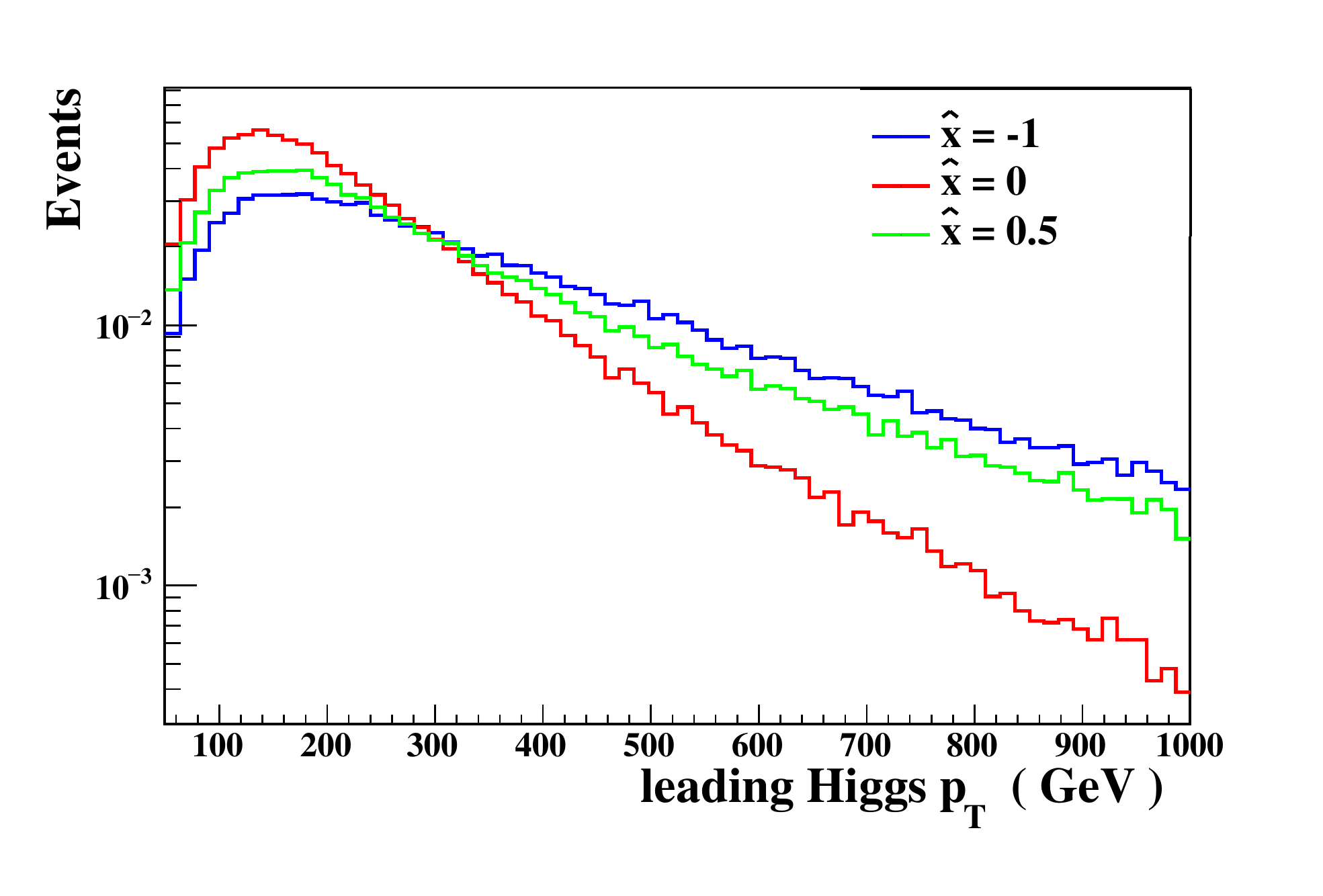}
\\[-5mm]
    \includegraphics[height=6.5cm,width=8cm]{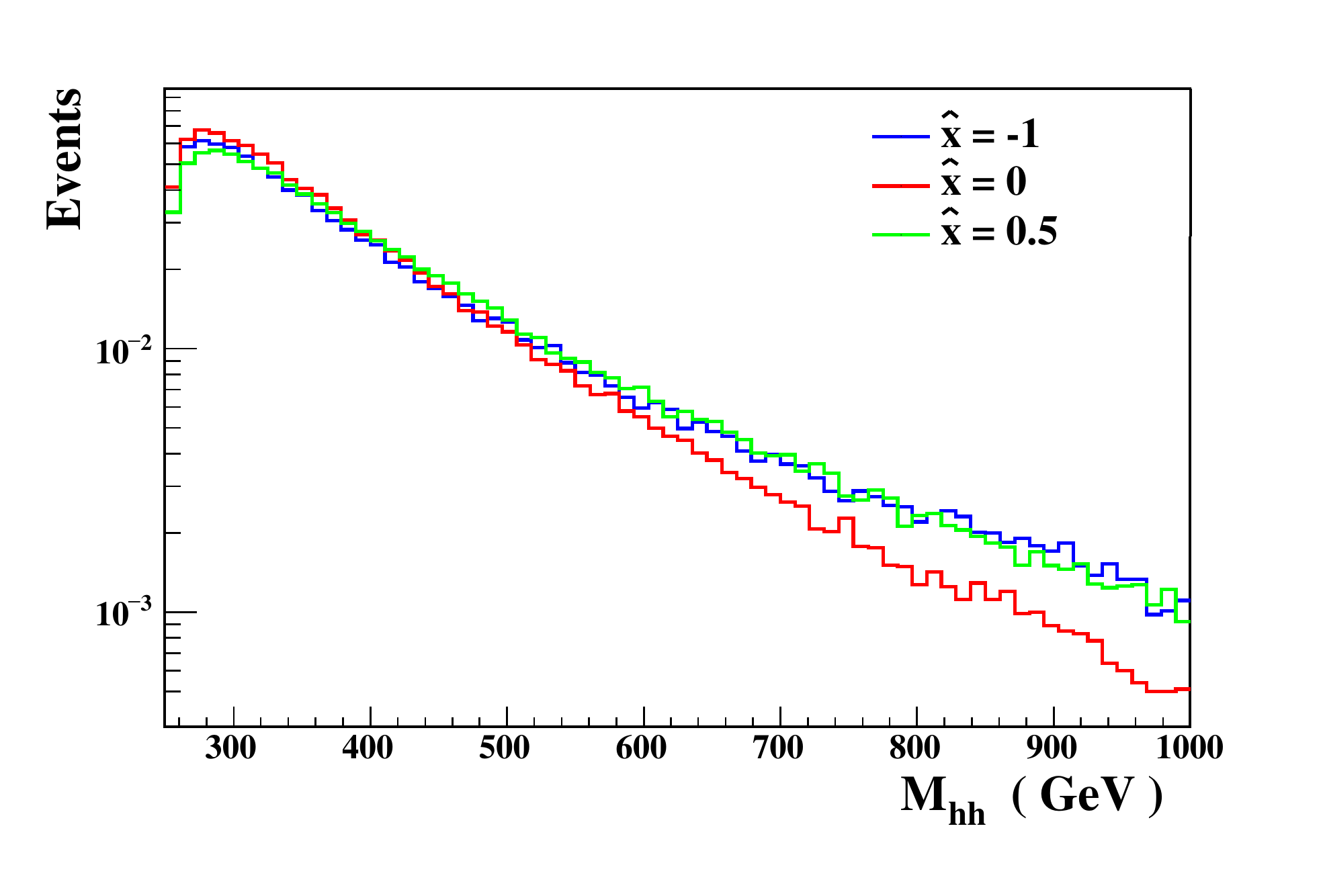}
    \includegraphics[height=6.5cm,width=8cm]{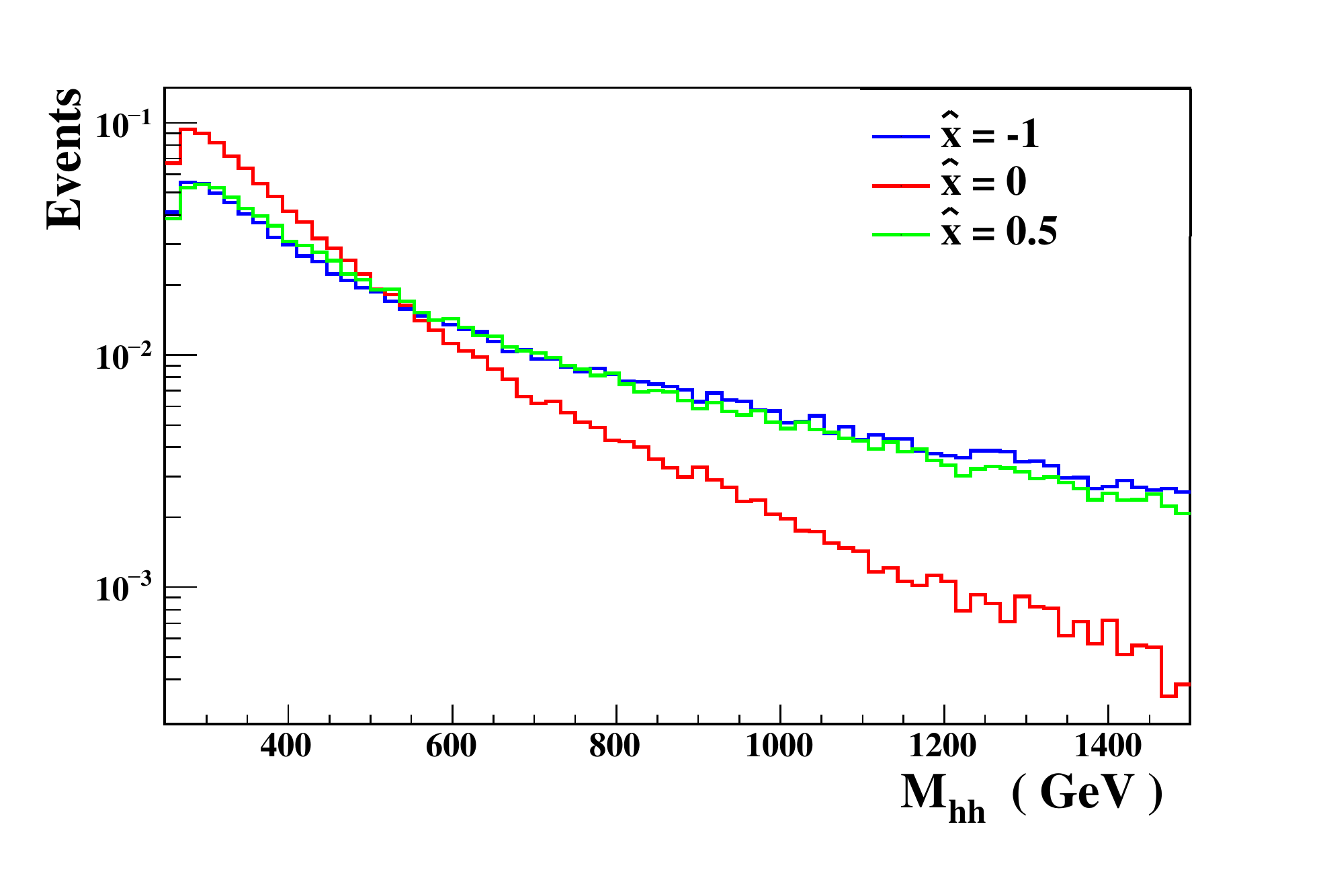}
\vspace*{-4mm}
\caption{Parton level distribution of $\,pp\to t\bar{t}hh$\, for the leading $p_T^{}$ distributions
 of Higgs boson (1st row), the invariant-mass distributions $M_{hh}^{}$ (2nd row)
 at the LHC(14TeV) (1st column) and the $pp$(100TeV) (2nd column).
 In each plot, we set $\,\rh =0\,$,\, and input $\,\xh=(-1,\, 0,\, 0.5)$\, which
 correspond to (blue,\,red,\,green) curves, respectively.}
\label{fig:pptthhparton}
\label{fig:6}
\vspace*{2mm}
\end{figure}

\vspace*{1mm}

Next, we consider the top-pair associated di-Higgs production.
The dependence of its cross section on
$(\xx,\, \xxx)$ can be reparametrized in terms of $(\xh,\,\rh)$,\,
and is similar to that of (\ref{eq:gghhXC}).\,
We generate the signal events by MadGraph\,5, and find the factor $\,K=1.2\,$ for
total cross sections at both the LHC\,(14TeV) and $pp$\,(100TeV) collider \cite{Frederix:2014hta}.
We perform numerical fits of total cross sections for
$\,-1\leqq\rh \leqq 1$\, and $\,-1\leqq\xh \leqq 0.5$\,,\,
which are summarized as follows,
\beqs
\label{eq:topXsecfit}
\begin{eqnarray}
\left.\frac{\sigma(pp\to \bar{t}thh)}
{\sigma(pp\to \bar{t}thh)_{\textrm{sm}}^{}}\right|_{14\textrm{TeV}}
&\!\!\!\!\!=\!\!&
(1\!-\!\xh )^2\!\left(1+0.23\,\rh -0.73\,\xh +0.04\,\rh^2\!+0.60\,\xh^2\!-0.26\,\rh\,\xh\right)\!,
\\[3mm]
\left.\frac{\sigma(pp\to \bar{t}thh)}
{\sigma(pp\to \bar{t}thh)_{\textrm{sm}}^{}}\right|_{100\textrm{TeV}}
&\!\!\!\!\!=\!\!&
(1\!-\!\xh)^2\!\left(1+0.23\,\rh -0.80\,\xh +0.07\,\rh^2\!+2.2\,\xh^2\!-0.54\,\rh\,\xh\right)\!.
\hspace*{14mm}
\end{eqnarray}
\eeqs
In comparison with the di-Higgs production via gluon fusion,
the cross section of top-pair associated production is less sensitive to
the change of either $\,\rh\,$ or $\,\xh\,$, due to the dominance of diagrams irrelevant to
Higgs self-interaction.\,
But, the $\xh$-dependence of top-pair associated production cross section
is much more sensitive to the increase of collision energy than that of
gluon fusion production, especially for the $\,\xh^2\,$ term.
We note that the derivative cubic Higgs coupling term
interferes destructively (constructively)
with the SM $t/u$-channel exchange of top for $\,\xh >0\,$ ($\,\xh<0\,$).
Hence, this process is complementary to gluon fusion production.
In Fig.\,\ref{fig:XSgghhxc}, we plot the total cross sections of
top-pair associated di-Higgs production by purple curves.
It is much suppressed in $\,\xh >0\,$ region due to the overall rescaling factor
$\,(1\!-\!\xh)^2=\zeta^4\,$.\,
For $\,\rh >0$,\, it adds positive contributions to that of the SM,
and makes the test of $\,\rh\,$ easier \cite{Englert:2014uqa}.

We present in Fig.\,\ref{fig:pptthhparton}
the normalized kinematic distributions for top-pair associated di-Higgs production at parton level.
The first row shows the leading \,$p_T^{}\,$ distribution of the Higgs boson,
and the second row depicts the di-Higgs invariant-mass ($M_{hh}^{}$) distribution,\,
at the LHC\,(14TeV) (in first column) and $pp$(100TeV) collider (in second column).
At the LHC, they are rather insensitive to the variation of
$\,(\xh ,\, \rh )$.\, However, the $pp$\,(100TeV) collisions significantly
improve the sensitivity to $\,\xh\,$.\,
In comparison with the di-Higgs production via gluon fusion in Fig.\,\ref{fig:gghhparton},
the top-pair associated production is more sensitive to the derivative
cubic Higgs coupling, with more signal events populated in the higher $\,p_T^{}$\,
and larger $\,M_{hh}^{}\,$ region.
To maintain perturbative unitarity,  we will require signal events to obey
$\,M_{hh} < \Lambda_{\textrm{U}2}^{}$\,,\, where  $\,\Lambda_{\textrm{U}2}^{}\,$
is derived in (\ref{eq:PUB2}).
We find that this bound at $\,\xh =0.5$\, is too weak to be relevant;
and there are $77\%$ ($97\%$) signal events passed this requirement
for  $\,\xh =-1$\, at $\,\sqrt{s}=100$\,TeV\,(14\,TeV).


\vspace*{2.5mm}
Finally, we turn to the di-Higgs production via vector boson fusion,
$\,pp\to V^*V^*jj\to hh jj$\,.\,
Its cross section depends on $(\xx,\, \xxx)$
through the overall rescaling factor $\zeta^4$,\,
the modified (SM-like) cubic Higgs coupling $\,\rh\,$,\,
and the new derivative cubic Higgs couplings $\,\xh\,$.\,
We generate signal events by Madgraph\,5 with electroweak process,
and apply the following VBF cuts to two tagging jets \cite{VBFcuts},
\beqs
\label{eq:topXsecfit}
\begin{eqnarray}
\textrm{14\,TeV:} &&
2<|\eta_j^{}|<5\,,\,~~ \eta_{j_1}^{}\!\!\cdot\eta_{j_2}^{}<0\,,\,~~
p_{T,j}^{} > 25\,\textrm{GeV},\,~~ M_{jj}^{} > 500\,\textrm{GeV};~~~~
\\
\textrm{100\,TeV:} &&
2<|\eta_j^{}|<5\,,\,~~ \eta_{j_1}^{}\!\!\cdot \eta_{j_2}^{}<0\,,\,~~
p_{T,j}^{}>50\,\textrm{GeV},\,~~  M_{jj}^{}>1000\,\textrm{GeV}.~~~~
\end{eqnarray}
\eeqs
We perform numerical fits to the total cross section for
$\,-1\leqq\rh \leqq 1\,$ and $\,-1\leqq\xh \leqq 0.5$\,,\,
and derive the following,\footnote{%
For the {\it ratio} between the VBF signal cross sections in Eq.\,\eqref{eq:VBFXsecfit},
we note that the QCD $K$-factors are largely cancelled out
and thus this ratio is very insensitive to the $K$-factors.}
\beqs
\label{eq:VBFXsecfit}
\begin{eqnarray}
\left.\frac{\sigma(pp\to hhjj)}{\sigma(pp\to hhjj)_{\textrm{sm}}}\right|_{14\textrm{TeV}}
&\!\!\!\!\!=\!\!\!&
(1\!-\!\xh )^2\!\left(1-0.86\,\rh +4.8\,\xh +0.59\,\rh^2\!+16\,\xh^2\! -4.6\,\rh\,\xh\right)\!,
\\[3mm]
\left.\frac{\sigma(pp\to hhjj)}{\sigma(pp\to hhjj)_{\textrm{sm}}}\right|_{100\textrm{TeV}}
&\!\!\!\!\!=\!\!\!&
(1\!-\!\xh )^2\!\left(1-0.47\,\rh +4.6\,\xh +0.42\,\rh^2\! +38\,\xh^2\! -4.1\,\rh\,\xh\right)\!.
\hspace*{14mm}
\end{eqnarray}
\eeqs
We find that the cross section of VBF channel is much more sensitive to $\,\xh\,$
than the other two processes discussed above.
After implementing VBF cuts, the cross section is dominated by longitudinal
weak boson scattering, and the amplitude has $E^2$ enhancement which greatly improves
the signal sensitivity to $\,\xh\,$ in $pp$(100TeV) collisions.
In Fig.\,\ref{fig:XSgghhxc}, we present the cross sections by red curves at
the LHC\,(14TeV) and $pp$(100TeV) collider.
These cross sections are normalized to the NLO SM prediction\,\cite{Frederix:2014hta}
at $(\rh,\, \xh)=(0,\,0)$.\,
The total cross sections become comparable to that of the gluon fusion production
over large negative $\,\xh\,$ region, but their dependence on $\,\rh\,$ is weaker.

\begin{figure}[t]
  \centering%
    \includegraphics[height=6.5cm,width=8cm]{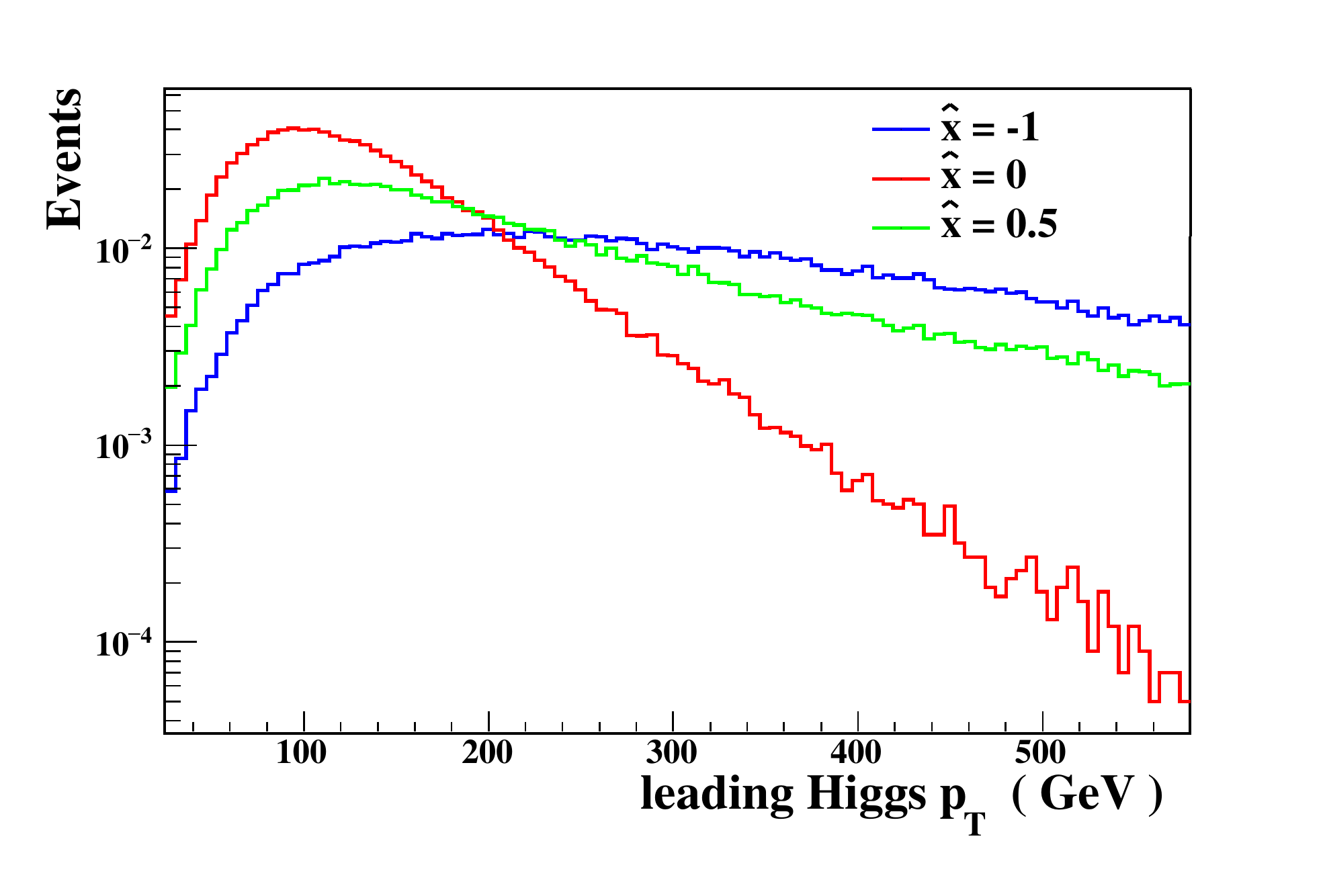}
    \includegraphics[height=6.5cm,width=8cm]{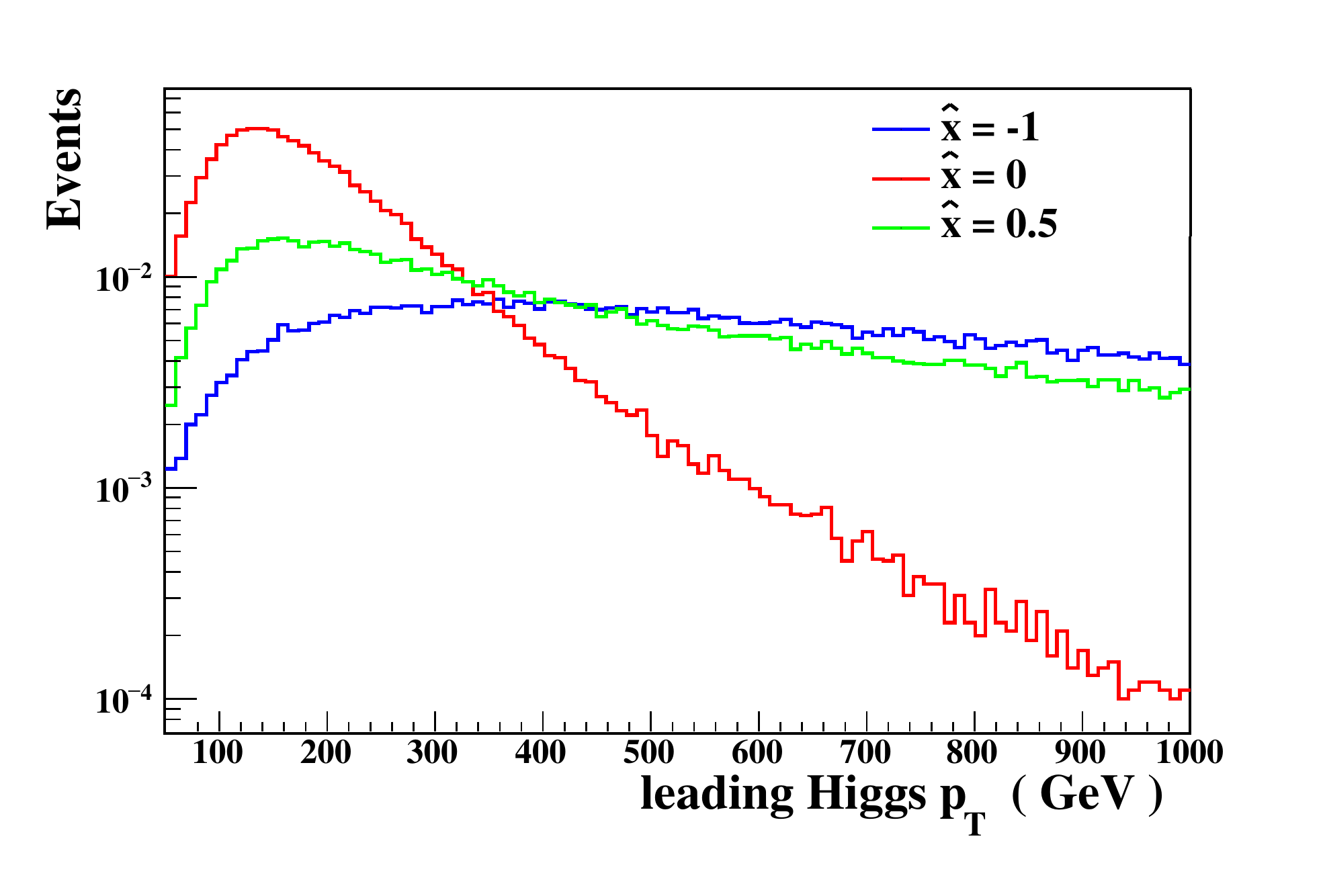}
  \\[-4mm]
    \includegraphics[height=6.5cm,width=8cm]{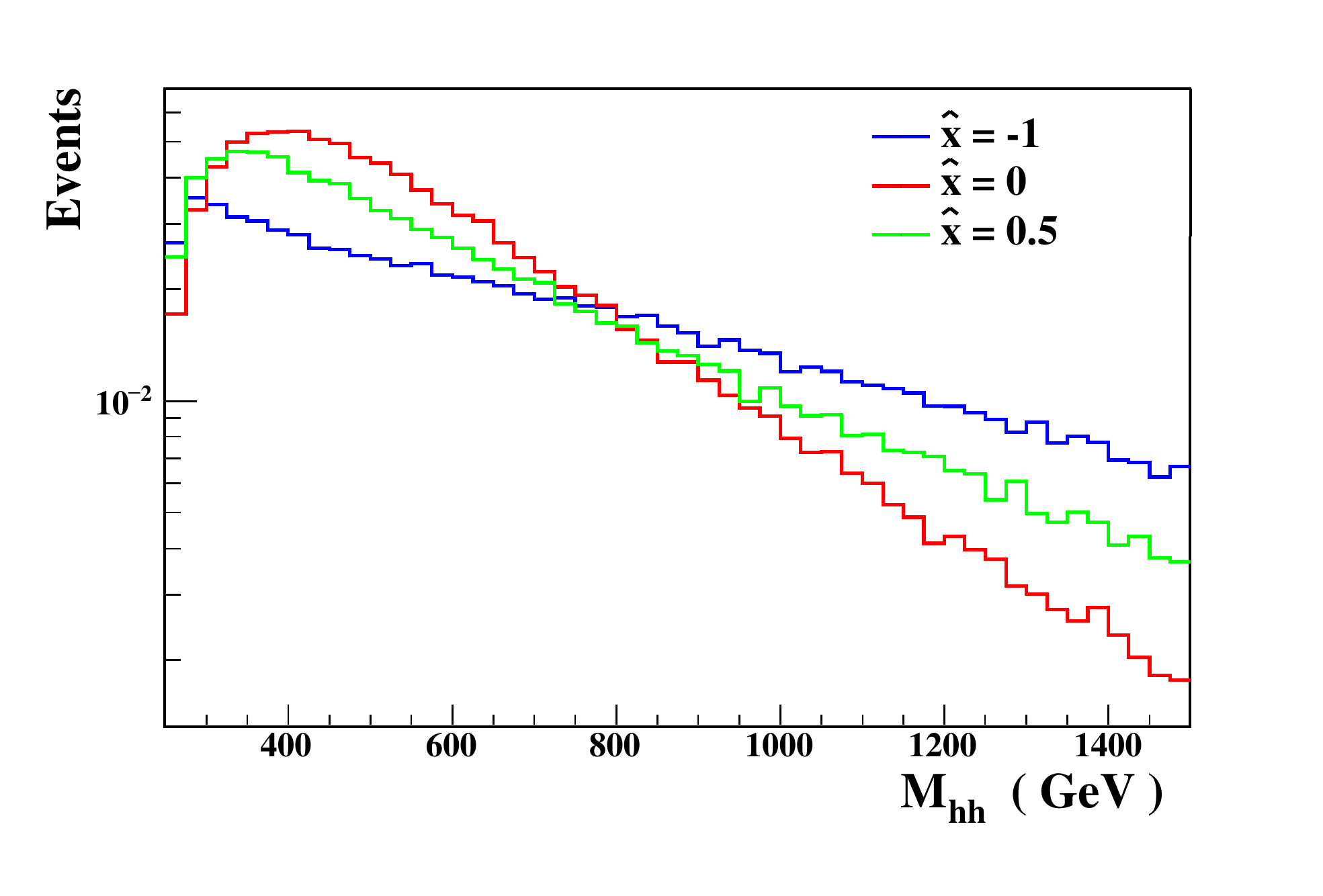}
    \includegraphics[height=6.5cm,width=8cm]{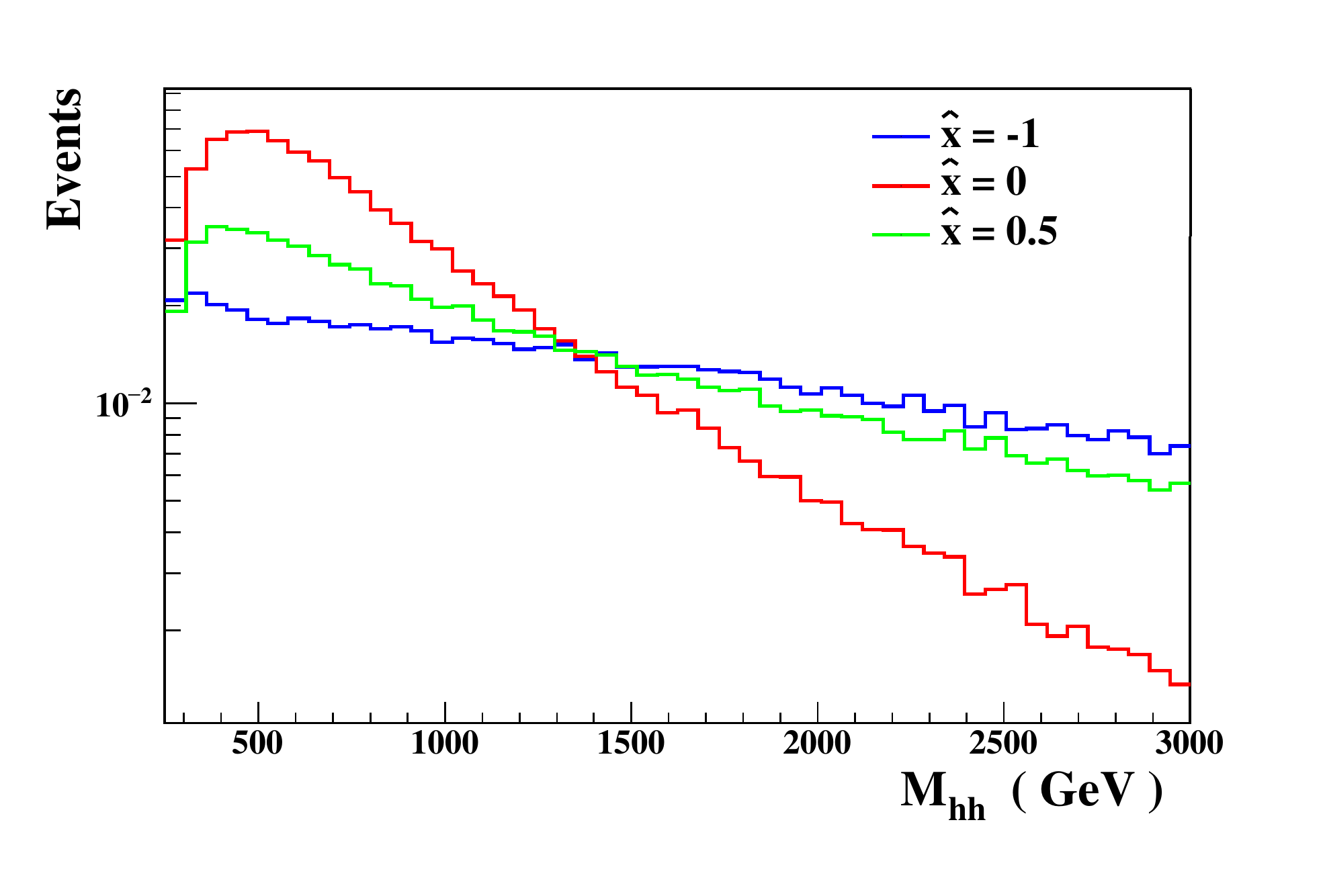}
\vspace*{-4mm}
\caption{Parton level distributions of $\,pp\to V^* V^*j j\to hhjj\,$
for the leading $p_T^{}$ of Higgs boson (first row),
the invariant-mass $M_{hh}^{}$ (second row) at LHC\,(14TeV) (first column),
and $pp$(100TeV) (second column). In each plot, we set $\,\rh=0\,$,\, and input
$\,\xh=(-1,\,0,\,0.5)$\, which correspond to (blue,\,red,\,green) curves.}
\label{fig:pphhjjparton}
\label{fig:7}
\end{figure}

In Fig.\,\ref{fig:pphhjjparton}, we present the distributions for the leading $p_T^{}$
of Higgs boson (first row), the di-Higgs invariant-mass $M_{hh}^{}$ (second row)
at the LHC\,(14TeV) (first column) and $pp$\,(100TeV) collider (second column).
In comparison with top-pair associated production of Fig.\,\ref{fig:pptthhparton},
more signal events are populated in the high $p_T^{}$ and $M_{hh}^{}$ regions
for $\,\xh\neq0$\,,\, which is notable even at the LHC\,(14TeV).
To further ensure the perturbative expansion of the present effective theory,
we will take into account the unitarity constraint.
We require signal events to obey the conservative bound
$\,M_{hh}^{}>\Lambda'_{\textrm{U}1}$\, in (\ref{eq:PUB1p}).
For $\,\sqrt{s}=14$TeV\,(100TeV) collisions,
this allows 84\% (31\%) signal events under $\,\xh =-1$\,,\,
and 97\% (62\%) signal events under $\,\xh =0.5$\,.

\vspace*{4.5mm}
\section{Full Analysis of ${gg\to hh\to b\bar{b}\gamma\gamma}$ at $\mathbf{pp}$(100TeV) Collider}
\label{sec:full}
\label{sec:4}
\vspace*{1.5mm}

In this section, we study di-Higgs production via gluon fusion by performing
a full analysis (including Delphes\,3 fast detector simulations) at the $pp$\,(100TeV) collider.
We will focus on the gluon fusion process $\,gg\to hh\to b\bar{b}\gamma\gamma$\,.\,
We construct four kinds of benchmark points, and study the sensitivities to probing
different regions of the parameter space of cubic Higgs interactions via this channel.
Our analysis extends the previous Snowmass study\,\cite{Yao:2013ika} by including
non-SM-like derivative cubic Higgs coupling via model-independent dimension-6 effective operators.
We also present a full background study which further includes jet-faking-photon backgrounds
and contributions of $jj\gaga$ due to mis-tagging $\,b$ or $\bar{b}$\,.\,
These improve the analysis of Ref.\,\cite{Yao:2013ika}.

\vspace*{2mm}
\subsection{Full Simulations for Signals and Backgrounds}
\vspace*{1mm}

For the present study, we generate the signal and background events by using
Madgraph\,5 and Pythia\,6.2 packages\,\cite{MG5}\cite{Pythia6},
which are then passed to Delphes\,3
for detector simulations \cite{Delphes3}.

We show the full list of backgrounds in Table\,\ref{tbl:background}.
All background processes include up to one extra parton with MLM matching
to avoid double-counting. We do not include $b\bar{b}jj$ background, since
after all selection cuts it is negligible compared with other faked backgrounds.
The detector responses are based on the
current performance of ATLAS and CMS.
The $b$-tagging operation point is chosen to have 75\%, 18.8\%, and 1\% for
bottom, charm, and light flavor jets in the central region
($E_T^{}>50$\,GeV and $|\eta|<2.5$), respectively.
The photon identification efficiency is about 80\% for photons with
$\,E_T^{}>50$\,GeV and $|\eta|<2.5$\,.\,
For the jet-faking-photon background, we assign a faking probability of
$\,f_j^{}=0.0093\exp (-E_T^{}/27)$\, as a function of $E_T^{}$ (in GeV) of the jet,
and scale the jet energy by $\,0.75\pm 0.12$\, as the photon energy~\cite{Atlas}.
The mass resolution is 2\,GeV for $\,h\rightarrow \gamma\gamma$\,
and 17\,GeV for $\,h\rightarrow b\bar b$\, at $\,M_h^{}=125$\,GeV.
To be consistent with the signal, we select two tagged $b$-jets
and two isolated photons in the final states,
where each object is required to have $\,E_T^{}>25$\,GeV and
$\,|\eta|<2.5$\,.\,

We further impose the mass-window cuts on the invariant-masses of
two photons and two $b$-jets.
Compared with the previous study\,\cite{Yao:2013ika}, we will narrow down
the diphoton invariant-mass window as $\,122\,\textrm{GeV}<M_{\gamma\gamma}<128\,$GeV.\,
This would kill another 40\% backgrounds beyond the previous case with
10\,GeV diphoton mass-window.
For two $b$-jets, we still impose  $\,85\,\textrm{GeV}<M_{b\bar b}<135\,$GeV.

\begin{table}[t]
\begin{center}
\caption{For signal and background processes, this table presents
$\,\sigma \times \textrm{Br}$\,,\, generated events, selected events, acceptance, and the
expected events at $pp$\,(100\,TeV) collider with an integrated luminosity of 3\,ab$^{-1}$.}
\begin{tabular}{c||c|c|c|c|c}
\hline \hline
&&&&&
\\[-3mm]
  Samples   & $\sigma\!\times\!\textrm{BR}$ (fb)
  & Generated Evt & Selected Evt & Accept & Expected   \\[1.2mm]
\hline
$h(b\bar{b})h(\gamma\gamma)\,$(SM) & 3.53 & 100000 & 3955 & 0.040& $418.8\pm 6.6$ \\
 \hline
$b\bar{b}h(\gamma\gamma)$ & 50.49 & 99611 & 78 & 0.00078& $118.6\pm 13.4$ \\
$Z(b\bar{b})h(\gamma\gamma)$ & 0.8756 & 68585 & 378 & 0.0055& $14.5 \pm 0.7$ \\
$t\bar{t}h(\gamma\gamma)$ & 37.26 & 63904 & 67 & 0.0010& $117.2\pm 14.3$ \\
$t\bar{t}\gamma\gamma$ & 335.8 & 150654 & 1 & $6.6\!\times\!\!10^{-6}$ & $6.75\pm 6.7$ \\
$t\bar{t}\gamma$ & 108400 & 285787 & 0.013 & $4.7\!\times\!\!10^{-8}$ & $15.2\pm 3.2$ \\
$b\bar{b}\gamma\gamma$ & 5037 & 763962 & 11 & $1.4\!\times\!\!10^{-5}$ & ~$217.6\pm 65.6$~ \\
$b\bar{b}j\gamma$ & 8960000 & 1119406 & 0.0051 & $4.6\!\times\!\!10^{-9}$ & $123.6\pm 31.9$ \\
$jj\gamma\gamma$ & 164200 & 813797 & 0.056 & $6.9\!\times\!\!10^{-8}$ & $33.9\pm 3.8$ \\
\hline
&&&&&
\\[-3.6mm]
 Total background &  $-$  &  $-$  &  $-$  &  $-$ & ~~$647.3\pm 76.0$~~ \\[1mm]
\hline
&&&&&
\\[-3.6mm]
 $S/\!\sqrt{B\,}$ ($S/\!\sqrt{B\!+\!S\,}$) & $-$ & $-$ &  $-$  &  $-$ & 16.5 (12.8)
 \\[1.2mm]
\hline \hline
\end{tabular}
\label{tbl:background}
\label{tab:1}
\end{center}
\end{table}

\begin{figure}
\vspace*{-2mm}
\centering%
    \includegraphics[height=6.5cm,width=8cm]{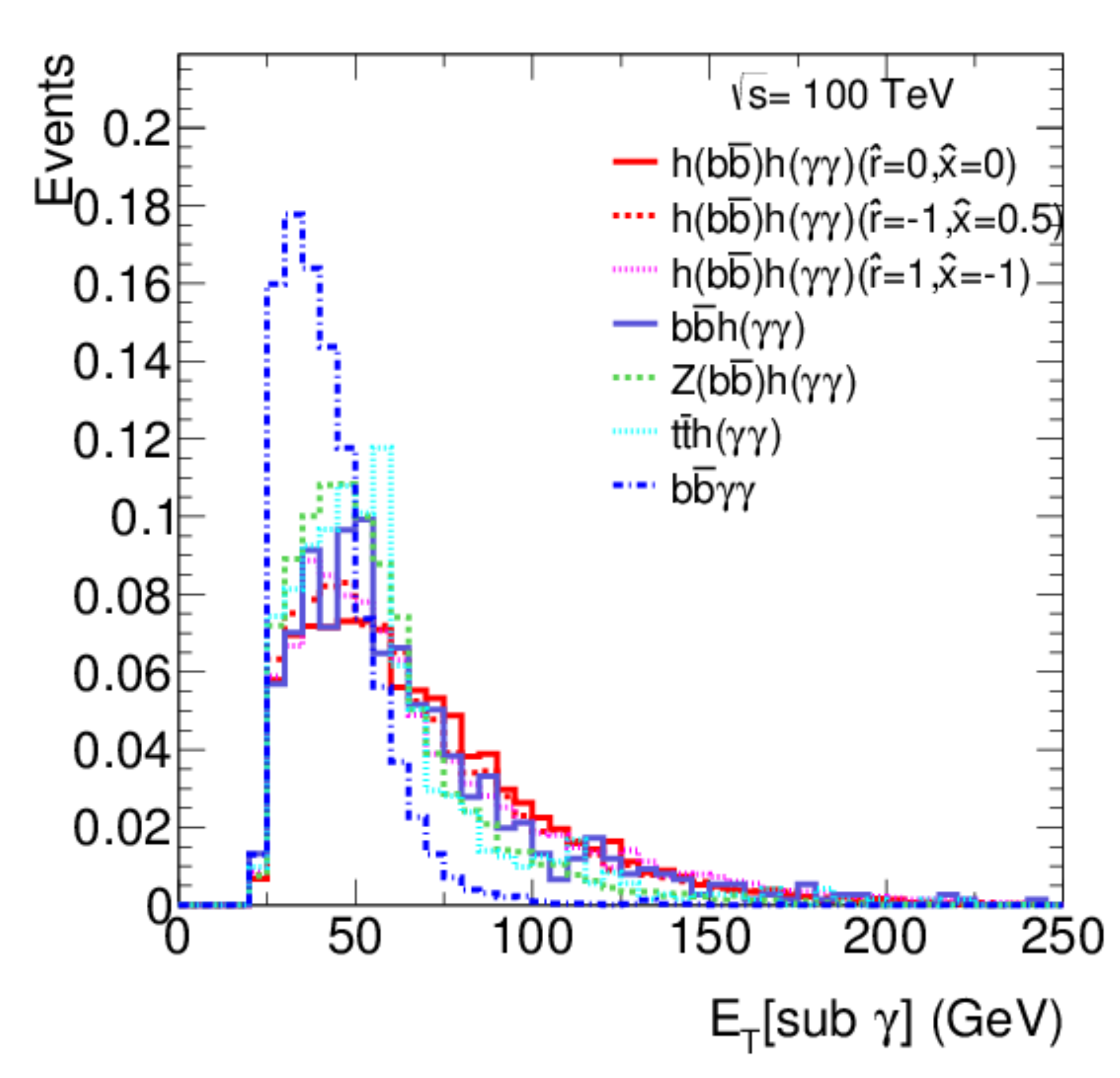}
    \includegraphics[height=6.5cm,width=8cm]{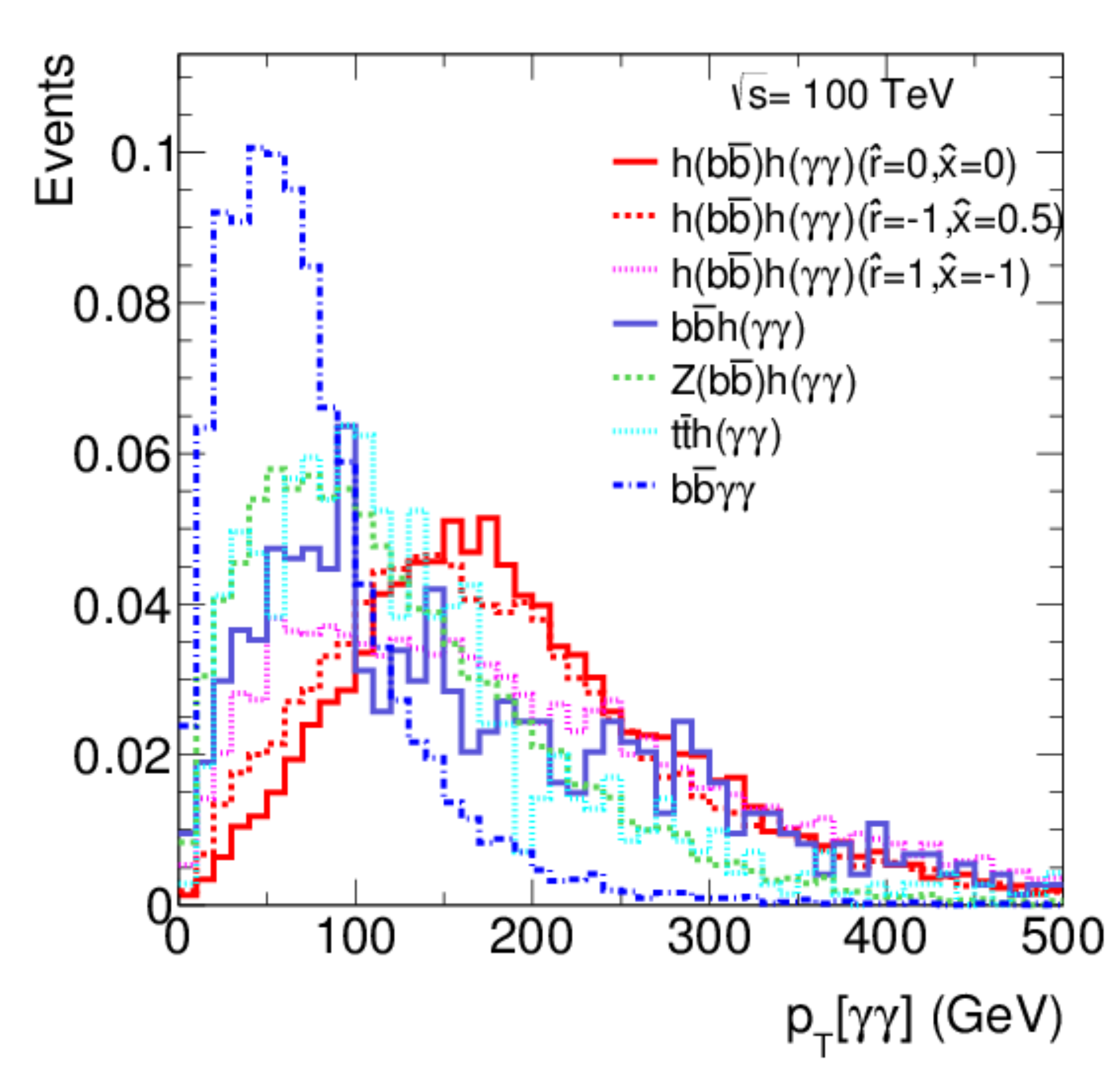}
\\[-2mm]
    \includegraphics[height=6.5cm,width=8cm]{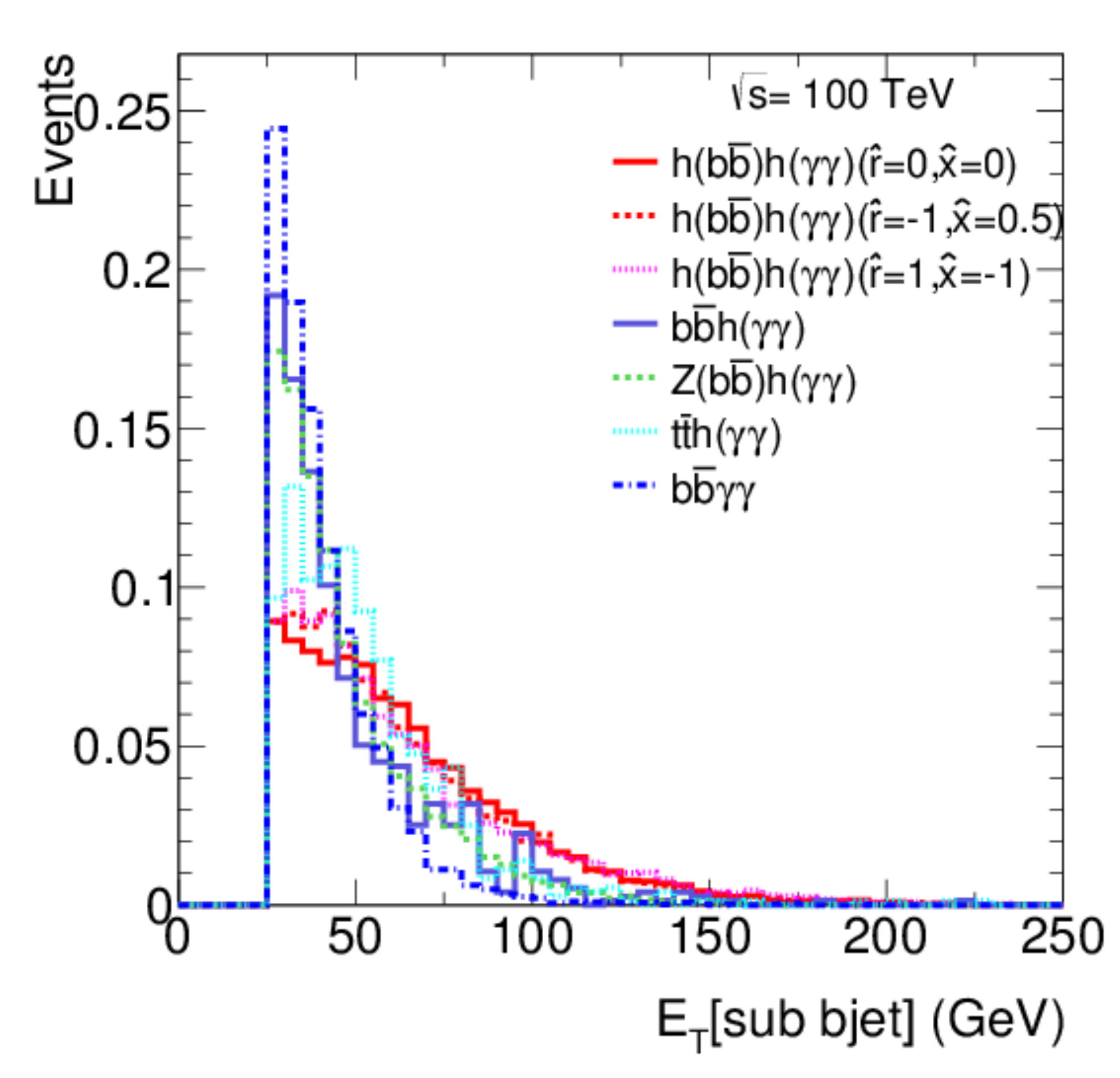}
    \includegraphics[height=6.5cm,width=8cm]{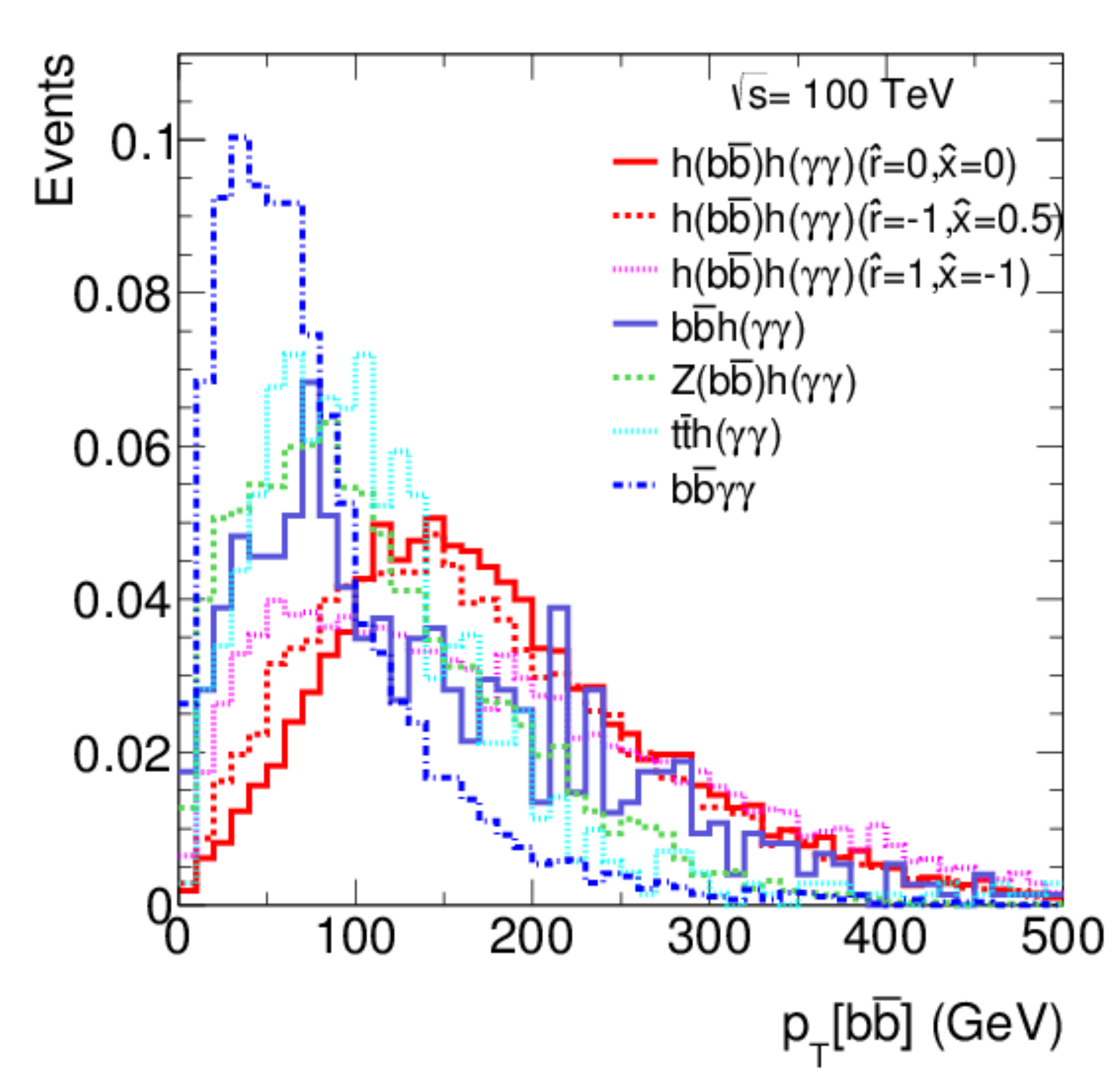}
\\[-2mm]
    \includegraphics[height=6.5cm,width=8cm]{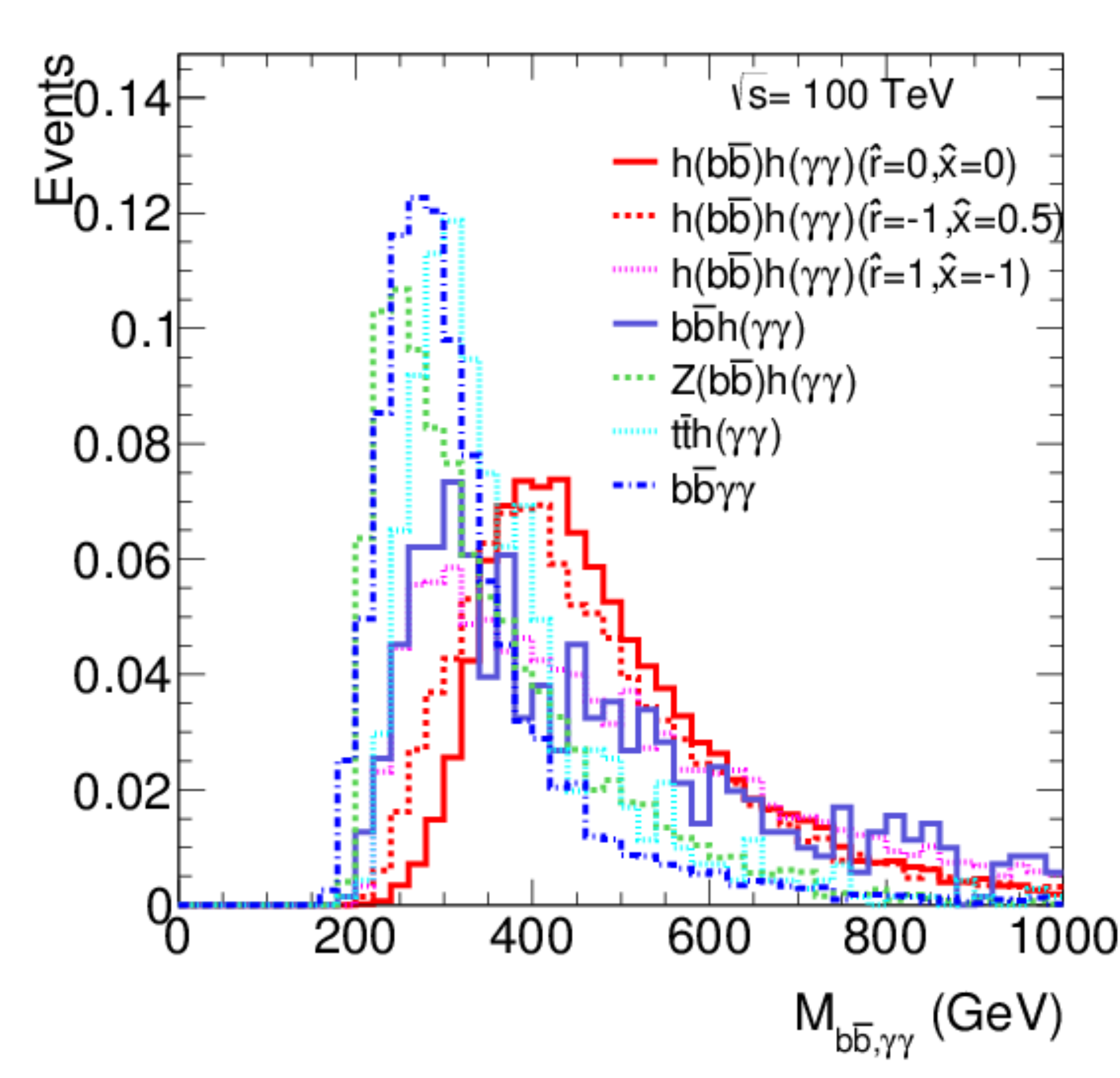}
\vspace*{-2mm}
\caption{Distributions of the sub-leading $E_T^{}[\text{sub}\,\ga]$
and $p_T^{}[\gaga]$ of selected diphotons
for the signal/background events are presented in the first row.
The distributions of $E_T^{}[\text{sub}\,b\text{-jet}]$ and
$p_T^{}[b\bar b]$ of the selected $b\bar b$ jets are depicted in the second row.
The invariant-mass distributions of the selected $\gaga b\bar b$ events
are plotted in the third row.}
\label{fig:hhkin}
\label{fig:8}
\end{figure}

Fig.\,\ref{fig:hhkin} shows the normalized distributions of the $p_T^{}$ and
the sub-leading $E_T^{}$ of two selected photons (or $b$-jets) in the first two rows.
The last plot of Fig.\,\ref{fig:hhkin} depicts the reconstructed di-Higgs invariant-mass
$\,M_{b\bar b\gamma\gamma}^{}$\, for both signals and backgrounds.
Here we only show the representative backgrounds.
The distributions of faked $\,b\bar{b}j\gamma$\, and $\,jj\gamma\gamma$\, are similar to
$\,b\bar{b}\gamma\gamma$,\, while $\,t\bar{t}\gamma\gamma$\, and $\,t\bar{t}\gamma$\,
have too few events after selection.
For illustration, we present distributions for the SM and two other cases
with new coupling inputs  $\,(\rh,\,\xh) =(-1,\,0.5)$\, and $\,(\rh,\,\xh) =(1,\,-1)$.\,
We find that including the new couplings $\,(\rh,\,\xh)\,$ does not significantly change
kinematic distributions after full simulation for the gluon fusion production,
as we have expected from the parton level analysis in Sec.\,\ref{sec:parton}.
Hence, for the rest of selections,
we use the same kinematical cuts as in the Snowmass study\,\cite{Yao:2013ika}.

We summarize these cuts as follows,
\begin{itemize}
\item Invariant-mass cut: $~M_{b\bar b \gamma\gamma}^{}>300$\,GeV\,;

\item $\Delta R$ cuts: ~$\Delta R_{\gamma\gamma}^{}<2.5$\,,~~~
      $\Delta R_{b\bar b}^{}<2.0$\,;

\item $p_T^{}$ cuts:
~$p_T^{}[\gamma],\,p_T^{}[b]>35$\,GeV,~~ $p_T^{}[\gamma\gamma],\, p_T^{}[b\bar b]>100$\,GeV\,;

\item Decay angle of $\,h\rightarrow \gamma\gamma$\, in the $hh$ rest frame:~  $|\cos\theta_h^{}|<0.8$\,;\footnote{The decay angle $\theta_h^{}$ is defined
    as the angle between one of the $h$ directions in the di-Higgs rest frame and
    the di-Higgs momentum in the lab frame.}

\item Total number $\,n$\, of jets, photons and leptons are required to be $\,n< 7\,$
      in each event.
\end{itemize}
%

%

We present the expected signal and background event numbers at $\,\sqrt{s}=100$\,TeV
and for an integrated luminosity $\,\mathcal{L}=3\,\textrm{ab}^{-1}$\,
in Table\,\ref{tbl:background}.
For the SM Higgs self-coupling of $(\rh,\,\xh)=(0,\,0)$,\, we find the
expected signal events to be 418.8\,.\,
The expected yield of total background events is 647.3\,,\,
with the largest contributions coming from $\,b\bar{b}\gamma\gamma$,\, $b\bar b j\gamma$,\,
$b\bar{b} h(\gamma\gamma)\,$ and $\,t\bar t h(\gamma\gamma)$\,.\,
The resultant signal statistic significance is about 16.5\,$\sigma$.\,
With some relaxation of kinematical cuts, we find that the sensitivity becomes
a bit worse due to increased background contributions, but the overall picture remains the same.
We have also compared our study with the recent analyses of $b\bar b\gamma\gamma$ channel
at $pp(100)$\,TeV in the literature \cite{Barr:1412.7154}\cite{Azatov:2015oxa}.
Ref.\,\cite{Barr:1412.7154} studied this channel for the SM cubic Higgs coupling,
and estimated 179 signal events with 447 background events
after all cuts and for the same luminosity.
Our study gives 418.8 signal events and 647.3 background events.
The difference is likely due to their more conservative assumptions for the detector performance,
especially the photon identification efficiency, which is lower than ours.
In the future, it would be helpful to directly compare the results by using the same assumptions
for detector performance. Ref.\,\cite{Azatov:2015oxa} estimated
$S/\!\sqrt{B}=15.2$ under all cuts and the same condition,
which is in good agreement with ours.

For the signal analysis, we perform full simulations
for parameters within the range $\,-1\leqq \rh\leqq 1$\, and $\,-1\leqq \xh\leqq 0.5\,$.\,
We find that the number of selected signal events can be fitted
by similar functions as in Eq.\,(\ref{eq:GFXsecfit}).
Under the above cuts, we deduce
\begin{eqnarray}
\label{eq:NevefitAll}
\left.\frac{\sigma}{\sigma_{\textrm{sm}}^{}}
\right|_{\textrm{All}}
\,=\, (1\!-\xh )^2\!\(1-0.55\,\rh +3.4\,\xh +0.11\,\rh^2+3.9\,\xh^2-1.2\,\rh\,\xh \).
\end{eqnarray}
Compared with the parton level fit (\ref{eq:GFXsecfit-b}), we see that
the cross section becomes less sensitive to the parameters $(\rh,\,\xh)$.\,
This is what we would expect from the contamination of parton shower,
hadronization, and detector simulation.

To further discriminate $\,\rh\,$ and $\,\xh\,$ dependence,
we can utilize distributions in different reconstructed 
di-Higgs invariant-mass bins\,\cite{Chen:2014xra,Azatov:2015oxa}, which include
different kinematic features of contributions from $\,\rh\,$ and $\,\xh\,$.\,
To efficiently suppress the background,
we choose $M_{hh}(=M_{b\bar b \gamma\gamma})$ bins as follows,
\begin{eqnarray}
\label{eq:Mhhbin}
\textrm{$M_{hh}$ bins (GeV):}~~~~
[300,500],\,[500,700],\,[700,900],\,[900,1100].
\end{eqnarray}
We note that for the $b\bar{b}\gamma\gamma$ final state, due to the small branching fraction
of $\,h\to\gaga\,$
and the fast decline of gluon parton distribution function, the probe of $\,M_{hh}^{}\,$
is not much higher than 1\,TeV even at the $pp$\,(100TeV) collider.
Since the derivative cubic Higgs coupling brings in more energy enhancement,
higher $M_{hh}^{}$ bin is more sensitive to $\,\xh$\,.\,
This can be seen from event fits in each bin as follows,
\beqs
\label{eq:NevefitMhhbin}
\begin{eqnarray}
\label{eq:NevefitMhhbin-a}
\left.\frac{\sigma}{\sigma_{\textrm{sm}}}\right|_{\textrm{bin\,}1}
&\!\!\!=\!\!&
(1\!-\xh )^2(1-0.82\,\rh+3.4\,\xh+0.17\,\rh^2+3.3\,\xh^2-1.5\,\rh\,\xh )\,,
\\
\left.\frac{\sigma}{\sigma_{\textrm{sm}}}\right|_{\textrm{bin\,}2}
&\!\!\!=\!\!&
(1\!-\xh )^2(1-0.42\,\rh +3.3\,\xh +0.06\,\rh^2+3.8\,\xh^2-0.95\,\rh\,\xh )\,,
\\
\left.\frac{\sigma}{\sigma_{\textrm{sm}}}\right|_{\textrm{bin\,}3}
&\!\!\!=\!\!&
(1\!-\xh )^2(1-0.14\,\rh +3.5\,\xh +0.04\,\rh^2+5.6\,\xh^2-0.85\rh\,\xh )\,,
~~~~~
\\
\left.\frac{\sigma}{\sigma_{\textrm{sm}}}\right|_{\textrm{bin\,}4}
&\!\!\!=\!\!&
(1\!-\xh )^2(1-0.03\,\rh +4.0\,\xh +0.03\,\rh^2+8.6\,\xh^2-0.65\,\rh\,\xh )\,.
\end{eqnarray}
\eeqs
With increasing $M_{hh}^{}$,\,
the coefficients of $\,\rh\,$ terms decrease, while $\,\xh\,$ terms become more important.
In passing, we clarify the difference of our analysis from Ref.\,\cite{Azatov:2015oxa}.
The paper \cite{Azatov:2015oxa} simplifies the computation by doing hadron-level analysis
for the SM case only, and infers the signal rate at other points by parton-level analysis
with rescaling of hadron-to-parton ratio for the SM, i.e., they assumed that the hadron-to-parton
cuts efficiency remains the same over the parameter space.
We test this assumption with our full analysis in the $\,\rh-\xh\,$ parameter space.
We find that it works well in lower $M_{hh}^{}$ bins, but would induce
$\,\mathcal{O}(10\%\!-\!100\%)$\, deviations in high mass bins\,\footnote{%
Depending on the luminosity, there could be $\mathcal{O}(10\%)$ statistical uncertainty
in the last $M_{hh}$ bin at 30\,ab$^{-1}$. But the statistical uncertainties in other bins
are much smaller.}.\,
For the inclusive rate, it is not a problem since it is dominated by low mass bins.
But, it could affect the conclusion of exclusive analysis (cf.\ Sec.\,\ref{sec:4.2}).
For later convenience, we summarize the numbers of selected background events for each
bin in Table\,\ref{tbl:backgroundbin}.

\vspace*{2mm}
\subsection{Probing Cubic Higgs Interactions via Parameter Space $(\rh,\,\xh)$}
\label{sec:4.2}
\vspace*{1.5mm}

In this section, we analyze the probe of $\,\rh-\xh\,$ parameter space at the
$pp$\,(100\,TeV) collider with a sample data from $3\,\textrm{ab}^{-1}$ ($30\,\textrm{ab}^{-1}$)
integrated luminosity.
As mentioned in Sec.\,\ref{sec:theory}, due to interferences
with other possible dimension-6 operators, the measurement of single Higgs productions cannot
uniquely constrain $\,\xh\,$.\,
Hence, it is important to independently probe the parameter space of
$\,\xh\,$ via di-Higgs production, which receives energy enhancement from
the derivative coupling induced by \,${\cal O}_{\Phi,2}^{}$\,.\,
To study sensitivities to different regions of the \,$(\rh,\,\xh)$\, parameter space,
we choose four kinds of benchmark points,
\begin{eqnarray}
&\hspace*{9.6mm} \textrm{Benchmark~A}:&~~ (\rh,\,\xh)_{\text{sm}}^{} =\, (0,\,0)\,;
\nonumber\\
&\textrm{Benchmarks}~\B_1,\,\B_2:&~~ (\rh,\,\xh) \,=\, (0,\,0.2),\,(0,\,0.5)\,;
\label{eq:ABC}
\\
&\textrm{Benchmarks}~\C_1,\,\C_2:&~~ (\rh,\,\xh) \,=\, (-0.5,\,0),\,(0.5,\,0)\,;
\nonumber\\
&\textrm{Benchmarks}~\D_1,\,\D_2:&~~ (\rh,\,\xh) \,=\, (-0.5,\,0.2),\,(0.5,\,-0.5)\,.
\nonumber
\end{eqnarray}
Benchmark\,A corresponds to the SM Higgs boson, and the sensitivity in this case
can be directly translated into a bound on the effective cutoffs of dimension-6 operators
$({\cal O}_{\Phi,2}^{},\,{\cal O}_{\Phi,3}^{})$.\,
We use Benchmarks $\B_1$ and $\B_2$ to represent the cases as predicted
by nonminimal coupling model with $\,\rh=0\,$ and $\,\xh>0\,$ (cf.\ Sec.\,\ref{sec:2.2}).\,
Benchmarks $\C_1$ and $\C_2$ correspond to nonzero $\,\rh\,$ and
vanishing derivative cubic Higgs coupling $\,\xh\,$.\,
The last two benchmarks $\D_1$ and $\D_2$ denote the general cases
with both $\,\rh\,$ and $\,\xh\,$ nonzero. For all non-SM benchmarks,
we choose \,$(\rh,\,\xh)$\, values corresponding to the effective cutoffs
$\,\tilde\Lambda_2^{},\tilde\Lambda_3^{}\gtrsim 500\,$GeV.
Note that the effective cutoff scale
$\,\tilde\Lambda_j^{}=\Lambda/\!\sqrt{f_{\Phi,j}^{}}\,$
is not exactly the mass scale of an underlying new particle (as the dimensionless coupling
$f_{\Phi,j}^{}$ could be larger than one and is usually less than about $4\pi$).
One example is the model of Higgs-gravity
interactions in Sec.\,\ref{subsec:NMCmodel} with dimension-6 operators
\eqref{eq:cut1-cut2}-\eqref{eq:cut12-xi}.
From the viewpoint of effective theory, the major issue is to make sure
that the energy scale is within the perturbative unitarity bound, so that the perturbative
analysis is valid. In Fig.\,\ref{fig:PUbound}, the plots (b)-(c) show that
for the effective cutoff  $\tilde\Lambda_2^{}\gtrsim 500\,$GeV,
the unitarity bounds on the scattering energy are well above 1\,TeV.
This justifies our perturbative analysis of signal events with di-Higgs invariant-mass
$\,M_{hh}^{}\lesssim 1.1\,$TeV.

\begin{table*}[t]
\vspace*{-3mm}
\begin{center}
\caption{Selected events in different $M_{hh}^{}$ bins for the SM signal and backgrounds
at the $pp$\,(100TeV) collider with an integrated luminosity of \,3\,ab$^{-1}$.}
\begin{tabular}{c||c|c|c|c}
\hline \hline
&&&&
\\[-3.5mm]
  $M_{hh}$ bins (GeV)   & [300, 500] & [500, 700] & [700, 900] & [900, 1100]
\\[1.2mm]
\hline
&&&&
\\[-3.8mm]
$h(b\bar{b})h(\gamma\gamma)$ (SM) & 200 & 170 & 52.5 & 11.1
\\[1mm]
\hline
&&&&
\\[-3.8mm]
$b\bar{b}h(\gamma\gamma)$ & 67.1 & 31.9 & 15.8 & 3.81 \\
$Z(b\bar{b})h(\gamma\gamma)$ & 11.2 & 2.77 & 0.46 & 0.04 \\
$t\bar{t}h(\gamma\gamma)$ & 97.5 & 15.9 & 3.22 & 0.58 \\
$t\bar{t}\gamma\gamma$ & 5.41 & 1.1 & 0.24 & 0.0 \\
$t\bar{t}\gamma$ & 13.9 & 1.09 & 0.16 & 0.05 \\
$b\bar{b}\gamma\gamma$ & 188 & 23.7 & 5.25 & 0.32 \\
$b\bar{b}j\gamma$ & 107 & 11.8 & 3.44 & 1.32 \\
$jj\gamma\gamma$ & 30.3 & 2.58 & 0.82 & 0.24 \\[1mm]
 \hline
 &&&&
\\[-3.8mm]
Total Backgrounds & 521 & 90.8 & 29.4 & 6.37\\[1.2mm]
\hline \hline
\end{tabular}
\label{tbl:backgroundbin}
\label{tab:2}
\end{center}
\vspace*{-2mm}
\end{table*}
\begin{figure}[t]
\centering%
\includegraphics[width=16cm]{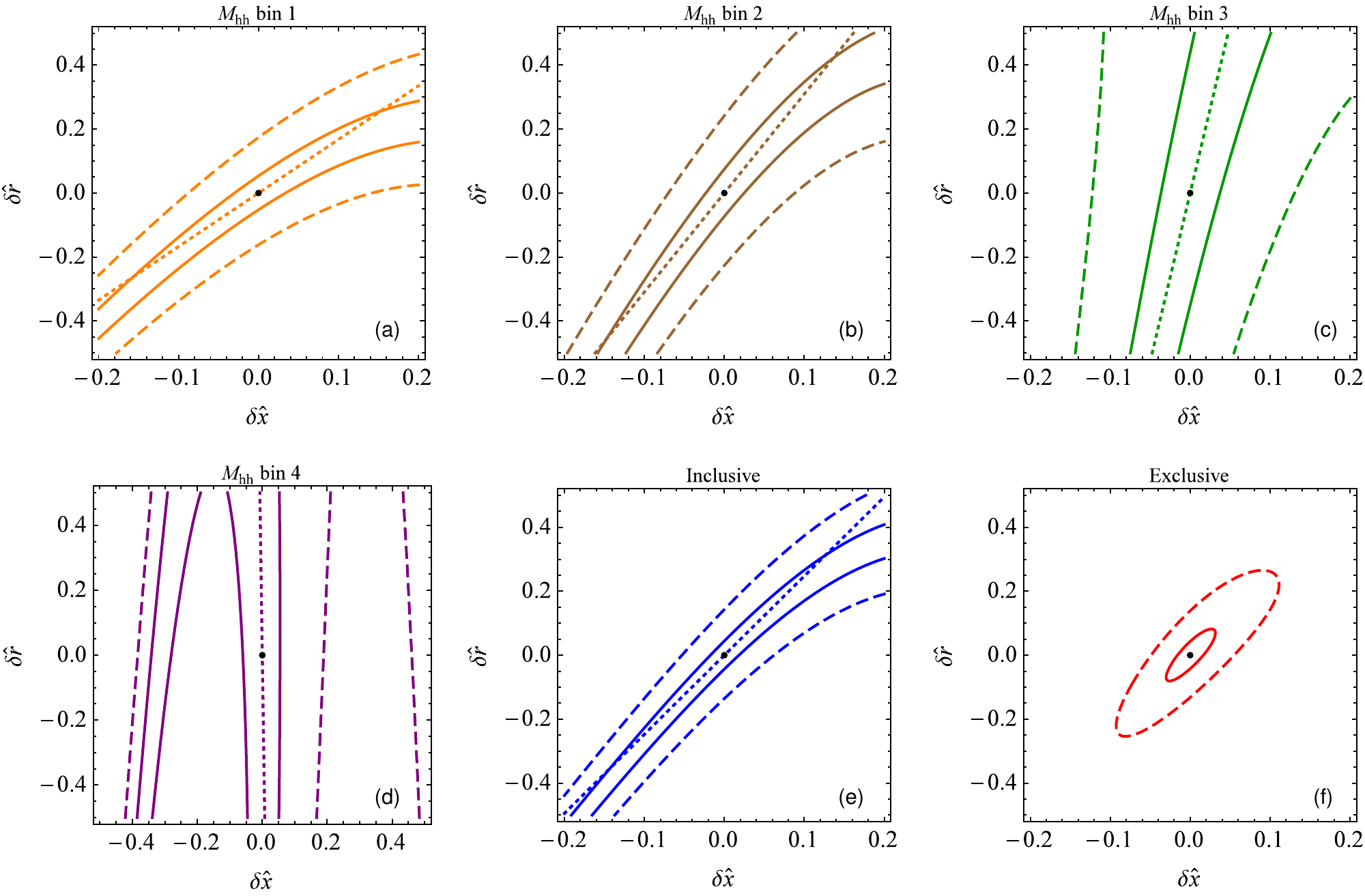}
\vspace*{-1.5mm}
\caption{Sensitivity to $\,\delta\rh -\delta\xh$\, plane around Benchmark\,A,
$(\rh,\,\xh)=(0,\,0)$.\, In each plot, the dashed (solid) curve depicts $68\%$\,C.L.\ contour
with 3\,ab$^{-1}$ (30\,ab$^{-1}$) integrated luminosity,
and the dotted line denotes the degenerate direction around the origin.
Plots (a)-(d) present the results for each $M_{hh}^{}$ bin.
Plots (e) and (f) show the inclusive sensitivity \eqref{eq:inclusive}
and exclusive sensitivity \eqref{eq:exclusive}, respectively.}
\label{fig:9}
\end{figure}

For each benchmark, we first analyze the sensitivities in different $M_{hh}^{}$ bins
as defined in (\ref{eq:Mhhbin}).
For a given set of $(\rh,\,\xh)$,\, the 68\%\,C.L.\ contour is defined as follows,
\begin{eqnarray}
\frac{\Delta S_i^{}(\rh,\,\xh)}{\sqrt{B_i^{}+S_i^{}(\rh,\,\xh)\,}\,} \,=\, 1\,,
\end{eqnarray}
where signal $\,S_i^{}$\, and background $\,B_i^{}$\,
in Table\,\ref{tbl:backgroundbin} denote the numbers of selected events
in a given bin $M_{hh}^{(i)}$, and
$\,\Delta S_i^{}(\rh,\,\xh)=|S_i^{}(\rh +\delta\rh,\, \xh+\delta\xh )- S_i^{}(\rh,\,\xh)|$.\,
The dependence of signal on the parameters $(\rh,\,\xh)$
is determined by the numerical fits in Eq.\,(\ref{eq:NevefitMhhbin}).
Around the origin of $(\rh,\,\xh)$, it is well approximated by the linear expansion,
$\,S_i^{} \simeq c_i^{} + a_i^{} \delta\rh+ b_i^{} \delta\xh$\,.\,
It means that the signal is only sensitive to the combination
$\,a_i^{} \delta\rh+ b_i^{} \delta\xh\,$,\,
but not the perpendicular direction $\,b_i^{} \delta\rh-a_i^{} \delta\xh$\,.\,
We call the later as ``degenerate direction", along which the signal remains constant
nearby the origin.  Using the fit (\ref{eq:NevefitAll}),
we further derive the sensitivity contour with inclusive data.
\begin{eqnarray}
\label{eq:inclusive}
\frac{\dis\sum_i\!\Delta S_i^{}(\rh,\,\xh)}
{\sqrt{\,\dis\sum_i \!\left[ B_i^{}+S_i^{}(\rh,\,\xh)\right]\,}\,} \,=\, 1\,.
\end{eqnarray}
Finally, to fully utilize the information of different $M_{hh}^{}$ bins, we can derive the
combined contour at 68\%\,C.L.,
\begin{eqnarray}
\label{eq:exclusive}
\sum_{i}\left(\frac{\Delta S_i^{}}{\sqrt{B_i^{}+S_i^{}}}\right)^{\!\!2} =\, 1 \,,
\end{eqnarray}
which we will call ``exclusive" sensitivity.
This is stronger than the ``inclusive" sensitivity \eqref{eq:inclusive}
which only uses the total rates.

\begin{figure}[t]
\vspace*{-4mm}
  \centering%
\includegraphics[height=7.7cm,width=8.7cm]{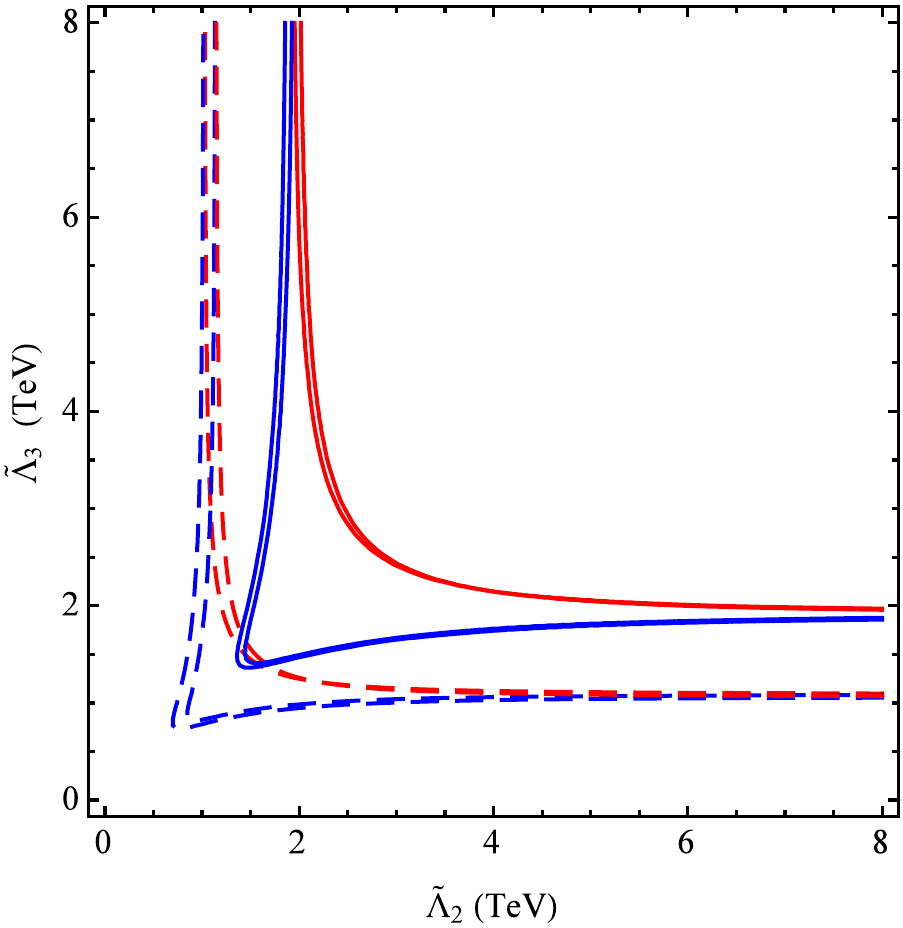}
\vspace*{-1mm}
\caption{Exclusive sensitivity contours (68\%\,C.L.) in $\,\cuttb -\cuttc$\, plane
for Benchmark\,A at the $pp$\,(100TeV) collider with an integrated luminosity
of 3\,ab$^{-1}$ (dashed curves) and 30\,ab$^{-1}$ (solid curves).\,
The two red and blue contours correspond to $\,x_2^{}x_3^{}>0\,$ and
$\,x_2^{}x_3^{}<0\,$,\, respectively.
The region on the right-hand-side of each contour (and above it) is allowed.}
\label{fig:10}
\end{figure}

In Fig.\,\ref{fig:9},
we analyze the sensitivity for Benchmark\,A, which corresponds to taking the
central values $(\rh,\,\xh)=(0,\,0)$ as in the SM. We present the 68\%\,C.L.\ contours
for each $\,M_{hh}^{}\,$ bin in plots (a)-(d).
Then, we show the inclusive sensitivity contour
\eqref{eq:inclusive} in plot-(e), and the
exclusive sensitivity contour \eqref{eq:exclusive} in plot-(f).
For each plot, the dashed (solid) curve depicts $68\%$\,C.L.\ contour
with 3\,ab$^{-1}$ (30\,ab$^{-1}$) integrated luminosity,
while the dotted line shows the degenerate direction around the origin.
The slope of dotted line varies for different bins of $\,M_{hh}^{}$.\,
It is clear that higher $M_{hh}^{}$ bins are more sensitive to $\,\xh\,$,\,
as also noted before \cite{Chen:2014xra,Azatov:2015oxa}.
However, the final sensitivity of a given bin also depends on
the number of selected events in this bin. Due to suppression in the tail region of
$\,M_{hh}^{}$ distribution, event number in the highest bin (purple) could be
quite small. This is the case with 3\,ab$^{-1}$ data in Fig.\,\ref{fig:9}(d),
where the sensitivity to $\,\xh\,$ is much lower than that in other bins.
Hence, the inclusive sensitivity is mainly determined by the first two bins.
For 30\,ab$^{-1}$ data, there are enough events
in the last bin to probe $\,\xh\,$ with a good accuracy.
Impressively, since various bins are sensitive to different combinations of
$\,\rh\,$ and $\,\xh\,$,\, the exclusive analysis \eqref{eq:exclusive} makes
a big improvement of the sensitivity, as shown in Fig.\,\ref{fig:9}(f).
Note that the exclusive analysis does not improve much of the sensitivity for each parameter
alone, but helps to break the degenerate direction in the 2-dimensional plane.
This demonstrates the important role played by the derivative cubic Higgs coupling $\,\xh\,$
in the di-Higgs production. It means that gluon fusion production
could probe both $\,(\rh,\,\xh)\,$ to a good accuracy. This is a new point.
For comparison, we derive the sensitivity to each parameter alone by fixing
the other parameter to its SM value.
We find that the exclusive sensitivity to $\,\delta\rh\,$ is about
$13\%$ (4.2\%), and that to $\,\delta\xh\,$ is about $5\%$ ($1.6\%$),
for the 3\,ab$^{-1}$ (30\,ab$^{-1}$)\, integrated luminosity.

\begin{figure}[t]
\vspace*{4mm}
  \centering%
\includegraphics[height=6.5cm,width=7.5cm]{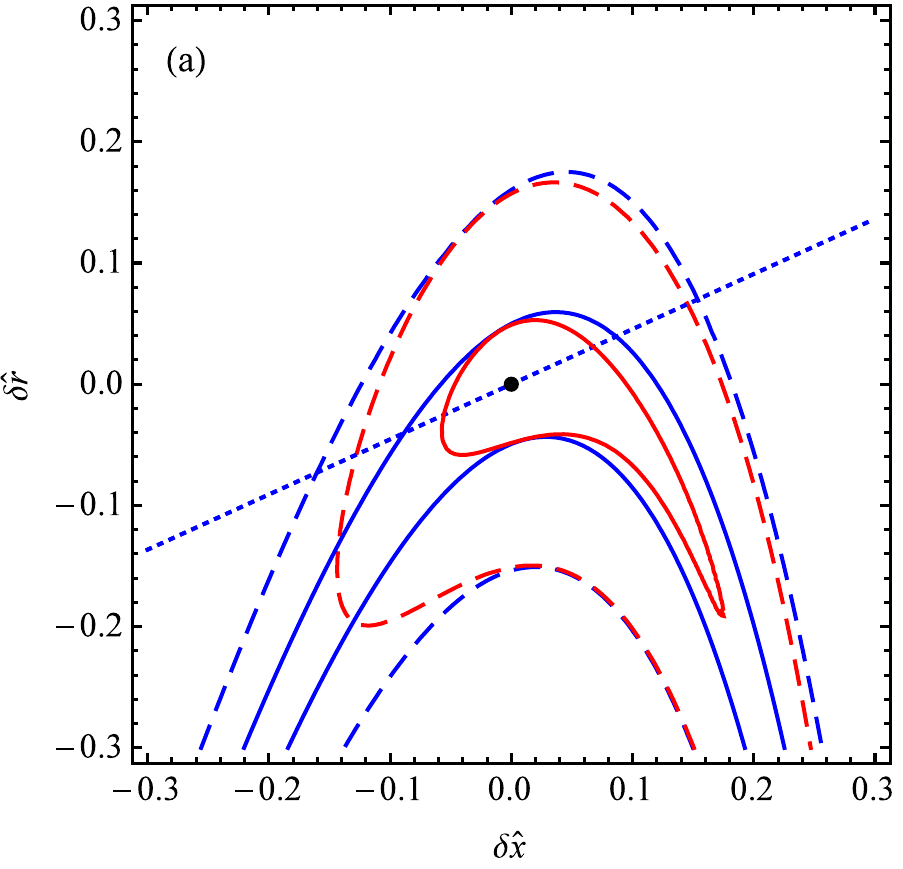}~~
\includegraphics[height=6.5cm,width=7.5cm]{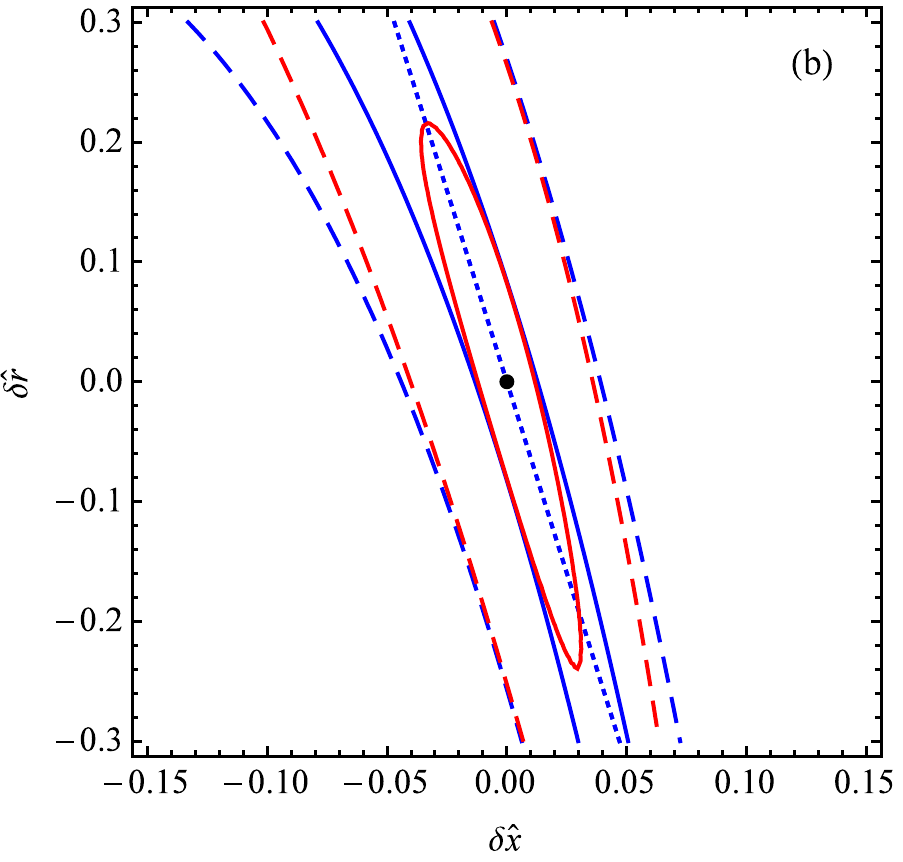}
\vspace*{-2mm}
\caption{Sensitivity contours in $\,\delta\rh -\delta\xh$\, plane
for Benchmark\,$\B_1$ with $\,(\rh ,\,\xh ) =(0,\,0.2)$\, as shown in plot-(a),
and for Benchmark\,$\B_2$ with $\,(\rh ,\,\xh ) =(0,\,0.5)$\, as shown in plot-(b).
In each plot, the dashed (solid) curve depicts $68\%$\,C.L.\ contour
with 3\,ab$^{-1}$ (30\,ab$^{-1}$) integrated luminosity,
and the dotted line denotes the degenerate direction around the origin.
The blue and red contours depict the inclusive sensitivity
\eqref{eq:inclusive} and exclusive sensitivity \eqref{eq:exclusive}, respectively.
}
\label{fig:11}
\vspace*{2mm}
\end{figure}

In Fig.\,\ref{fig:10}, we present the exclusive sensitivity contours (68\%\,C.L.)
in $\,\cuttb -\cuttc$\, plane for Benchmark\,A
at the $pp$\,(100TeV) collider with an integrated luminosity of
3\,ab$^{-1}$ (dashed curves) and 30\,ab$^{-1}$ (solid curves).\,
The region on the right-hand-side of each contour (and above it) is allowed.
The cases of $\,x_2^{}x_3^{}>0\,$ ($\,\xh^{}\,\rh^{}<0\,$)
and $\,x_2^{}x_3^{}<0\,$ ($\,\xh^{}\,\rh^{}>0\,$) are shown
by the two red and blue contours, respectively.
For each contour, the asymptotically flat or vertical behavior gives
the sensitivity to one operator (when the other is absent),
which can be read from the intersection of 68\%\,C.L.
sensitivity contour in Fig.\,\ref{fig:9}(f)
with each axis. The sensitivities of probing the two operators are comparable,
$\,\cuttb,\cuttc \gtrsim 1\,$TeV with 3\,ab$^{-1}$,\,
and $\,\cuttb,\cuttc \gtrsim 2\,$TeV with 30\,ab$^{-1}$.\,
For the blue contours, the cusps correspond to the end points of
ellipse long axis in Fig.\,\ref{fig:9}(f).
These cusp regions give the weakest 2d sensitivities,
$\,\cuttb,\cuttc \gtrsim 0.75\,$TeV
for 3\,ab$^{-1}$ data, and
$\,\cuttb,\cuttc \gtrsim 1.4\,$TeV for 30\,ab$^{-1}$ data.
For the red contours, the 2d sensitivity is always stronger.

\begin{figure} 
\vspace*{2mm}
\centering%
\includegraphics[height=6.5cm,width=7.5cm]{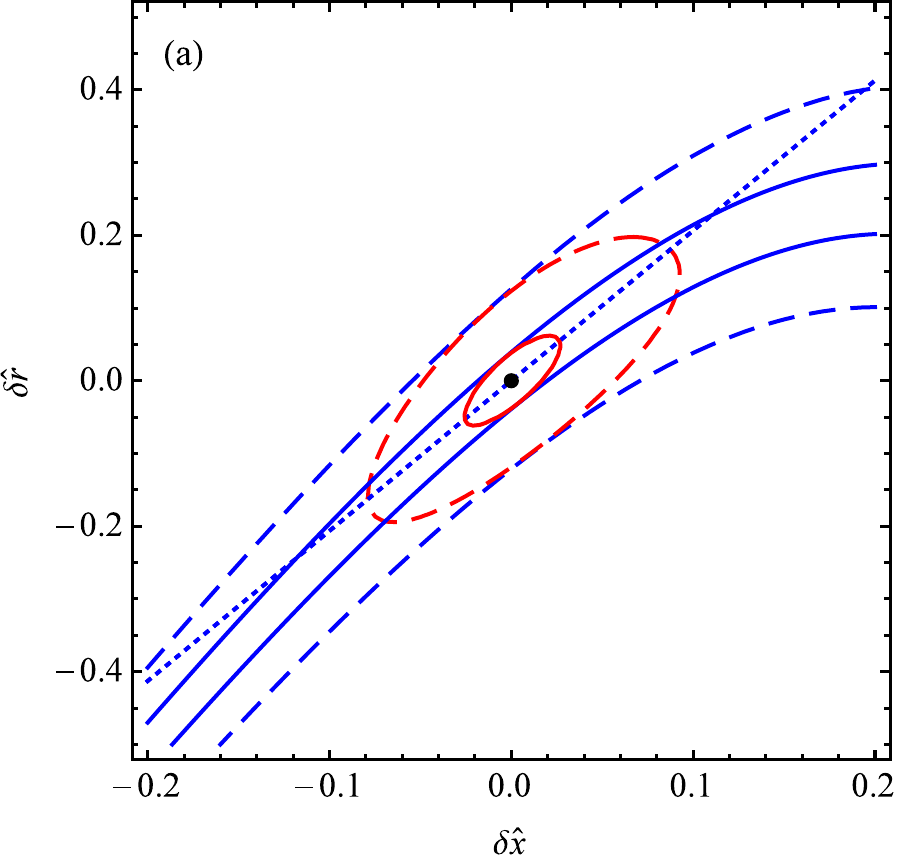}~~
\includegraphics[height=6.5cm,width=7.5cm]{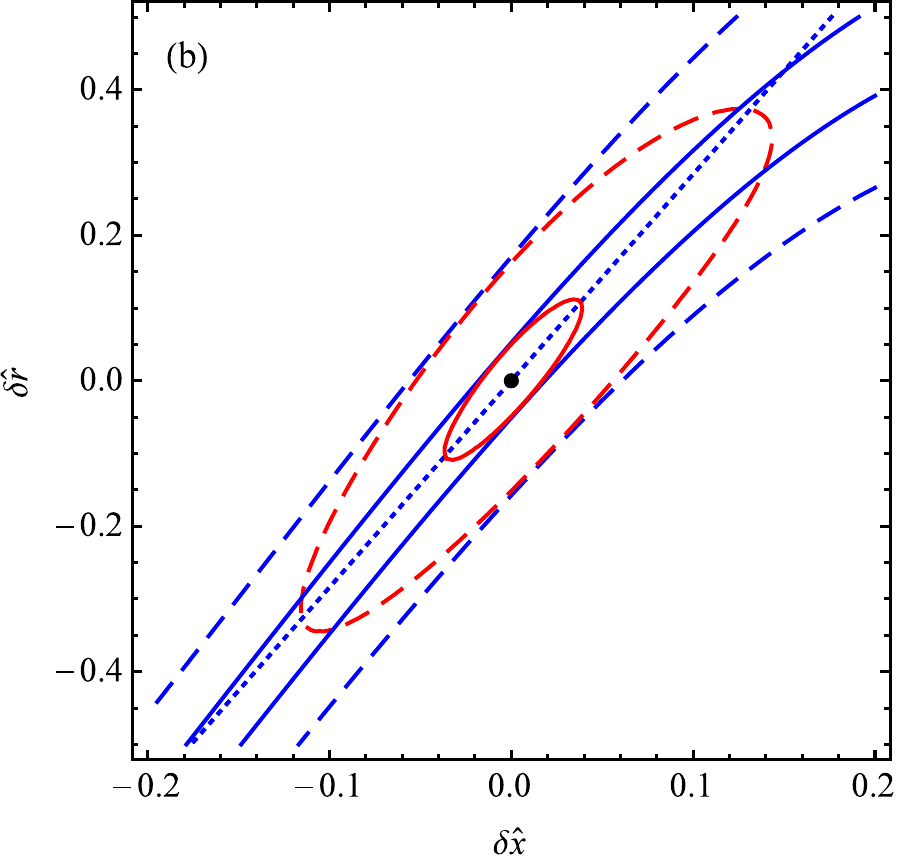}
\caption{Sensitivity contours in $\,\delta\rh -\delta\xh$\, plane
for Benchmark\,$\C_1$ with $\,(\rh ,\,\xh ) =(-0.5,\,0)$\, as shown in plot-(a),
and for Benchmark\,$\C_2$ with $\,(\rh ,\,\xh ) =(0.5,\,0)$\, as shown in plot-(b).
In each plot, the dashed (solid) curve depicts $68\%$\,C.L.\ contour
with 3\,ab$^{-1}$ (30\,ab$^{-1}$) integrated luminosity,
and the dotted line denotes the degenerate direction around the origin.
The blue and red contours depict the inclusive sensitivity
\eqref{eq:inclusive} and exclusive sensitivity \eqref{eq:exclusive}, respectively.
}
\label{fig:12}
\end{figure}
\begin{figure} 
\vspace*{3mm}
  \centering%
\includegraphics[height=6.5cm,width=7.5cm]{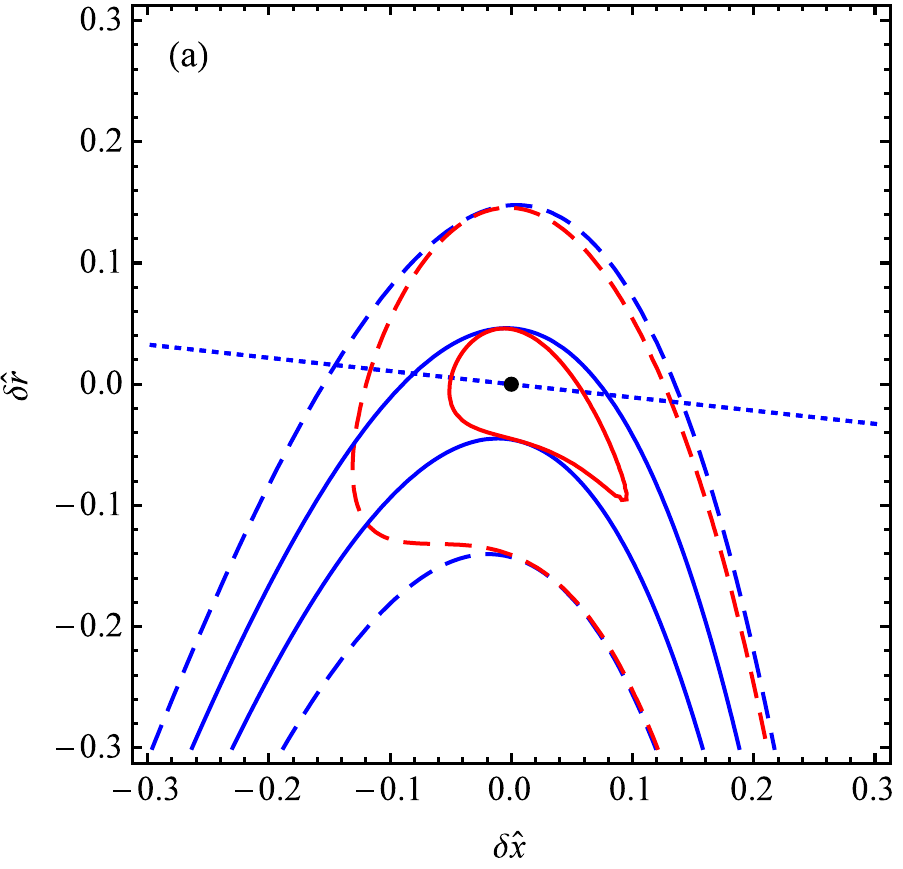}~~
\includegraphics[height=6.5cm,width=7.5cm]{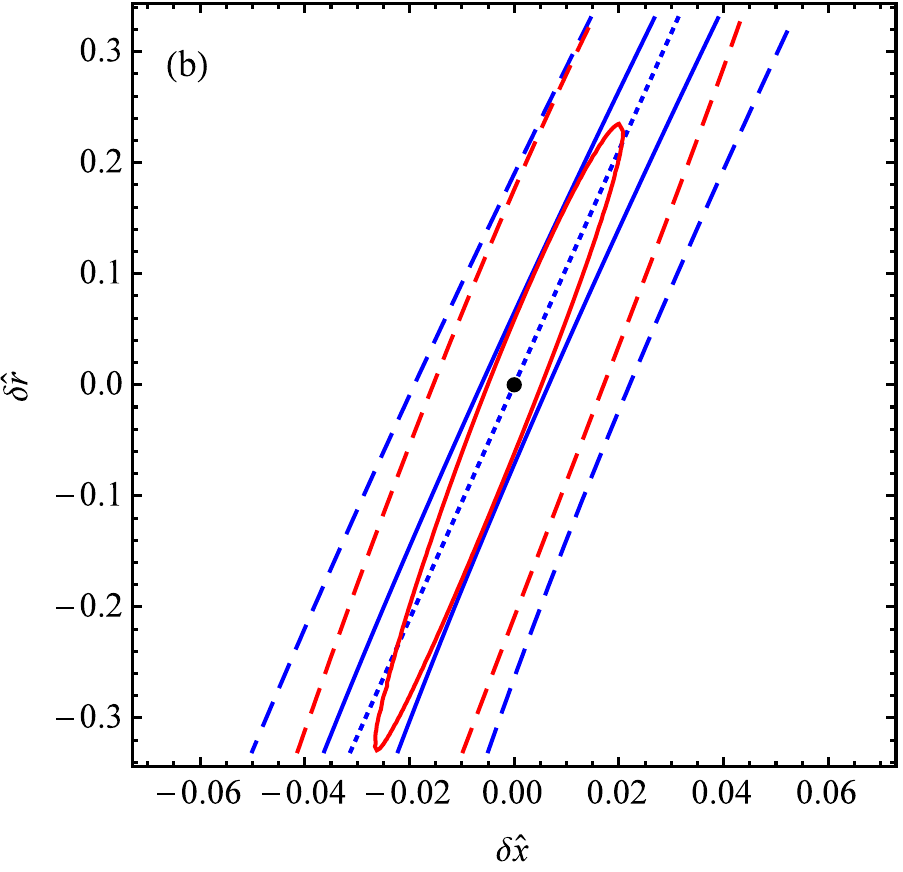}
\caption{Sensitivity contours in $\,\delta\rh -\delta\xh$\, plane
for Benchmark\,$\D_1$ with $\,(\rh ,\,\xh ) =(-0.5,\,0.2)$\, as shown in plot-(a),
and for Benchmark\,$\D_2$ with $\,(\rh ,\,\xh ) =(0.5,\,0.5)$\, as shown in plot-(b).
In each plot, the dashed (solid) curve depicts $68\%$\,C.L.\ contour
with 3\,ab$^{-1}$ (30\,ab$^{-1}$) integrated luminosity,
and the dotted line denotes the degenerate direction around the origin.
The blue and red contours depict the inclusive sensitivity
\eqref{eq:inclusive} and exclusive sensitivity \eqref{eq:exclusive}, respectively.
}
\label{fig:13}
\end{figure}

In Fig.\,\ref{fig:11}, we present the inclusive sensitivity \eqref{eq:inclusive}
and exclusive sensitivity \eqref{eq:exclusive} for Benchmark\,$\B_1$ [plot-(a)]
and Benchmark\,$\B_2$ [plot-(b)] by blue and red contours, respectively.
The dashed (solid) curve depicts $68\%$\,C.L.\ contour with 3\,ab$^{-1}$ (30\,ab$^{-1}$)
integrated luminosity, and the dotted line denotes the degenerate direction around the origin.
The Higgs gravitational interaction predicts $\,\rh =0\,$ and $\,\xh>0\,$.\,
As shown in plots (a) and (b), the sensitivity contours,
including slope of the degenerate direction,
strongly depend on the explicit value of $\,\xh\,$.\,
Fig.\,\ref{fig:12} demonstrates the sensitivities for
Benchmark\,$\C_1$ [plot-(a)] and Benchmark\,$\C_2$ [plot-(b)],
where $\,\xh=0\,$ and two nonzero $\,\rh\,$ values take opposite signs.
We find that their shape and sensitivity range are quite similar to that
of Benchmark\,A (the SM case). This is expected given the fact that
the di-Higgs total cross section and invariant-mass ($M_{hh}$) distribution
are much more sensitive to $\,\xh\,$ than $\,\rh\,$.

In Fig.\,\ref{fig:13}, we present the inclusive sensitivity \eqref{eq:inclusive}
and exclusive sensitivity \eqref{eq:exclusive}
for Benchmark\,$\D_1$ [plot-(a)] and Benchmark\,$\D_2$ [plot-(b)]
to illustrate the features for $(\rh,\,\xh)$ both nonzero.
Benchmark\,$\D_1$ represents the case that the signals in the first two bins of
$M_{hh}^{}$ are quite insensitive to $\,\delta\xh\,$ at the linear order,
and the shape of the $68\%$ sensitivity contour is mainly determined by quadratic terms.
Although the last two bins still have strong dependence on $\,\xh\,$,\,
the inclusive sensitivity is determined by the first two bins (due to their large rates)
with parabola-like shape.
The exclusive sensitivity is largely improved, especially with 30\,ab$^{-1}$ data.\,
Fig.\,\ref{fig:13}(b) presents the sensitivity contours ($68\%$\,C.L.)
for Benchmark\,$\D_2$, where all $M_{hh}^{}$ bins have strong dependence on $\,\xh\,$.\,
The sensitivity to $\,\xh\,$ is significantly enhanced
as compared to other Benchmarks.\footnote{%
Note that the plot range of $\,\delta\xh\,$ in Fig.\,\ref{fig:13}(b) is much smaller than
that in Fig.\,\ref{fig:13}(a).}\,
Since the sensitivity has little change among different bins,
the $68\%$ contour is only slightly improved
by the exclusive analysis \eqref{eq:exclusive}.

In summary, the qualitative feature of sensitivity contours in the
$\,\delta\rh -\delta\xh\,$ plane can vary significantly for different benchmarks.
In some cases (such as Benchmarks A, $\B_1$, $\C_1$, $\C_2$, and $\D_1$),
the exclusive analysis of different $M_{hh}^{}$ bins makes big improvements.
In particular, it can break the possible degenerate direction around the origin,
and impose much stronger constraints on the 2d parameter space even with
the di-Higgs production measurement alone.
For some other cases (such as Benchmarks $\B_2$ and $\D_2$),
the parameter-dependence of signals appears quite similar in different bins.
Thus, both the exclusive and inclusive analyses give comparable sensitivities.

\vspace*{3mm}
\section{Conclusions}
\label{sec:conclusion}
\label{sec:5}
\vspace*{1.5mm}

Despite the LHC Higgs discovery,
the Higgs boson self-interaction is fully untested so far.
It is the key ingredient of Higgs potential,
and plays vital roles for electroweak symmetry breaking, vacuum stability,
electroweak phase transition, and Higgs inflation. This is a most likely place
to encode new physics beyond the standard model (SM).

In this work, we studied the probe of cubic Higgs interactions
via di-Higgs production at hadron colliders.
We parametrized the new physics of Higgs self-interactions in terms of
model-independent dimension-6 effective operators in section\,\ref{sec:2}.
We take the nonminimal Higgs-gravity interaction as an explicit example to motivate
such effective operators. The contributions of the two dimension-6 operators
(\ref{eq:dim6scalar2}) to cubic Higgs couplings have different kinematic structures
as shown in Eq.\,(\ref{eq:Chhh}). They give different kinematic distributions
in various di-Higgs production channels due to the different energy-dependence.
This is demonstrated in Figs.\,\ref{fig:5}--\ref{fig:7} of section\,\ref{sec:3}.
We also analyzed the weak boson scattering and $t\bar{t}$ scattering at high energies,
and derived perturbative unitarity constraints on the parameter space in Fig.\,\ref{fig:2}.
Among the three channels of di-Higgs production, top-pair associated production
and vector boson fusion (VBF) production are more sensitive to the
energy-enhancement in high energy collisions, though their cross sections are generally smaller
than the gluon fusion production (Fig.\,\ref{fig:4}).

In section\,\ref{sec:4}, we performed systematical Monte Carlo analysis of di-Higgs production
$\,gg\to hh\,$ in the decay channel $\,hh\to b\bar{b}\gamma\gamma$\,
by using Delphes\,3 fast detector simulations.\,
We computed both signals and full SM backgrounds at the $pp$\,(100TeV) collider
with a 3\,$\textrm{ab}^{-1}$ integrated luminosity,
as summarized in Table\,\ref{tab:1} and Fig.\,\ref{fig:8}.
This channel shows a good potential of discovering cubic Higgs couplings
in $pp$\,(100TeV) collisions.
Our derived significance is in main agreement with the literature\,\cite{Azatov:2015oxa},
while a difference from \cite{Barr:1412.7154} appears
due to the different assumptions about detector performance.
We further studied the probe of new physics effects 
in the $\,\rh-\xh$\, parameter space with
full simulations.  Since different bins of the di-Higgs invariant-mass $M_{hh}^{}$ exhibit distinctive
kinematical features, we used them to discriminate the two dimension-6 operators.
We did an exclusive analysis to incorporate such kinematical information
and obtained a big improvement of sensitivity.
We further identified four kinds of representative benchmarks \eqref{eq:ABC}
for the parameter space of cubic Higgs coupling, which have qualitatively different features.
For each benchmark, we studied the sensitivity to the 2d parameter space
of $\,(\rh,\,\xh)$,\,
via both the inclusive analysis \eqref{eq:inclusive} and exclusive analysis \eqref{eq:exclusive}.
For comparison, we used two sample integrated luminosities (3\,ab$^{-1}$ and 30\,ab$^{-1}$)
of the $pp$\,(100TeV) collider.
For Benchmark\,A (SM case), the exclusive analysis breaks the degeneracy
in the 2d plane and makes it possible to probe both \,$(\rh,\, \xh)$\,
to a good accuracy by the di-Higgs measurement alone.
This is demonstrated in Fig.\,\ref{fig:9}--\ref{fig:10}.
For one-parameter analysis, we found that
with a 3\,ab$^{-1}$ (30\,ab$^{-1}$)\, integrated luminosity,
the exclusive sensitivity to $\,\rh\,$ and $\,\xh\,$ are about
\,$13\%$ (4.2\%)\, and  \,$5\%$ ($1.6\%$),\, respectively.
Fig.\,\ref{fig:11} presented Benchmarks $\B_1$ and $\B_2$ with $\rh=0$ and $\xh>0$,
as motivated by the nonminimal Higgs-gravity interaction.
We found that the sensitivity contours strongly depend on the size of $\,\xh\,$.\,
Fig.\,\ref{fig:12} analyzed Benchmarks $\C_1$ and $\C_2$
with $\,\xh=0$\, and different values of $\,\rh$\,.\,
As expected, the dependence on the change of $\,\rh\,$ is pretty weak.
For general regions with $(\rh,\, \xh)$ both nonzero,
we found that the sensitivity contours behave qualitatively different from
the SM case of $\,(\rh,\, \xh)=(0,\,0)$,\,
as shown in Fig.\,\ref{fig:13}(a)-(b) for Benchmarks $\D_1$ and $\D_2$.\,
In the case where the parameter-dependence of signals in different bins is similar,
such as Benchmark\,$\D_2$ in Fig.\,\ref{fig:13}(b),
the improvement of the exclusive analysis over the inclusive analysis
becomes rather modest.

\vspace*{4mm}
\appendix

\section{\,Redundancy of Dimension-6 Operators}
\label{app:dim6EOM}

In this appendix, we discuss the redundancy of dimension-6 operators
in Eqs.\,(\ref{eq:dim6scalar}) and (\ref{eq:dim6sf}).
In the SM action, the Higgs sector contains the following terms,
\begin{eqnarray}
S_{\textrm{sm}}^{}~\supset~
\int\!\! d x^4\!
\left[(D^\mu H)^\dag (D_\mu H)+ \mu^2H^\dag H
-\lambda(H^\dag H)^2 - y_f^{} \overline{L}H f_R^{}+ \text{h.c.}\right]\!,
\hspace*{9mm}
\end{eqnarray}
where $\,L=(f_L^{u},\,f_L^{d})^T\,$ denotes the $SU(2)_L$ doublet, and $f_R^{}$
the $SU(2)_L^{}$ singlet.
Then, we can derive EOM for the Higgs field,
$\,(D^2H)^\dag=\mu^2H^\dag-2\lambda(H^\dag H)H^\dag-y_f^{}\overline{L}f_R^{}$\,,\,
and its hermitian conjugate.
After integration by part, we can rewrite the operator
$\,\mathcal{O}_{\Phi,2}\,$ as
\begin{eqnarray}
2\mathcal{O}_{\Phi,2}^{} &\!=\!&
\partial^{\mu}(H^\dag H)\partial_{\mu}(H^\dag H)
= -(H^\dag H)\partial_{\mu}\partial^{\mu}(H^\dag H)+\textrm{(total\,derivative)}
\nonumber\\
&\!=\!&
-(H^\dag H)\left[2(D^{\mu}H)^\dag(D_{\mu}H)+H^\dag D^2H+(D^2H)^\dag H\right]
\nonumber\\
&\!=\!&
-2\mathcal{O}_{\Phi,4}^{} -2\mu^2(H^\dag H)^2
+12\lambda\mathcal{O}_{\Phi,3}^{}+
\left( y_f^{}\mathcal{O}_{\Phi,f}^{}+\textrm{h.c.}\right)\!,
\label{eq:O2-re}
\end{eqnarray}
where $\,\partial^{\mu}(H^\dag H)=D^{\mu}(H^\dag H)$.\,
In the above, we have neglected the total derivative term.
We have also implemented the SM EOM in the last step,
since we only keep operators up to dimension-6.
With the relation \eqref{eq:O2-re}, we may replace $\,\mathcal{O}_{\Phi,4}^{}\,$
by other operators,
\begin{eqnarray}
\label{eq:dim6EOM}
&& \frac{1}{\Lambda^2}\Big[f_{\Phi,2}\mathcal{O}_{\Phi,2}^{}
+f_{\Phi,3}^{}\mathcal{O}_{\Phi,3}^{}+f_{\Phi,4}^{}\mathcal{O}_{\Phi,4}^{}
+f_{\Phi,f}(\mathcal{O}_{\Phi,f}^{}+\text{h.c.})\Big]
\nonumber\\[1.5mm]
&=& -\frac{\,\mu^2\!f_{\Phi,4}^{}\,}{\Lambda^2}(H^\dag H)^2
+\frac{f_{\Phi,2}^{}\!-\!f_{\Phi,4}^{}}{\Lambda^2}\mathcal{O}_{\Phi,2}^{}
+ \frac{f_{\Phi,3}^{}\!+\! 6\lambda f_{\Phi,4}^{}}{\Lambda^2}
\mathcal{O}_{\Phi,3}^{}+
\left(\!\frac{\,2f_{\Phi,f}^{}\!+\!y_f^{}f_{\Phi,4}^{}\,}{2\Lambda^2}
\mathcal{O}_{\Phi,f}^{}+\text{h.c.}\!\right)\hspace*{10mm}
\nonumber\\
& \to &
\frac{1}{v^2}\left\{(x_2^{}\!-\!x_4^{})\mathcal{O}_{\Phi,2}^{}
+\left(\! x_3^{}\!+\!x_4^{}\frac{3M_h^2}{v^2}\right)\!
\mathcal{O}_{\Phi,3}^{}+\left[\left(\! x_f^{}\!+\!\frac{y_f^{}}{2}x_4^{}\!\right)
\!\mathcal{O}_{\Phi,f}^{}\!+\textrm{h.c.}\right]\right\}\!.
\end{eqnarray}
Note that this also shifts the quartic Higgs coupling in the original Higgs potential,
but can be absorbed by a coupling redefinition,
$\,\lambda\to \lambda - {\mu^2\!f_{\Phi,4}^{}}/\Lambda^2\,$.\,
At the order of $\Lambda^{-2}$,\,
the coupling $\,\lambda\,$ in front of $\,f_{\Phi,4}^{}$\,
can be replaced by the leading order relation
$\,\lambda=M_h^2/2v^2$.\,
Hence, for on-shell physical amplitudes, we can organize
their dependence on $\,(f_{\Phi,2}^{}, f_{\Phi,3}^{}, f_{\Phi,4}^{}, f_{\Phi,f}^{})$\,
via the three combinations in Eq.\,(\ref{eq:dim6EOM}).

\vspace*{3mm}
\section{\,Loop Functions for Triangle and Box Diagrams}
\label{app:loopfactors}

For the analyses of Sec.\,\ref{sec:3}--\ref{sec:4}, we need to compute
cross sections of the di-Higgs production via gluon fusion
$\,g(p_a^{})g(p_b^{})\to h(p_c^{})h(p_d^{})$,\, which invoke
loop functions of triangle and box diagrams \cite{Plehn:1996wb}.
The triangle loop function is given by
\beqs
\begin{eqnarray}
F_\triangle^{} &=&\tau_f\left[1+(1-\tau_f) f(\tau_f)\right]\,,
\\[1.5mm]
f(\tau_f)&=&\left\{\begin{array}{ll}
\textrm{arcsin}^2\frac{1}{\sqrt{\tau_f^{}\,}\,}\,, &~~~ \tau_f\geqq 1 \,,
\\[2.2mm]
-\frac{1}{4}\left[\log \frac{1+\sqrt{1-\tau_f}}{1-\sqrt{1-\tau_f}}-\ii \pi\right]^2,
&~~~ \tau_f<1 \,,
\end{array}\right.
\end{eqnarray}
\eeqs
where $\,\tau_f^{}\equiv 4m_f^2/\hat{s}$,\,
and $\,\hat{s}$ is the partonic center of mass energy.
The box loop functions are defined as follows,
\beqs
\begin{eqnarray}
F_\Box^{} &=& \frac{1}{S^2}
\left[4S+8m_f^2SC_{ab}^{}-2m_f^4S(S+2\rho -8)
(D_{abc}^{}+D_{bac}^{}+D_{acb}^{})\right]
\nonumber\\
&& +m_f^2(2\rho -8)\left[\,\TB (C_{ac}^{}\!+C_{bd}^{})+\UB (C_{bc}^{}\!+C_{ad}^{})
-m_f^2(TU-\rho^2)D_{acb}^{}\right]\!,
\hspace*{12mm}
\\[1.5mm]
G_\Box^{} &=& \frac{1}{S(TU-\rho^2)}
\bigg\{ m_f^2(T^2+\rho^2-8T)
\left[ SC_{ab}^{}+\TB (C_{ac}^{}\!+C_{bd}^{})-m_f^2STD_{bac}^{}\right]
\nonumber\\
&& +\,m_f^2(U^2+\rho^2-8U)\left[ SC_{ab}^{}
+\UB (C_{bc}^{}\!+C_{ad}^{})-m_f^2SUD_{abc}^{}\right]
\nonumber\\
&& -\,m_f^2(T^2+U^2-2\rho^2)(T+U-8)C_{cd}^{}
\nonumber\\
&& \left. -2m_f^4(T+U-8)(TU-\rho^2)(D_{abc}^{}\!+D_{bac}^{}\!+D_{acb}^{})
\right\}\!,
\end{eqnarray}
\eeqs
where $\,\rho =M_{h}^2/m_f^2$,\, $S=\hat{s}/m_f^2$,\,
$T=\hat{t}/m_f^2$,\, $\,U=\hat{u}/m_f^2$,\,
$\TB =T-\rho\,$\, and $\,\UB =U-\rho\,$.\,
The two scalar integrals are given by
\beqs
\begin{eqnarray}
C_{ij} &\!\!\!\!=\!\!\!\!& \int\!\!\!\frac{\,d^4q\,}{\,\ii\pi^2\,}
\frac{1}{(q^2\!-\!m_f^2)\left[(q\!+\!p_i^{})^2\!-\!m_f^2\right]
\left[(q\!+\!p_i^{}\!+\!p_j^{})^2\!-\!m_f^2\right]},
\\
D_{ijk} &\!\!\!\!=\!\!\!\!&
\int\!\!\!\frac{\,d^4q\,}{\ii\pi^2}\dis
\frac{1}{(q^2\!-\!m_f^2)\left[(q\!+\!p_i^{})^2\!-\!m_f^2\right]
\left[(q\!+\!p_i^{}\!+\!p_j^{})^2\!-\!m_f^2\right]
\left[(q\!+\!p_i^{}\!+\!p_j^{}\!+\!p_k^{})^2\!-\!m_f^2\right]}.
\hspace*{13mm}
\end{eqnarray}
\eeqs
In the low energy limit $\,\hat{s}\ll m_f^2$\,,\, the loop functions behave as
\begin{eqnarray}
F_\triangle^{} = \frac{2}{3}+\mathcal{O}\!\left(\!\frac{\hat{s}}{m_f^2}\!\right)\!,
\quad~~~
F_\Box^{} = -\frac{2}{3}+\mathcal{O}\!\left(\!\frac{\hat{s}}{m_f^2}\!\right)\!,
\quad~~~
G_\Box^{} = \mathcal{O}\!\left(\!\frac{\hat{s}}{m_f^2}\!\right)\!.
\end{eqnarray}
In the high energy limit $\,m_f^2\ll\hat{s}$\,,\, they take the asymptotical forms,
\begin{eqnarray}
\label{eq:Fd-Fb-bigE}
F_\triangle^{} = -\frac{m_f^2}{\hat{s}}\left[\log\frac{m_f^2}{\hat{s}}+\ii\pi\right]^2
\!\!+\mathcal{O}\!\left(\!\frac{m_f^2}{\hat{s}}\!\right)\!,\quad~~~
F_\Box^{} = \mathcal{O}\!\left(\!\frac{m_f^2}{\hat{s}}\!\right)\!,\quad~~~
G_\Box^{} = \mathcal{O}\!\left(\!\frac{m_f^2}{\hat{s}}\!\right)\!.
\end{eqnarray}

\vspace*{6mm}
\noindent
{\Large\bf Acknowledgments}
\\[1.5mm]
We thank Nima Arkani-Hamed, Matthew Reece, and Matthew Strassler for discussions
during the finalization of this work. We also thank Florian Goertz, Margarete Muhlleitner,
Andreas Papaefstathiou, Tilman Plehn, Michael Spira, Li Lin Yang and Jose Zurita 
for correspondences.
HJH was supported in part by National NSF of China, 
and 
by the visiting grants of Harvard University and IAS Princeton.
JR was supported in part by the International Postdoctoral Exchange Fellowship Program of China.
WY was supported in part by the Office of Science, Office of High Energy Physics, of the
U.S.\ Department of Energy under contract DE-AC02-05CH11231.

\end{document}